\documentclass[prd,aps,twocolumn,a4paper,notitlepage,showkeys,nofootinbib]{revtex4-1}

\usepackage{graphicx,psfrag}
\usepackage{mathrsfs}
\usepackage{amsmath,amsfonts,amssymb}
\usepackage{multirow}
\usepackage{comment}
\usepackage{hyperref}
\usepackage{enumitem}
\usepackage{tcolorbox}
\usepackage[T1]{fontenc}
\usepackage[utf8]{inputenc}

\newcommand{\be}{\begin{equation}}
\newcommand{\ee}{\end{equation}}
\newcommand{\bea}{\begin{eqnarray}}
\newcommand{\eea}{\end{eqnarray}}
\newcommand{\bel}{\begin{align}}
\newcommand{\eel}{\end{align}}

\newcommand{\athena}[1]{\texttt{{#1}Athena++}}
\newcommand{\pittnull}{\texttt{PittNull}}
\newcommand{\teob}[1]{\texttt{TEOBResumS#1}}
\newcommand{\dali}{\teob{-Dal\'{i}}}

\newcommand{\orcid}[1]{\href{https://orcid.org/#1}{
\includegraphics[width=10pt]{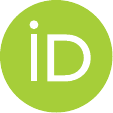}
}}

\DeclareSymbolFontAlphabet{\mathrsfs}{rsfs}
\DeclareMathAlphabet{\mathcal}{OMS}{cmsy}{m}{n}
\DeclareSymbolFontAlphabet{\mathrsfs}{rsfs}
\DeclareMathAlphabet\mathbfcal{OMS}{cmsy}{b}{n}

\def\l{\ell}
\def\lm{{\ell m}}

\def\non{\nonumber}                     
\def\half{\frac{1}{2}}

\def\i{{\rm i}}

\def\Msun{{\rm M_{\odot}}}
\def\Mo{{\Msun}}

\def\GMc2{{\rm G M_{\odot} c^{-2}}}

\def\half{\frac{1}{2}}
\def\de{\partial}
\def\be{\begin{equation}}
\def\ee{\end{equation}}
\def\beq{\begin{eqnarray}}
\def\eeq{\end{eqnarray}}

\def\t{\theta}

\def\gs{\sigma}
\def\gM{{}^{(4)}{g}}
\def\gMb{{}^{(4)}\overset{o}{g}}
\def\ie{{\it i.e.}}
\def\eg{{\it e.g.}}
\def\cf{{\it c.f.}}
\def\scri{{\mathcal{I}^+}}

\usepackage{pifont} 

\usepackage{color}
\definecolor{cyan}{rgb}{0,0.9,0.9}
\definecolor{orange}{rgb}{0.9,0.5,0}
\definecolor{magenta}{rgb}{1,0,1}
\definecolor{purple}{rgb}{0.8,0.4,0.8}
\definecolor{gray}{rgb}{0.8242,0.8242,0.8242}
\definecolor{light-gray}{gray}{0.95}

\begin{document}

\title{Covariant and Gauge-invariant Metric-based \\
  Gravitational-waves Extraction in Numerical Relativity} 

\author{Joan \surname{Fontbuté}$^1$\orcid{0009-0004-7893-7386}}\email{joan.fontbute@uni-jena.de}
\author{Sebastiano \surname{Bernuzzi}$^1$\orcid{0000-0002-2334-0935}}\email{sebastiano.bernuzzi@uni-jena.de}
\author{Simone \surname{Albanesi}$^{1,2}$\orcid{0000-0001-7345-4415}}
\author{David \surname{Radice}$^{3,4,5}$\orcid{0000-0001-6982-1008}}
\author{Alireza \surname{Rashti}$^{3,4}$\orcid{0000-0003-3558-7684}}
\author{William \surname{Cook}$^1$\orcid{0000-0003-2244-3462}}
\author{Boris \surname{Daszuta}$^1$\orcid{0000-0001-6091-2827}}
\author{Alessandro \surname{Nagar}$^{2,6}$\orcid{0000-0001-7998-2673}}

\affiliation{$^1$ Theoretisch-Physikalisches Institut, Friedrich-Schiller-Universit{\"a}t Jena, 07743, Jena, Germany}
\affiliation{$^2$ INFN Sezione di Torino, Via P. Giuria 1, 10125 Torino, Italy}
\affiliation{$^3$ Institute for Gravitation and the Cosmos, The Pennsylvania State University, PA 16802, USA}
\affiliation{$^4$ Department of Physics, The Pennsylvania State University, PA 16802, USA}
\affiliation{$^5$ Department of Astronomy \& Astrophysics, The Pennsylvania State University, PA 16802, USA}
\affiliation{$^6$ Institut des Hautes Etudes Scientifiques, 91440 Bures-sur-Yvette, France}

\date{\today}

\begin{abstract}
  We revisit the problem of gravitational-wave extraction in numerical
  relativity with gauge-invariant metric perturbation theory of
  spherical spacetimes.  
  Our extraction algorithm allows the computation of 
  even-parity (Zerilli-Moncrief) and odd-parity (Regge-Wheeler)
  multipoles of the strain from a (3+1) metric without the
  assumption that the spherical background is in Schwarzschild coordinates.
  The algorithm is validated with a comprehensive suite of 3D problems 
  including fluid ($f$-modes) and spacetime ($w$-modes) perturbations
  of neutron stars, gravitational collapse of 
  rotating neutron stars, circular binary black holes mergers 
  and black hole dynamical captures and binary neutron star mergers. 
  We find that metric extraction is robust in all the considered
  scenarios and delivers waveforms of overall quality similar to
  curvature (Weyl) extraction. Metric extraction is particularly valuable
  in identifying waveform systematics for problems in which the
  reconstruction of the strain from the Weyl multipoles is ambiguous. 
  Direct comparison of different choices for the gauge-invariant
  master functions show very good agreement in the even-parity
  sector. Instead, in the odd-parity sector, 
  assuming the background in Schwarzschild coordinates can minimize
  gauge effects related to the use of the $\Gamma$-driver shift.
  Moreover, for optimal choices of the extraction radius, a simple 
  extrapolation to null infinity can deliver waveforms compatible to 
  Cauchy-characteristic extrapolated waveforms.
\end{abstract}

\maketitle

\section{Introduction}
\label{sec:intro}

Numerical relativity is the main approach for the computation
of gravitational waves (GWs) from dynamical spacetimes with black holes
and neutron stars. Most of the successfull simulation methods rely on
the (3+1) formulation of general relativity, which turns Einstein
equations into a Cauchy problem (with constraints) for the description of
globally hyperbolic spacetimes, \eg~\cite{Gourgoulhon:2007ue}. The
computation of GWs from these numerically generated spacetimes is not
straightforward as it requires specific algorithms to identify the
gauge-invariant multipoles of the radiation from the metric fields. 
These procedures introduce part of the systematic uncertainties that
affect numerical waveforms.

Waveform extraction in (3+1) numerical-relativity~\footnote{
We strictly refer to simulations with Cauchy foliations.}
can be implemented using the metric-based perturbation theory of a
spherically symmetric spacetimes proposed by
\citet{Regge:1957td,Zerilli:1970se} and
\citet{Moncrief:1974am,Moncrief:1974bis}. The original formalism
adopts a background metric in Schwarzschild
coordinates and the Regge-Wheeler (RW) gauge for the
perturbations. Gauge-invariant perturbations were introduced in 
\cite{Moncrief:1974am,Moncrief:1974bis}. A covariant formalism
(spherical background in generic coordinates) 
perturbations was later proposed by \citet{Gerlach:1979rw} and further
developed by other authors
\cite{Gundlach:1999bt,MartinGarcia:2000ze,Sarbach:2001qq,Martel:2005ir}.

Gauge-invariant metric-based GW extraction algorithms have been
extensively used in numerical relativity, see
\eg~\cite{Abrahams:1995gn,Camarda:1998wf,Pazos:2006kz,Pollney:2007ss,Baiotti:2008nf,Buonanno:2009qa,Witek:2010xi,Reisswig:2012nc}.
Most of these algorithms assume Schwarzschild coordinates for the
background metric~\cite{Abrahams:1995gn,Camarda:1998wf},
although the latter do not necessarily match the coordinates of the
numerically generated spacetimes at finite extraction radii.
However, waveform extraction algorithms can be
formulated in {\it any} coordinate system for the background metric by
identifying the areal radius $r$ of the 2-spheres in the numerically
generated spacetime \cite{Sarbach:2001qq,Martel:2005ir}. 
\citet{Pazos:2006kz} consider a metric-based extraction algorithm
based on gauge-invariant and coordinate-independent \emph{odd parity}
perturbations (axial sector). Their (3+1)D simulations of a perturbed
Kerr-Schild black hole demonstrate that the coordinate-independent approach
has significant advantages over the use of Schwarzschild background. \citet{Baiotti:2008nf} consider a similar algorithm based on
gauge-invariant and coordinate-independent \emph{even parity}
perturbations (polar sector) and a preliminary application in (3+1)D
simulations of a perturbed non-rotating neutron star.
\citet{Buonanno:2009qa} (see their Appendix) present a binary black hole merger
waveform computed with a gauge-invariant and coordinate-independent
metric-based algorithm, but no details are given on the implementation.  
To our knowledge, algorithms based on the gauge-invariant and
coordinate-independent formalism and including both even and odd
perturbations have not been assessed for other problems.

Following the binary black hole
breakthrough~\cite{Pretorius:2005gq,Baker:2005vv,Campanelli:2005dd},
other GW extraction methods have gained more 
popularity than metric-based algorithms.
Algorithms using the Newmann-Penrose (NP, curvature-based) formalism 
\cite{Newman:1961qr} are routinely used to extract GWs at finite
coordinate radii in the ``wave zone'' of the simulated
spacetime, \eg~\cite{Baker:2001sf,Campanelli:2005ia,Brugmann:2008zz,Scheel:2008rj,Reisswig:2012nc,Cook:2016qnt}.
Direct comparisons of waveforms extracted with metric- and
curvature-based algorithms are discussed in
\eg~\cite{Pollney:2007ss,Baiotti:2008nf,Buonanno:2009qa,Reisswig:2010cd,Reisswig:2012nc,Cook:2017fec}
for various problems including isolated neutron stars, gravitational
collapse and black hole mergers. These works find that the two approaches
broadly agree, although curvature-based algorithms appear more robust
than metric-based algorithms. The latter are more susceptible to numerical
noise and gauge effects leading to waveforms with larger spurious
high-frequency oscillations. Curvature-based algorithms also have 
drawbacks. They compute the multipoles of Weyl's $\psi_4$ pseudoscalar,
which are asymptotically connected to the strain multipoles by
$\ddot{h}(t) = \psi_4(t)$. The time-integration of the latter
equation is known to be problematic: various sources of numerical
error in the simulations contribute to large secular non-linear drifts
in the integrated strain data (randomwalk effects for time-domain
integrations; spectral leakage in the frequency domain.) 
The frequency-domain integration with high-pass filters proposed
in~\cite{Reisswig:2010di} can significantly reduce spurious
effects in the oscillatory modes, but it requires the identification
of a cutoff frequency. Such frequency is easily found in terms of the
initial data orbital frequency for simulations of circular binary
mergers, but its determination for non-circular mergers and
binary scattering waveforms is ambiguous~\cite{Albanesi:2024xus}. As a result, the
uncertainty introduced by the choice of the cutoff frequency significantly
affects the waveform morphology in quantitative modeling studies
\cite{Albanesi:2024xus}. 
Similarly, nonoscillatory modes like those
dominating waveforms from gravitational collapse can be challenging
to integrate \cite{Reisswig:2012nc,Dietrich:2014wja}.

A common issue of the above metric- and curvature-based 
approaches is that the numerical 
waveforms are not rigorously computed at future null infinity ($\scri$).
Waveforms extracted at 2-spheres of finite coordinated radius should
be extrapolated to $\scri$~\footnote{
In general, the extrapolation procedure allows one to estimate an 
uncertainty of the finite-radius extraction.}
using an approximate retarded time,
\eg~\cite{Scheel:2008rj,Lousto:2010qx,Bernuzzi:2011aq,Nakano:2015pta}.  
Alternatively, GW extraction at $\scri$ can be performed with the
Cauchy-characteristic extraction (CCE) algorithm~\cite{Bishop:1996gt,Bishop:1997ik,Babiuc:2008qy,Reisswig:2006nt,Reisswig:2009rx,Moxon:2020gha}.
In this approach, (3+1)D data from a worldtube in the interior of the
simulated Cauchy domain is propagated to null infinity using a 
characteristic evolution (null-cones formulation) of Einstein equations.
Note that also CCE data retain a dependence on the
extraction radius of the worldtube data and is affected by systematics introduced by the timelike boundary of
the Cauchy evolution, \eg~\cite{Rashti:2024yoc}.

More consistent approaches to the computation of GWs 
require to directly incorporate $\scri$ in the computational domain,
thus adding significant mathematical and numerical complexity to the problem.
Numerical-relativity simulations with the Cauchy-characteristic
matching~\cite{dInverno:1996tcu,Bishop:1996gt,Bishop:1998uk,Reisswig:2009us,Ma:2023qjn}
achieve this goal by joining an inner world tube with a characteristic
evolution that provides both waveforms and boundary
conditions at $\scri$, see \cite{Winicour:2008vpn} for a review. While
significant progress has been made with Cauchy-characteristic 
matching, simulations of astrophysical scenarios are still limited. 
Yet another option is to consider evolution schemes for Einstein
equations with hyperboloidal foliations~\cite{Friedrich:1983,Frauendiener:1997zc,Frauendiener:1997ze,Frauendiener:1998yi,Frauendiener:2000mk,Zenginoglu:2008pw}.
Recent efforts in this direction appear
promising~\cite{Hilditch:2016xzh,Gasperin:2019rjg,Duarte:2022vxn,Peterson:2024bxk} 
but obtaining a robust hyperboloidal
framework for (3+1) numerical relativity remains an open problem.

In this paper we revist the problem of gravitational-wave extraction 
with the gauge-invariant and coordinate-independent metric-based
formalism. The motivation behind our work is threefold.
First, metric-based algorithms can be a valuable option to study
systematics effects in numerical waveforms especially for problems in
which $\psi_4$ integration is a main source of uncertainty and more
sophisticated methods may not be yet available or sufficiently robust.
Second, the performances of coordinate-independent metric-based
extraction in various computational problems relevant in astrophysical
simulations have not yet been studied in detail.
Third, the GW multipole data extracted here offer a  
way to investigate GW extraction at $\scri$ based on
hyperboloidal foliations. A companion paper describes a first effort
that employs a perturbative hyperboloidal evolution of worldtube data
to extrapolate GW to $\scri$.

The paper is organized as follows.
Section~\ref{sec:dec} gives a comprehensive description of gauge-invariant
coordinate-independent metric-based algorithms. General expressions
for the odd and even parity master functions are explicitely reported
for a background in generic coordinates and including time dependence.
Section~\ref{sec:algo} describes the wave extraction algorithm and the
different choices of the areal radius.
Section~\ref{sec:sim} describes our implementation in the \athena{GR-}
code and the setup of the (3+1)D simulations.
The following sections discuss numerical results for a suite of (3+1)D
problems: 
perturbations of neutron stars (Sec.~\ref{sec:tov}), gravitational collapse of
rotating neutron stars (Sec.~\ref{sec:coll}), circular
and eccentric binary black holes mergers (Sec.~\ref{sec:bbh}), and binary neutron star
mergers (Sec.~\ref{sec:bns}). 
The paper concludes with a summary of our findings and recommendations
on the use of metric-based extraction.

The paper uses geometric units $c=G=1$ unless explicitely specified.
All conventions are summarized in Appendix~\ref{app:not}.
Appendix~\ref{app:num} provides further details on the numerical
implementation.
Appendix~\ref{app:rextrap} provides further results on a simple and
yet effective extrapolation of metric waveforms to $\scri$.

\section{Multipolar decomposition of wave-zone metric and master functions}
\label{sec:dec}

In the metric-based extraction formalism, the wave-zone metric is decomposed
according to the gauge-invariant and coordinate-independent
perturbation theory of spherical spacetimes of
Regge-Wheeler-Zerilli-Moncrief and others (hereafter, RWZ). 
Here, we assume that the spherical background metric 
has been identified (see Sec.~\ref{sec:algo}) and 
discuss how to build the gauge-invariant Zerilli-Moncrief
(even-parity, $\Psi^{\rm (e)}_\lm$) and Regge-Wheeler (odd-parity,
$\Psi^{\rm (o)}_\lm$) master functions from a numerically generated
(3+1)-metric. 
The master functions are independent on the coordinates in which the
background spacetime is expressed. Gravitational waves are expressed
as a $s=-2$ spin weighted spherical harmonics expansion with the even
and odd parity master functions: 
\begin{align}
\label{eq:gw:rwz}
  h_+ - i h_\times &=\frac{1}{D_L}\sum_{\ell\geq2}\sum_{m=-\ell}^{\ell} h_\lm{}^{(-2)}Y_\lm\nonumber\\
   &=\frac{1}{D_L}\sum_{\ell\geq2}\sum_{m=-\ell}^{\ell}
  \sqrt{\frac{(\ell+2)!}{(\ell-2)!}}
  \left(\Psi^{\rm (e)}_\lm + i \Psi^{\rm (o)}_\lm\right){}{}^{(-2)}Y_\lm\ .
\end{align}

The 4-metric $g_{\mu\nu}$ of a perturbed spacetime
is given by the spherical background $\gMb_{\mu\nu}$ plus a
nonspherical perturbation, $h_{\mu\nu}$.
The background metric can be time-dependent and it is
fully characterized by the mass of the central object $M$, which is a
time-independent quantity, see Appendix~\ref{app:not}. 
The perturbation is decomposed in spherical tensor harmonics and, due
to the rotational symmetry of the background, decouples in even- and
odd-parity parts: 
\be\label{eq:pert_metric}
\gM_{\mu\nu} = \gMb_{\mu\nu} +
\sum_{\ell=2}^{\infty}\sum_{m=-\ell}^{\ell}\left( h^{(\lm)}_{{\rm
    even}\ \mu\nu} + h^{(\lm)}_{{\rm odd}\ \mu\nu}\right)\ . 
\ee
The even-parity perturbation is 
\begin{eqnarray}
\label{even:metric}
  h_{{\rm even}\ \mu\nu}^{(\lm)}
  =\begin{pmatrix}
   h_{AB}Y_{{\ell m}} & h_A Y^{(\lm)}_c \\ 
   h_A Y^{(\lm)}_c & r^2\left(K\gs_{ab}Y_\lm+ G Y^{(\lm)}_{ab}\right)\\
  \end{pmatrix}\ ,
\end{eqnarray}
and $h_{AB}$, $h_A$, $K$ and $G$ are the {\it even-parity multipoles}
with indeces $A,B=0,1$.
The suffix $(\lm)$ in these fields is omitted for brevity.
Each multipole is a function of the background coordinates $x^A$
of the Lorentzian background 2-metric $g_{AB}$. If one choses, for example, the areal (Schwarzschild) radius, then $x^1=r$.
We refer the reader to Appendix~\ref{app:not} for the definition of
the even-parity tensor spherical harmonics $Y_{\lm}, Y_a^{(\lm)}, Y_{ab}^{(\lm)}$
with indeces $a,b=2,3$.

The notation here mainly follows \citet{Gundlach:1999bt} and
\citet{Martel:2005ir}. The functions $h_A$ that are indicated as $j_A$
in \citet{Martel:2005ir}, while we follow here the original notation of
\citet{Regge:1957td}. Working in Schwarzschild coordinates,
\citet{Regge:1957td} use for the multipoles the symbols $H_0, H_1,
H_2$. The latter are related to ours via $h_{00}\equiv N H_0$,
$h_{01}\equiv H_1$ and $h_{11}\equiv H_2/N$, where $N$ is the
background function defined in Appendix~\ref{app:not}. The function
$K$ differs from the one used by \citet{Regge:1957td},
\be
K_{\rm RW} := K + \frac{\lambda}{2}G \ ,
\ee
because of the
different convention used there for the $Y_{ab}^{(\lm)}$ basis.
Note that the Regge-Wheeler gauge corresponds to the choice $h_A=0=G$. 

The odd-parity perturbation is 
\be
\label{odd:metric}
  h_{{\rm odd}\ \mu\nu}^{(\lm)}
  =\begin{pmatrix}
    0  & H_{A} S_c^{(\lm)} \\
    H_{A} S_c^{(\lm)} & H S_{ab}^{(\lm)}\\ 
\end{pmatrix}\ ,
\ee
where $H_A$ and $H$ are {\it odd parity multipoles}, and again $(\lm)$
is omitted.
We refer the reader to Appendix~\ref{app:not} for the definition of
the odd-parity tensor spherical harmonics $S_a^{(\lm)}, S_{ab}^{(\lm)}$.

The notation used here is close to \citet{Gundlach:1999bt} but uses
uppercases to avoid clashes with the even-parity functions and additional
superscripts. Comparing our notation to \citet{Martel:2005ir}, those
authors use the symbols $h_A$ and $h_2$ for our $H_A$ and $H$
respectively. In turn, \citet{Martel:2005ir} notation follows closely
the notation of \citet{Regge:1957td} except that the function $h_2$ in
the latter paper differs of a minus sign due to a sign difference in
the definition of $S_a^{(\lm)}$ (see Eq.~\eqref{eq:vspharm:S}.) Note that
the Regge-Wheeler gauge corresponds to the choice $H=0$. 

The multipoles $h_{AB},h_A,K,G,H_A,H$ are in general gauge
dependent. We detail below the construction of gauge-invariant
multipoles and the RWZ master functions.

\subsection{Gauge-invariant multipoles}
\label{sec:dec:multi}

For $\ell\geq 2$ it is always possible to build gauge-invariant metric
multipoles starting from the multipolar decomposition given
above. Using the notation originally introduced by
\citet{Gerlach:1979rw} (see also \citet{Gundlach:1999bt}), the
gauge-invariant \emph{even-parity} metric perturbations are given by ($(\lm)$
omitted) 
\begin{align}
\kappa_{AB} &:= h_{AB}-(\nabla_A p_B + \nabla_B p_A) \\
\kappa     &:= K_{\rm RW} - 2\dfrac{\nabla_Ar}{r} p_B g^{AB} \ ,
\end{align}
with the quantity
\begin{equation}
p_A:=h_A-\dfrac{1}{2}r^2\nabla_A G \ .
\end{equation}

In the most general case of a \emph{time-dependent
background}, the gauge-invariant multipoles can be expressed in terms
of those defined by Eq.~\eqref{even:metric}, 
\begin{widetext}
\begin{subequations}
\label{eq:k:even}
\begin{align}
\kappa      &= K_{\rm RW}-\dfrac{1}{r}\bigg\{\dot{r}\left(2h_0-r^2\de_tG\right)g^{00}+\left[\dot{r}\left(2h_1-r^2\de_rG\right)+\left(2h_0-r^2\de_tG\right)\right]g^{01}
+\left(2h_1-r^2\de_r G\right)g^{11}\bigg\}\ , \\
\kappa_{00} &= h_{00}-2\de_th_0+2\left(\Gamma^0_{00}h_0+\Gamma^1_{00}h_1\right)+2r\de_t r \de_t G + r^2\left(\de^2_tG-\de_tG\Gamma^0_{00}-\Gamma^1_{00}\de_rG\right)\ , \\
\kappa_{01} &= h_{01} - \de_th_1-\de_rh_0+r\dot{r}\de_rG+r\de_tG+2\left(\Gamma^0_{01}h_0+\Gamma^1_{01}h_1\right)+r^2\left(\de_t\de_rG-\Gamma^0_{01}\de_tG-\Gamma^1_{01}\de_r G\right)\ , \\
\kappa_{11}&=
h_{11}-2\de_rh_1+2\left(\Gamma^0_{11}h_0+\Gamma^1_{11}h_1\right)+2r\de_r
G + r^2\left(\de^2_r G-\Gamma^0_{11}\de_t G-\Gamma^1_{11}\de_r
G\right) \ , 
\end{align}
\end{subequations}
\end{widetext}
where $t$ is the global time coordinate, $r$ is the areal radius,
$\dot{r}=dr/dt$ and $\Gamma_{BC}^C$ are the Christoffel's
symbols of the background 2-metric $g_{AB}$. %
For completeness, we
explicitely write the gauge-invariant multipoles for the case of a
\emph{time-independent background}, 
\begin{equation}
\label{eq:k:time_indep}
\kappa = K_{\rm
  RW}-\dfrac{2}{r}\left[\left(h_1-\dfrac{1}{2}r^2\de_rG\right)g^{11}+\left(h_0-\dfrac{1}{2}r^2\de_t
  G\right)g^{01}\right] \ , 
\end{equation}
and
\begin{widetext}
\begin{subequations}
\label{eq:kAB}
  \begin{align}
\label{eq:k00}
\kappa_{00}&=h_{00}-2\de_t h_0 + 2\left(\Gamma^0_{00}h_0+\Gamma^1_{00} h_1\right)+ r^2\left(\de^2_t G-\Gamma^0_{00}\de_t G-\Gamma^1_{00}\de_r G \right) \ , \\
\label{eq:k01}
\kappa_{01}&=h_{01}-\de_rh_0-\de_th_1+2\left(\Gamma^0_{01}h_0+\Gamma^1_{01}h_1\right)
       + r\de_t G + r^2\left(\de_t\de_r G-\Gamma^0_{01}\de_t G-\Gamma^1_{01}\de_r G \right)\, \\
\label{eq:k11}
\kappa_{11}&= h_{11}-2\de_rh_1+2\left(\Gamma^0_{11}h_0+\Gamma^1_{11}h_1\right)+2r\de_r G + r^2\left(\de^2_r G-\Gamma^0_{11}\de_t G-\Gamma^1_{11}\de_r G\right) \ .
\end{align}
\end{subequations}
\end{widetext}
The tensor $\kappa_{AB}$ is called $\hat{H}_{ab}$ in
\cite{Sarbach:2001qq}, see \eg~their Eq.~(16), and $\tilde{h}_{ab}$ in \citet{Martel:2005ir}
(note also the different convention for the indeces). 

The gauge-invariant \emph{odd-parity} perturbation can be constructed
from the odd parity scalar $H$ and one-form $H_A$ perturbations
\cite{Gerlach:1979rw,MartinGarcia:2000ze},  
\be\label{eq:kA}
\kappa_A := H_A - \frac{1}{2}\nabla_A H + H\frac{\nabla_A r}{r} \ .
\ee
In terms of the multipoles of Eq.~\eqref{odd:metric}, the
gauge-invariant odd-parity multipoles are 
\begin{subequations}\label{eq:kodd}
\begin{align}
\label{eq:k0}
\kappa_0 &= H_0 - \frac{1}{2}\partial_t H + H \frac{\dot{r}}{r} \ ,\\
\label{eq:k1}
\kappa_1 &= H_1 - \frac{1}{2}\partial_r H + \frac{H}{r}\ .
\end{align}
\end{subequations}

\subsection{Even-parity master function}
\label{sec:dec:Psie}

We discuss here the computational of the Zerilli-Moncrief master function from the even-parity multipoles $h_{AB}$, 
$h_{A}$, $G$, $K$.
The even-parity master function implemented in most of the literature
follows the computation of
\citet{Moncrief:1974am,Moncrief:1974bis}. The function is
gauge invariant, but not coordinate independent, since its derivation
assumes that the background metric is explicitly written in
Schwarzschild coordinates. We briefly review its expression below.
In the most general case, the construction of the gauge-invariant
master function is explained in \citet{Sarbach:2001qq,Martel:2005ir},
which are the only papers where
such invariant expression is reported. Here we additionally present,
for the first time, its explicit expression in terms of the multipoles
$h_{AB},h_A,K,G,H_A,H$. These multipoles are those directly computed
from the 3-metric, as detailed in Sec.~\ref{sec:algo}.

For a background in Schwarzschild coordinates, the background
2-metric is $g_{01}=0$, $g_{00}\equiv-S:=-(1-2M/r)$ and
$g_{11}=-1/g_{00}$. The mass is then (\cf~Eq.~\eqref{eq:mass})
\be\label{eq:mass:Schw}
M=\dfrac{r}{2}\left(1-S\right) \ .
\ee
The even-parity master function reads
\begin{widetext}
\be\label{eq:Q+}
Q_+ =\dfrac{1}{\Lambda}\sqrt{\dfrac{2(\ell-1)(\ell+2)}{\ell(\ell+1)}}\bigg\{\ell(\ell+1)S\left(r^2\de_rG-2h_1\right)+ 2rS\left( Sh_{11}-r\de_rK\right)+r\Lambda K\bigg\} \ ,
\ee
where
\end{widetext}
\be
\Lambda=\left(\ell-1\right)\left(\ell+2\right)+\dfrac{6M}{r} \ 
\ee
Only the multipoles $h_{11}$, $h_1$, $G$ and $K$ are needed to compute
the even-parity master $Q_+$ function. Note that our $Q_+$ is
indicated as $Q$ in Refs.~\cite{Moncrief:1974am,Moncrief:1974bis}. 

The even-parity master function in generic coordinates and in terms of the
gauge-invariant multipoles is \begin{equation}
\label{eq:psi:even_a}
\Psi^{(\rm
  e)}=\dfrac{r}{\lambda}\left[\kappa+\dfrac{2r}{(\lambda-2)r+6M}\left(r^Ar^B
  \kappa_{AB}-r\, r^A\de_A \kappa\right)\right] \ ,
\end{equation}
where the suffix $(\lm)$ is omitted and the vector $r^A$ is defined in
Appendix~\ref{app:not}. This equation can be found in
\citet{Martel:2005ir} but it is equivalent to the combination of
Eqs.~(20),~(25),~(26) and~(27) of \citet{Sarbach:2001qq}. In the
latter reference, the even-parity function is simply called
$\Psi$. The latter differs of a minus sign with respect to ours,
\ie~$\Psi=-\Psi^{(\rm e)}$.
Since $r^A=\left(g^{01},g^{11}\right)$, Eq.~\eqref{eq:psi:even_a} can also be written as  
\begin{widetext}
\begin{equation}
\label{eq:psi:even_a_ext}
\Psi^{(\rm e)}=\dfrac{r}{\lambda}\left\{\kappa +\dfrac{2r}{(\lambda-2)r+6M}
\left[\left(g^{11}\right)^2\kappa_{11}+2g^{11}g^{01}\kappa_{01}+\left(g^{01}\right)^2\kappa_{00}-rg^{01}\de_t \kappa-rg^{11}\de_r \kappa\right]\right\} \ .
\end{equation}
\end{widetext}
The final expression for $\Psi^{(\rm e)}$ is obtained by replacing the
gauge-invariant variables $k$ and $\kappa_{AB}$ with the multipoles $h_{AB}$,
$h_{A}$, $G$, $K$. For a \emph{time-independent background}
and using Eq.~\eqref{eq:k:time_indep} one obtains
\begin{widetext}
\begin{align}
\label{eq:psi:even:stat}
\Psi^{(\rm e)}_{\rm static}&=\dfrac{r}{\lambda}\bigg\{K_{\rm RW}-\frac{2}{r}\left[\left(h_1-\dfrac{1}{2}r^2\de_r G\right)g^{11}
+\left(h_0-\dfrac{1}{2}r^2\de_t G\right)g^{01}\right] \nonumber \\
&+\dfrac{2r}{(\lambda-2)r+6M}
\bigg[\left(g^{01}\right)^2h_{00}+2g^{01}g^{11}h_{01}+\left(g^{11}\right)^2 h_{11}\nonumber\\
&+\dfrac{1}{r}\left[-r\left(g^{01}\right)^3\de_rg_{00}+2rg^{11}\left(g^{00}g^{11}\de_rg_{01}+\de_rg^{01}\right)
+g^{01}g^{11}\left(-2+2rg^{00}\de_r g_{00}+rg^{11}\de_r g_{11}\right)\right]h_0 \nonumber \\
&+\dfrac{1}{r}\left[2\left(g^{11}\right)^2\left(-1+rg^{01}\de_r g_{01}\right)
 +r \left(g^{11}\right)^3\de_rg_{11}+rg^{11}\left(g^{01}\right)^2\de_r g_{00}+2rg^{11}\de_r g^{11}\right]h_1\nonumber \\
&-\dfrac{1}{2}rg^{11}\left[2g^{11}\left(-1+rg^{01}\de_rg_{01}\right)+r\left(g^{11}\right)^2\de_rg_{11}
+r\left(g^{01}\right)^2\de_rg_{00}+2r\de_rg^{11}\right]\de_r G\nonumber \\
&+\dfrac{1}{2}r\left[r\left(g^{01}\right)^3\de_r g_{00}-2rg^{11}\left(g^{00}g^{11}\de_r g_{01}+\de_r g^{01}\right)
-g^{01}g^{11}\left(-2+2rg^{00}\de_rg_{00}+rg^{11}\de_r g_{11}\right)\right]\de_t G \nonumber \\
&-r g^{11}\de_r K_{\rm RW}-r g^{01}\de_t K_{\rm RW}
\bigg]\bigg\} \ .
\end{align}
\end{widetext}
Note that one needs to compute just first order derivatives  of the multipoles, \ie~$\de_r G$, 
$\de_t G$, $\de_r K$ and $\de_t K$, and the inverse background metric.
In the most general case of a \emph{time-dependent background}, the
even-parity master function reads
\begin{widetext}
  \begin{align}
    \label{eq:psi:even}
\Psi^{(\rm e)}&=\Psi^{(\rm e)}_{\rm static} \nonumber \\
& + \dfrac{r}{\lambda}\Bigg\{ -\frac{2\dot{r}}{r} \left[ \left( h_0 - \frac{1}{2}r^2\de_t G \right)g^{00} + \left( h_1 - \frac{1}{2}r^2\de_r G \right)g^{01}\right] \nonumber \\
& + \dfrac{2r}{(\lambda-2)r+6M} \Bigg[ \dot{r}\left[ \dot{r}\left(g^{00}\right)^2 + 2g^{00}g^{01} \right]h_{00} +2\dot{r}\left[ \dot{r}g^{00}g^{01}+\left(g^{01}\right)^2 +g^{00}g^{11}\right] h_{01} + \dot{r}\left[ \dot{r}\left(g^{01}\right)^2 + 2g^{01}g^{11} \right]h_{11} \nonumber \\
& + \dfrac{1}{r}\bigg[ 2r\left(g^{01}\right)^3 \de_t g_{01} + 2rg^{01}\de_tg^{01} + \left[ -r g^{00}\left( g^{11} \right)^2 + 2r g^{11}\left( g^{01}\right)^2 \right]\de_t g_{11} + r\left( g^{01} \right)^2 g^{00} \de_t g_{00} \bigg] h_0 \nonumber \\
& + g^{01}\bigg[\left(g^{01}\right)^2\de_tg_{00}+2g^{01}g^{11}\de_tg_{01}+\left(g^{11}\right)^2\de_tg_{11}\bigg]h_1 \nonumber \\
& - \dfrac{r^2g^{01}}{2}\bigg[2g^{01}g^{11}\de_tg_{01}+\left(g^{01}\right)^2\de_tg_{00}+\left(g^{11}\right)^2\de_tg_{11}\bigg]\de_r G\nonumber \\
& + \dfrac{r^2}{2}\bigg[-2\left(g^{01}\right)^3\de_tg_{01}-2g^{01}\de_tg^{01}-\left(g^{01}\right)^2g^{00}\de_tg_{00} +g^{11}\left(g^{11}g^{00}-2\left(g^{01}\right)^2\right)\de_tg_{11}\bigg]\de_t G \nonumber \\
& + \dfrac{\dot{r}}{r} \bigg[ r\dot{r} \left(g^{00} \right)^3 \de_t g_{00} + 2rg^{11}\de_r g^{00} + 2r g^{01} \left( \de_r g^{01} + \de_t g^{00} + \dot{r}\de_r g^{00} \right) + r\left(g^{01}\right)^3\left(2\de_t g_{11} + \dot{r}\de_r g_{11} \right)  + 2r g^{00} \de_t g^{01} \nonumber \\
& + 2\left(g^{01}\right)^2 (-2 + r g^{11}\de_r g_{11}) + \left(g^{00}\right)^2 \left[ -2\dot{r}^2 + r\dot{r} g^{01}\left( \de_r g_{00} + 2\de_t g_{01} \right) + 2r\left( g^{01} \de_t g_{00} + g^{11} \de_r g_{00} \right) \right] -2g^{00}g^{11}  \nonumber \\
& + \dot{r}g^{00} \left[ 2r\de_t g^{00} - 6g^{01} + r\left( g^{01} \right)^2 \left( 2\de_r g_{01} + \de_t g_{11} \right) \right] + 4rg^{00}g^{01}\left( g^{01}\de_t g_{01} + g^{11}\de_r g_{11} \right) \bigg] \left( h_0 - \dfrac{1}{2}r^2 \de_t G \right) \nonumber \\
& + \dfrac{\dot{r}}{r}\bigg[ -2\dot{r}^2 g^{00}g^{01} + 2r\left(g^{01} \right)^3 \de_r g_{00} + 2r\left(g^{01}\right)^2\left( g^{00} \de_t g_{00} +2g^{11}\de_r g_{01} \right) + 2rg^{11} \left( \de_r g^{01} + g^{00}g^{11}\de_t g_{11} \right) \nonumber \\
& + 2g^{01} \left[ r \left( \de_r g^{11} + \de_t g^{00} \right) + g^{11}\left( -3  + 2r g^{00}\de_t g_{01} \right) + r\left(g^{11}\right)^2 \de_r g_{11} \right] +r\dot{r}\left(g^{01}\right)^3\left( 2\de_r g_{01} - \de_t g_{11} \right)  \nonumber \\
& + r\dot{r}\left(g^{00}\right)^2 \left[ g^{01}\de_t g_{00} - g^{11}\left( \de_r g_{00} - 2\de_t g_{01} \right) \right] + \dot{r} \left( g^{01} \right)^2\left( -4 +rg^{11}\de_r g_{11}\right) + 2r\dot{r} g^{01} \de_r g^{01} \nonumber \\
& + 2\dot{r}g^{00} \left[ r\de_t g^{01} + r\left( g^{01} \right)^2 \de_r g_{00} -g^{11} + rg^{01}g^{11}\de_t g{11} \right] \bigg] \left(h_1 - \dfrac{1}{2}r^2 \de_r G \right) \nonumber \\
& - r\dot{r}\left( g^{00}\de_t K_{\rm RW} + g^{01}\de_r K_{\rm RW} \right) \nonumber \\
& + 2\ddot{r}\left(\dot{r}g^{00} + g^{01} \right)\left[ \left( h_0 - \frac{1}{2}r^2\de_t G \right)g^{00} + \left( h_1 - \frac{1}{2}r^2\de_r G \right)g^{01}\right] \Bigg] \Bigg\}
\end{align}
\end{widetext}
This $\Psi^{(\rm e)}$ is the ``natural'' invariant function, being
proportional to the multipole of the GW strain and differs from the
$Q_+$. The two master functions are related by a normalization factor
\be\label{eq:Q+:norm_psi}
Q_+=\sqrt{2(\ell+2)(\ell+1)\ell(\ell-1)}\Psi^{(\rm e)} =\sqrt{2\lambda(\lambda-2)}\Psi^{(\rm e)} \ .
\ee
Note also that $\Psi^{(\rm e)}$ can be computed in two ways.
Either one uses expression Eq.~\eqref{eq:psi:even}, where just first
order derivative (in time and space) of the metric multipoles $G$ and
$K$ appear, or one uses expression Eq.~\eqref{eq:psi:even_a}, where
the computation of the second time derivative of the metric multipoles
$G$ is required. In our implementation we opt for the first approach
and we set the terms proportional to $\ddot{r}$ to zero (see
Appendix~\ref{app:num}). 

\subsection{Odd-parity gauge-invariant multipoles}
\label{sec:dec:Psio}

We discuss here the computation of the Regge-Wheeler 
master function from the odd-parity multipoles $H_{A}$, $H$. 
The first calculation is presented in~\citet{Moncrief:1974am} and it is
specialized in Schwarzschild coordinates. Other definitions were given \eg~in
\citet{Cunningham:1978zfa}; see \citet{Martel:2005ir} and below for a
comparison of definitions. 

For a background expressed in Schwarzschild coordinates, the 
odd-parity master function of \citet{Moncrief:1974am} is 
\be\label{eq:Qx}
Q_\times := g^{AB}\kappa_B\frac{\nabla_A r}{r} =\frac{1}{r}\left(1-\frac{2M}{r}\right)\left[H_1 - \frac{r^2}{2}\partial_r\left(\frac{H}{r^2}\right)\right]\ .
\ee
Note that this quantity is called $Q$ in \cite{Moncrief:1974am}.

The odd-parity master function in terms of the gauge-invariant odd-parity multipoles is
\be\label{eq:psi:odd_a}
\Psi^{\rm (o)} = \frac{r^3}{\lambda-2} \epsilon^{BA}\nabla_A\left(\frac{\kappa_B}{r^2}\right) \ ,
\ee
where the suffix $(\lm)$ is omitted. By explicitly expanding the
covariant derivative, one obtains 
\be\label{eq:psi:odd_a_ext}
\Psi^{\rm (o)} = \frac{1}{\lambda-2} \frac{1}{\sqrt{g}}\left[
  r \left( \partial_t \kappa_1 - \partial_r \kappa_0\right)
  + 2\left(\kappa_0 - \kappa_1\dot{r} \right)\right] \ .
\ee
Interestingly, one finds the same expression in terms of the odd-parity multipoles used in the direct decomposition of the spacetime, 
\be\label{eq:psi:odd}
\Psi^{\rm (o)} = \frac{1}{\lambda-2} \frac{1}{\sqrt{g}}\left[ r \left(\partial_t H_1 - \partial_r H_0\right) + 2 \left(H_0 - H_1 \dot{r}\right)\right]\ .
\ee
In Schwarzschild coordinates~\footnote{
Or any time-independent background with $g=1$, \eg~Schwarzschild with the retarted/advanced time coordinates, $t\pm r$.}
Eq.~\eqref{eq:psi:odd} reduces to known result \cite{Nagar:2005ea}
\be\label{eq:psi:odd:stat}
\Psi^{\rm (o)}_{\rm static} = \frac{1}{\lambda-2} r\left[\partial_t H_1 - r^2 \partial_r \left(\frac{H_0}{r^2}\right)\right]\ .
\ee
This $\Psi^{(\rm o)}$ is the invariant function proportional to the
multipole of the GW strain.
Moncrief's $Q_\times$ is related to $\Psi^{\rm (o)}_{\rm static}$ by the equation
\be\label{eq:Qx:norm_psi}
\partial_t \Psi^{\rm (o)} = - Q_\times + \frac{16\pi}{\lambda-2}\frac{r}{\gMb_{rr}} T_1^{(\lm)}\ ,
\ee
where $T_1^{(\lm)}$ is the component of the odd-parity matter-source
vector, see Ref.~\cite{Nagar:2005ea} for more details. In Regge-Wheeler
gauge ($H\equiv0$), the $Q_\times$ master function coincides with the
original choice of \citet{Regge:1957td}. Note that $\Psi^{\rm (o)}$
and $Q_\times$ have different dimensions and satisfy Regge-Wheeler
equations that differ in the source terms. 
Yet another choice is the master function
$\tilde{\psi}\equiv\lambda(\lambda-2)\Psi^{\rm (o)}$ introduced in
\citet{Cunningham:1978zfa}. 

The master function $\Psi^{(\rm o)}$ can be computed in two
ways. Either using expression Eq.~\eqref{eq:psi:odd}, where just first
order derivative of the metric multipoles $H_0$ and
$H_1$ (in space and time respectively) appear, or Using
expression Eq.~\eqref{eq:psi:odd_a_ext}, where the computation of the
second time derivatives of the metric multipoles $\kappa_A$ is
required. In our implementation we opt for the first approach.

\section{Wave extraction algorithm}
\label{sec:algo}

We describe here the wave extraction algoritm to compute the master
functions for the GW strain, Eq.~\eqref{eq:gw:rwz}.
Assuming a numerical spacetime generated by a $(3+1)$D
numerical-relativity simulation in Cartesian coordinates
$X^i=(x,\,y,\,z)$, a wave-zone is first identified in the computational
domain. The wave zone is populated with extraction spheres at various
radii $R=\sqrt{x^2+y^2+z^2}$. The radial coordinate $R$ is
asymptotically interpreted as 
the isotropic radius of the asymptotically flat spacetime, and is not
the areal radius. The main steps of the algorithm are:
\begin{enumerate}[label=\alph*),nosep]
\item Interpolate the (3+1)-metric to the extraction sphere and
  transform to coordinates $(R,\theta,\phi)$; 
\item Compute the areal radius $r$ and perform the coordinate
  transformation to $(r,\theta,\phi)$;
\item Compute the background metric $g_{AB}$ at the extraction sphere;
\item Compute the even- and odd- parity (gauge-dependent) multipoles
  as tensor spherical harmonics projections of the (3+1)-metric at the
  extraction sphere;
\item Compute the gauge-invariant master functions from
  Eqs.~\eqref{eq:psi:even}, \eqref{eq:psi:even:stat}, \eqref{eq:Q+},
  \eqref{eq:psi:odd}, \eqref{eq:psi:odd:stat}, \eqref{eq:Qx}.
\end{enumerate}
Details about steps a)-d) are given in the following.
Tab.~\ref{tab:masterfuns} summarizes notation and properties of the
master functions considered in step e).

\begin{table}[t]
  \caption{\label{tab:masterfuns} Summary of RWZ master functions.}
\begin{center}
\begin{ruledtabular}
\begin{tabular}{l|cc|cc}
  Background & Even & Eq. & Odd &  Eq. \\
\hline
Schwarzschild & $Q_+$ & \eqref{eq:Q+} & $Q_\times$ & \eqref{eq:Qx}\\
Static & $\Psi^{\rm (e)}_{\rm static}$ & \eqref{eq:psi:even:stat} & $\Psi^{\rm (o)}_{\rm static}$ & \eqref{eq:psi:odd:stat}\\
General & $\Psi^{\rm (e)}$ & \eqref{eq:psi:even} & $\Psi^{\rm (o)}$ & \eqref{eq:psi:odd}
\\
\end{tabular}
\end{ruledtabular}
\end{center}
\end{table}

\subsection{(3+1)-metric at the extraction sphere}
\label{sec:algo:a}

The (3+1)-metric $\gamma_{ij}$, shift vector $\beta_i$ and lapse function
$\alpha$ in Cartesian coordinates are interpolated to the extraction
sphere with coordinates $(R,\theta,\phi)$ together with their
spatial Cartesian derivatives and time derivatives. The
components are then transformed to spherical polar coordinates using the
usual transformation matrix
\begin{equation}
O = \begin{pmatrix}
\sin\theta\cos\phi &  R\cos\theta\cos\phi & -R\sin\theta\sin\phi \cr
\sin\theta\sin\phi &  R\cos\theta\sin\phi &  R\sin\theta\cos\phi \cr
\cos\theta         & -R\sin\theta         &  0                   \cr
\end{pmatrix} \ .
\end{equation}
This gives the components
\be\label{eq:metric_comp_iso}
\alpha, \, \beta_R, \beta_\theta,\beta_\phi,\, 
\gamma_{RR}, \gamma_{R\theta}, \gamma_{R\phi}, \gamma_{\theta\theta}, \gamma_{\theta\phi},\gamma_{\phi\phi},\,
\ee
at the extraction sphere.
The radial derivatives with respect to $R$ are computed from the
Cartesian derivatives by defining the unit vector $n_i=X_i/R$ and, for
each function $f \equiv f(x,y,z)$, 
\begin{align}
&\de_R f =\delta^{ij}n_i\de_j f =\hat{n}_x\de_x f + \hat{n}_y\de_y f+\hat{n}_z\de_z f \ ,  \non\\ 
&\de^2_R f = \delta^{ik}\delta^{jl}n_in_j\de_{k}\de_l f \ . 
\end{align} 

\subsection{Areal radius}
\label{sec:algo:b}

Different options are available to define the areal radius
coordinate from the (3+1)-metric at the extraction sphere.
A first choice, which we simply call \emph{areal radius}, is to
consider the area of the 2D manifold endowed with metric
$\gamma_{ab}\equiv(\gamma_{\theta\theta},\gamma_{\theta\phi},\gamma_{\phi\phi})$,
\be\label{eq:r:areal}
r_{\rm areal}^2 = \frac{1}{4\pi}\int
\sqrt{\gamma_{\theta\theta}\gamma_{\phi\phi}-\gamma_{\theta\phi}^2}
d\theta d\phi \ . 
\ee
The expression assumes the manifold has the topology of a
sphere. Since this is not guaranteed, the argument of the square root
can become negative. Alternatively, \citet{Abrahams:1995gn} suggest to
consider just the diagonal component of the metric in the squared root
above, which we call \emph{areal-simple}.
Note, however, that this choice sets to zero terms $(\de_R
\gamma_{\theta\phi})^2$ in the Hessian (see later Eq.~\ref{eq:Hessian}).
Consequently, the even parity master function $\Psi^{\rm (e)}$
calculation gives unphysical amplitude already for Schwarzschild background. 

A second choice is the \emph{average Schwarzschild} radius defined in
\citet{Camarda:1998wf}, which we motivate in detail. The ``spherical
part'' of the 3-metric can be written as
\be
\gamma_{ab}\equiv r^2
\gs_{ab}+\sum_{\ell\geq1}\sum_{m=-\ell}^\ell\alpha_{\ell m}
Y_{ab}^{\ell m} \ ,
\ee
where $ \gs_{ab}$ is the 2-sphere metric (Appendix~\ref{app:not}) and 
the coefficients $\alpha_{\ell m}$ are functions of the radius
$r$. The average Schwarzschild radius is found by integrating on the
2-sphere and using $\gs_{ab}\gs^{ab}=2$ and orthogonality relations, 
\be\label{eq:r:avg}
r_{\rm avg}^2=\dfrac{1}{8\pi}\int
\left(\gamma_{\theta\theta}+\dfrac{\gamma_{\phi\phi}}{\sin^2\theta}\right)d\Omega
\ . 
\ee
This choice give rise to two other obvious options, which are each of the terms that 
conform the expression above. We call them $g\theta\theta$ and $g\phi\phi$:
\be\label{eq:r:gthth}
r_{g\theta\theta}^2=\dfrac{1}{4\pi}\int
\gamma_{\theta\theta}d\Omega
\ , 
\ee
\be\label{eq:r:gphph}
r_{g\phi\phi}^2=\dfrac{1}{4\pi}\int
\dfrac{\gamma_{\phi\phi}}{\sin^2\theta}d\Omega
\ . 
\ee
The performance of these choices will be tested for various spacetimes in Sec.~\ref{sec:sim}.

Having identified the coordinate $r$, the 3-metric components on the
extraction sphere, Eq.~\eqref{eq:metric_comp_iso} are transformed to
coordinates $(t,r,\theta,\phi)$, 
\begin{subequations}
\begin{align}
&\beta_r  = \dfrac{\de R}{\de r}\beta_R \, , 
\gamma_{rr}     = \dfrac{\de R}{\de r}\dfrac{\de R}{\de r}\gamma_{RR} \, , \non\\
&\gamma_{r\theta} = \dfrac{\de R}{\de r}\gamma_{R\theta} \, , \
\gamma_{r\phi}   = \dfrac{\de R}{\de r}\gamma_{R\phi} \, , 
\end{align}
together with the spatial derivatives, 
\begin{align}
&\de_r \gM_{00}  =\dfrac{\de R}{\de r}\de_R \gM_{00} \, , \
\de_r\beta_r   =\dfrac{\de^2 R}{\de r^2}\beta_R+\left(\dfrac{\de
  R}{\de r}\right)^2\de_R\beta_R \, , \non \\
&\de_r\gamma_{rr} = 2\left(\dfrac{\de R}{\de r}\right)\dfrac{\de^2R}{\de r^2}\gamma_{RR}+\left(\dfrac{\de R}{\de r}\right)^3\de_R\gamma_{RR} \ .
\end{align} 
and the time derivatives
\begin{align}
&\de_t\beta_r =\de_t\left(\dfrac{\de R}{\de r}\right)\beta_r + \dfrac{\de R}{\de r}\de_t \beta_R\non\\
&\de_t\gamma_{rr} = \left(\dfrac{\de R}{\de r}\right)^2\left[-2\dfrac{\de R}{\de r}\de_t\left(\frac{\de r}{\de R}\right)\gamma_{RR}+\de_t\gamma_{RR}\right] \ , \non\\
&\de_t\gamma_{r\theta} = \de_t\left(\dfrac{\de R}{\de r}\right)\gamma_{R\theta} + \dfrac{\de R}{\de r}\de_t\gamma_{R\theta} \ .
\end{align}
\end{subequations}
The above transformations require the inverse Jacobian and Hessian of
the $r(R)$ transformation as well as to keep track of the time
dependence in $r$. We illustrate these calculations for the average
Schwarzschild radius (for which we drop the label hereafter); similar
ones are performed for the other definitions of $r$. From
Eq.~\eqref{eq:r:avg}, 
one gets 
\begin{subequations}
  \begin{align}
  \dfrac{\de r}{\de R} &=\dfrac{1}{16\pi r}\int \left(\de_R\gamma_{\t\t}+\dfrac{\de_R \gamma_{\phi\phi}}{\sin^2\t}\right)d\Omega \\
  \dfrac{\de^2 r}{\de R^2} &=-\dfrac{1}{r}\left(\dfrac{\de r}{\de R}\right)^2+
  \dfrac{1}{16\pi r}\int \left(\de^2_R\gamma_{\t\t}+\dfrac{\de^2_R \gamma_{\phi\phi}}{\sin^2\t}\right)d\Omega \ \label{eq:Hessian}, 
  \end{align}
  and thus
\be
\dfrac{\de^2 R}{\de r^2}=-\left(\dfrac{\de R}{\de r}\right)^3\dfrac{\de^2 r}{\de R^2} \ .
\ee
For the time derivatives one computes
\be
\de_t\left(\dfrac{\de r}{\de R}\right) = -\dfrac{\dot{r}}{r}\dfrac{\de r}{\de R}+\dfrac{1}{16\pi r}\int\left(\de_t\de_R\gamma_{\theta\theta}+\dfrac{\de_t\de_R\gamma_{\phi\phi}}{\sin^2\theta}\right)d\Omega \ ,
\ee
and thus 
\be
  \de_t\left(\dfrac{\de R}{\de r}\right) = -\left(\dfrac{\de R}{\de r}\right)^2\de_t\left(\dfrac{\de r}{\de R}\right) \ .
\ee
The derivatives of $r(t)$ are
\begin{align}
\dot{r} &=\dfrac{1}{16\pi r}\int\left(\de_t\gamma_{\theta\theta}+\frac{\de_t\gamma_{\phi\phi}}{\sin^2\theta}\right)d\Omega \\
\ddot{r} &= -\dfrac{(\dot r)^2}{r}+\dfrac{1}{16\pi r}\int\left(\de^2_t\gamma_{\theta\theta}+\dfrac{\de^2_t\gamma_{\phi\phi}}{\sin^2\theta}\right)d\Omega \ .
\end{align}
\end{subequations}

\subsection{Background 2-metric}
\label{sec:algo:c}

Given the (3+1)-metric in spherical coordinates $(t,r,\theta,\phi)$ at
the extraction sphere, the background 2-metric is computed by
averaging on the sphere, \ie extracting the monopoles:
\begin{subequations}
\begin{align}
g_{00} &= -\dfrac{1}{4\pi}\int \left(\alpha^2-\beta_i\beta^i\right)\,d\Omega \ , \\
g_{01} &= \dfrac{1}{4\pi}\int \beta_r     \,d\Omega \ , \\
g_{11} &= \dfrac{1}{4\pi}\int \gamma_{rr} \,d\Omega \ .
\end{align}
\end{subequations}

\subsection{Projection of (3+1)-metric}
\label{sec:algo:d}

Given the (3+1)-metric and the background 2-metric in coordinates
$(t,r,\theta,\phi)$ it is possible to compute the multipoles at the
extraction sphere.
The even-parity multipoles in Eq.~\eqref{even:metric} are computed from the following projections 
\begin{widetext}
\begin{subequations}
\label{eq:multipoles:even}
\begin{align}
h_{00}^{(\lm)} & =-\int \left(\alpha^2-\beta_i\beta^i\right)Y_{\ell m}^*d\Omega \ , \\
h_{01}^{(\lm)} & = \int \beta_r Y_{\ell m}^* d\Omega \ ,\\
h_{11}^{(\lm)} & = \int \gamma_{rr} Y_{\ell m}^* d\Omega \,  \\
h_0^{(\lm)}    &=  \frac{1}{\lambda}\int\left(\beta_{\theta}\de_{\theta}Y^*_{\ell m}+\dfrac{\beta_{\phi}}{\sin^2\theta}\de_\phi Y^*_{\ell m} \right)d\Omega \ , \\
h_1^{(\lm)}    &= \frac{1}{\lambda}\int\left(\gamma_{r\theta}\de_\theta 
Y^*_{\ell m}+\dfrac{\gamma_{r\phi}}{\sin^2\theta}\de_{\phi}Y^*_{\ell m} \right)d\Omega \ , \\
K^{(\lm)}_{{\rm RW}} & = \dfrac{1}{2r^2}\int\left(\gamma_{\theta\theta}+\dfrac{\gamma_{\phi\phi}}{\sin^2\theta}\right) Y^*_{\ell m} d\Omega
+ \dfrac{1}{2r^2(\lambda-2)}\int \left[\left(\gamma_{\theta\theta}-\dfrac{\gamma_{\phi\phi}}{\sin^2\theta}\right)W^*_{\ell m}+\dfrac{2}{\sin^2\theta}\gamma_{\theta\phi}X^*_{\ell m}\right]d\Omega \ , \\
G^{(\lm)}  & = \dfrac{1}{r^2\lambda(\lambda-2)}\int \left[\left(\gamma_{\theta\theta}-\dfrac{\gamma_{\phi\phi}}{\sin^2\theta}\right)W^*_{\ell m}+\dfrac{2}{\sin^2\theta}\gamma_{\theta\phi}X^*_{\ell m}\right]d\Omega \ .
\end{align}
\end{subequations}
\end{widetext}
where we have defined the functions \cite{Regge:1957td} 
\begin{widetext}
\begin{subequations}
\begin{align}
  W_{\ell m}&:=
  \frac{1}{\sin\theta}\nabla_{(\phi}S^{(\lm)}_{\theta)}
  = \de_{\theta}^2Y_{\ell m} - \cot\theta\de_{\theta}Y_{\ell m} - \dfrac{1}{\sin^2\theta}\de^2_{\phi}Y_{\ell m}
  = \left(2 \de_{\theta}^2 +\lambda\right)Y_{\ell m} 
  \ , \\
  X_{\ell m}&:= - \sin\theta\left(\nabla_\theta S_\theta^{(\lm)}-\frac{\nabla_\phi S_{\phi}^{(\lm)}}{\sin^2\theta}\right)
  = 2\left(\de^2_{\theta\phi}Y_{\ell m}-\cot\theta \de_{\phi}Y_{\ell m}\right) \ .
\end{align}
\end{subequations}
\end{widetext}
The odd-parity multipoles in Eq.~\eqref{odd:metric} are computed from the following projections 
\begin{widetext}
\begin{subequations}
  \label{eq:multipoles:odd}
\begin{align}
  H_0^{(\lm)}    &= \frac{1}{\lambda}\int \frac{1}{\sin\theta}  \left( %
  - \beta_\theta \partial_\phi Y_{\lm}^{*} + \beta_\phi \partial_\theta Y_{\lm}^{*}\right) d\Omega \ , \\
  H_1^{(\lm)}    &= \frac{1}{\lambda}\int \frac{1}{\sin\theta}  \left( %
  - \gamma_{r\theta} \partial_\phi Y_{\lm}^{*} + \gamma_{r\phi} \partial_\theta Y_{\lm}^{*}\right) d\Omega\ , \\
  H^{(\lm)} &= \frac{1}{\lambda(\lambda-2)}\int\left[
    \frac{1}{\sin\theta}\left(-\gamma_{\theta\theta}+\frac{\gamma_{\phi\phi}}{\sin^2\theta}\right) X_\lm^*
    + \frac{2\gamma_{\theta\phi}}{\sin\theta} W_\lm^*
  \right]d\Omega \ .
\end{align}
\end{subequations}
\end{widetext}
Similar projections are performed on the time and radial derivatives
of the (3+1)-metric in order to obtain time and radial derivatives of
the multipoles. The latter are required to construct the master functions
in Eqs.~\eqref{eq:psi:even}, \eqref{eq:psi:even:stat}, \eqref{eq:Q+},
\eqref{eq:psi:odd},\eqref{eq:Qx}.
Note that we do \emph{not} subtract the background metric. Although
this may improve the multipole computation in simulations at low
resolutions, we rely exclusively on the orthogonality of spherical
harmonics for the projections.

\section{Simulations}
\label{sec:sim}

We implement the above algorithm in \athena{GR-}~\cite{Daszuta:2021ecf,Cook:2023bag,Daszuta:2024ucu}.
\athena{GR-} implements modules for solving the Z4c free-evolution
scheme of Einstein equations~\cite{Bernuzzi:2009ex,Hilditch:2012fp} 
coupled to radiation magnetohydrodynamics equations in Eulerian (3+1)
conservative formulation. In this work, we consider simulations involving only hydrodynamics and metric evolution; for the gauge sector we adopt the moving puncture 
gauge implemented as described in~\cite{Daszuta:2021ecf}.

\athena{GR-} uses a Cartesian mesh and a block-based adaptive mesh
refinement (AMR) strategy, in which the computational domain is
divided into mesh block (MB). The MBs can be individually refined,
splitting into (in 3D) 8 new MBs with double linear resolution. Each
MB exists on a single refinement level. Our simulations are peformed
in 3D without symmetries. Therefore, the grid is defined by the number
of points along each dimension for the coarsest level $N_M$, 
the number of refinement levels $0,...,N_L$,
and the number of points along each dimension in the MB. The
latter is fixed to $N_B=16$ for all problems.
We use the cell-centered representation for spacetime
variables and, for binary problems, the AMR criterion described in
\citet{Rashti:2023wfe} ($L_2$-norm based, used for black hole cases) and \citet{Daszuta:2024ucu}
($L_\infty$-norm based, used for matter cases).

Metric variables are discretized using finite differencing operators
of fourth order accuracy for problems involving neutron stars and
sixth order accuracy for problems with binary black holes.
Kreiss-Oliger dissipation operators are utilized to suppress numerical
noise with parameter $\sigma$.
Finite volume (high resolution shock capturing) methods based on the
local Lax-Friedrichs flux and WENOZ reconstruction are used for
the discretization of hydrodynamics equations following
closely~\cite{Cook:2023bag}. Time integration of all equations is performed explicitely via the
method of lines, using a low storage fourth-order Runge-Kutta with
fixed timestep for all levels and subject to a Courant (CFL) condition.  

Gravitational waves are extracted at coordinate spheres. \athena{GR-}
implements both spherical geodesic grids and 
latitude-longitude spheres. The former are employed for the extraction
of the Weyl $\psi_4$ pseudoscalar using a 
resolution of 9002 vertices ($n_Q=30$) \cite{Daszuta:2021ecf}.
The latter are used for the RWZ extraction.
Output for CCE is also dumped at the same
spheres.
Waveforms and other quantities are shown as a function of the
retarded time $u = t - r_*$, where $r_*=r_*(r(R))$ is the tortoise
coordinate associated to the isotropic radius of the extraction sphere
and the Arnowitt Deser Misner (ADM) mass of the spacetime is employed
in the transformation.

The latitude-longitude grids for the RWZ extraction are
evenly spaced in the longitude coordinate $\phi$ while use
Gauss-Legendre nodes for the coordinate $\theta$. 
For the simulations presented here, we choose $N_\theta=N_\phi/2=128$
nodes in each direction in order to obtain good quality waveforms for
all the different problems considered at a sufficiently large
extraction distance. %
Integrals on the 2-sphere are computed with Gauss-Legendre
quadratures; other details are given in Appendix~\ref{app:num}.
CCE waveforms are computed using the \pittnull~code
\cite{Bishop:1998uk,Babiuc:2010ze}. These CCE data are in terms of
Weyl's pseudoscalar or News function, and thus require integration to
compute the RWZ multipoles.
The strain multipoles from both $\psi_4$ multipoles and the CCE
News multipoles is calculated using both the fixed frequency
integration (FFI)~\cite{Reisswig:2010di} method, or the direct time
integration (DTI), \eg~\cite{Berti:2007fi,Baiotti:2008nf}.
The former method requires the identification of a cutting frequency
$f_0$ for the high-pass filters applied during the frequency domain
integration.
The latter method introduces a drift in the integrated multipole that
requires a polynomial corrections. 
These choices are detailed and discussed below for each problem considered.

The mesh setups in this work are specified by prescribing a base 
cubic grid $\Omega=[x_{\min},x_{\max}]^3$ (refinement level zero)
covered by $N_M$ points per direction and with resolution 
$h_0=(x_{\max}-x_{\min})/N_M$ in each direction. 
$N_L$ static or adaptive mesh refinements are added from the
beginning of the simulation;
the highest resolution is achieved at the highest refinement level,~$h_L$.
Tab.~\ref{tab:sim:setup} summarizes the \athena{GR-} setup for all
the problems discussed in this work. The simulated spacetimes are: 
Tolman-Oppenheimer-Volkoff (TOV, static and spherically symmetric)
stars with a fluid or spacetime perturbation, gravitational collapse
of a rotating neutron star (RNS), binary black hole (BBH) circular
merger and dynamical capture and a binary neutron star (BNS) circular
merger.

\begin{table*}[t]
  \caption{\label{tab:sim:setup} Simulations setup. For each problem
    the columns report: the extension of refinement level zero $\Omega$ (lenghts are in units of $\Mo$),
    the number of points per direction $N_M$,
    the number of refinement levels $N_L$,
    the wave-extraction radii $R$ (only where the CCE data is dumped, in which all the algorithms can be compared),
    the corresponding resolution at the extraction radii $\delta R$,
    the CFL factor and the Kreiss-Oliger dissipation parameter $\sigma$.}
\begin{center}
\begin{ruledtabular}
\begin{tabular}{lccccccc}
  Problem & $\Omega$ & $N_M$ & $N_L$ & $R$ $[\Mo]$& $\delta R$ $[\Mo]$ & CFL & $\sigma$ \\
\hline
TOV & $[-512,512]^3$ & $64,96,128$ & 6 & $150,250,350,450$ & $4,4,8,8$ & 0.25 & 0.5 \\
RNS & $[-512,512]^3$ & $128$ & 7 & $150,250,350,450$ & $8,8,8,8$ & 0.2 & 0.1 \\
BBH & $[-1536,1536]^3$ & $128$ & 10 & $100,140,220,400$ & $3,3,6,12$ & 0.25 & 0.2 \\
BNS & $[-1024,1024]^3$ & $160$ & 6 & $200,400,600,800$ & $3.2,6.4,12.8,12.8$ & 0.25 & 0.5 \\
\end{tabular}
\end{ruledtabular}
\end{center}
\end{table*}

\section{Perturbed TOV spacetimes}
\label{sec:tov}

As a first test of the extraction algorithm we consider the evolution
of a TOV star which is 
perturbed in either the even or the odd sector using a fluid or a 
spacetime perturbation respectively. The initial perturbation
excites the star's proper modes and the emission of GWs. We compare
systematically waveforms computed with the metric extraction algorithm
to those from the NP algorithm and CCE waveforms. 
The initial data coordinates are Schwarzschild isotropic coordinates
and the puncture gauge keeps the coordinates very close to the initial
ones during the evolution. Therefore, this test also allows to probe 
the consistency of the different RWZ master functions in the scenario
of a static background in known coordinates.

The \athena{GR-} setup for this test is as follows (see also Tab.~\ref{tab:sim:setup}).
The refinement level zero covers the domain 
$\Omega=[-512 M_\odot,512 M_\odot]^3$ with $N_M=(64,96,128)$ points per direction. 
Six static mesh refinement (SMR) levels are employed to resolve the star 
up to a maximum resolution $h_6=0.125 M_\odot$ over the entire star.
The CFL factor is set to $0.25$ and the Kreiss-Oliger dissipation
parameter is $\sigma=0.5$. The extraction spheres for the RWZ and NP
algorithms are located at 
$R=(75,100,150,200,250,300,350,400,450)M_\odot$, 
while the world-tube output for the CCE is dumped 
at $R=(150,250,350,450)M_\odot$.

\subsection{Even-parity perturbations}
\label{sec:tov:even}

We consider a TOV star of gravitational mass $M_{\rm TOV}=1.36M_{\odot}$
and whose matter is described by the piecewise polytropic fit to
the SLy equation of state (EOS) \cite{Read:2008iy}. The isotropic radius of the
star is $R_{\rm TOV}=7.76M_\odot$.
Even-parity modes are excited by superimposing an axisymmetric pressure 
perturbation to the initial TOV data~\cite{Baiotti:2008nf}, 
\be\label{eq:tov:pert_even}
\delta p(R,\theta) = (p + e)A\sin \left(\frac{\pi R}{2R_{\rm TOV}}\right)Y_{\ell 0}(\theta) \ , 
\ee
where $p$ ($e$) is the TOV pressure (energy density) and we choose 
$A=0.005$ and $\ell=2$. 

\begin{figure*}[t]
  \centering
  \includegraphics[width=\textwidth]{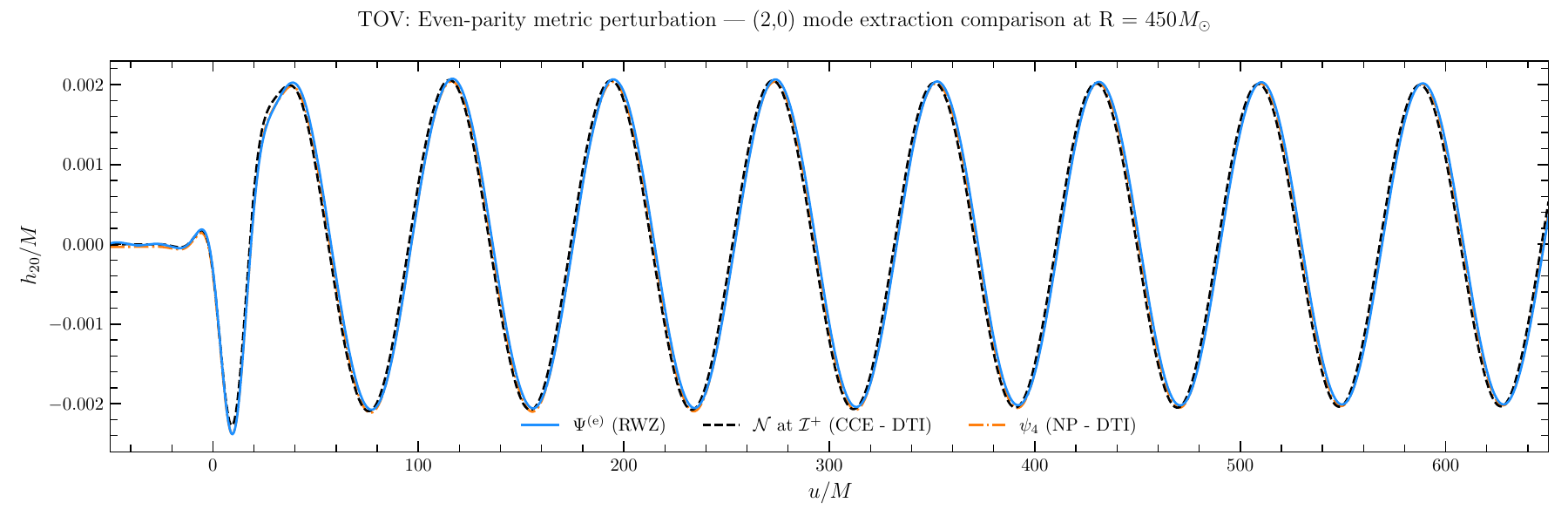}
  \caption{Strain $(2,0)$ mode for the TOV problem with pressure 
           perturbation (Eq.~\eqref{eq:tov:pert_even})
           extracted at $R=450M_\odot$. Different lines refer to 
           different extraction algorithms: RWZ (solid blue), CCE (dashed black) and NP 
           (dotted-dashed orange). The mode is dominanted by the
           $f$-mode frequency of the star. Data refer to the highest
           resolution simulation.}
  \label{fig:tov:even_pert_leading}
\end{figure*}

The GWs from these simulations are approximately monocromatic signals
of frequencies corresponding to the $f$-modes excitations.
Fig.~\ref{fig:tov:even_pert_leading} shows the dominant $(2,0)$ mode
of the strain as computed from RWZ and NP multipoles extracted at
$R=450 M_\odot$ and compared to CCE data evolved from the same
extraction radius.
Throughout this paper we will represent the dynamical background in general coordinates RWZ 
wave in solid blue, the CCE in dashed black and the NP in
dotted-dashed orange.
The figure shows a good agreement between the different extraction
methods; CCE waveforms are slightly dephased with respect to finite
extraction. This dephasing increases at smaller extraction radii (not
shown) and fixed resolution.
We mention that, due to quasi-monocromatic character of the signal,
the strain from the $\psi_4$ and CCE News waveforms can be
obtained using both the FFI (with $f_0\lesssim f$) and DTI method. In
the following, we only consider DTI integrated waveforms using a linear
polynomial subtraction applied to the entire  signal.

\begin{figure}[t]
  \centering
  \includegraphics[width=0.5\textwidth]{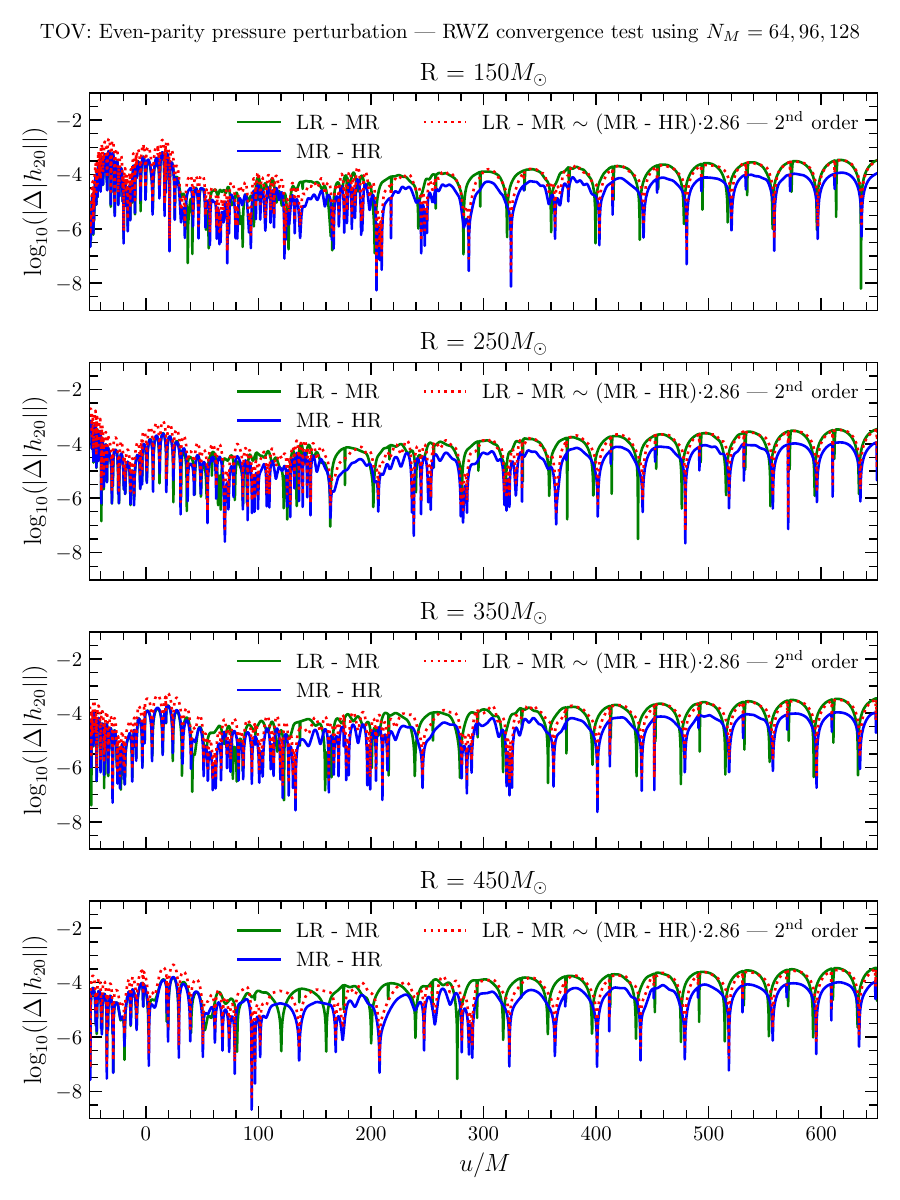}
  \caption{Convergence test of the RWZ $(2,0)$ mode for the TOV problem with pressure 
           perturbation (Eq.~\eqref{eq:tov:pert_even}).
           Different solid lines refer to differences between data from
           simulations at successive resolutions $N_M=64,96,128$
           labelled LR, MR, HR. The dotted red line is the 
           blue line rescaled by second order convergence.
           Different panels show representative extraction radii.}
  \label{fig:tov:even_pert_convergece}
\end{figure}

We first demonstrate convergence of the RWZ waveform by
considering simulations with 
$N_M = 64, 96, 128$~points which correspond to a maximal resolutions
$h_6 = 0.25, 0.167, 0.125M_\odot$. The three meshes are indicated as
LR, MR and HR. Focusing on the $(2,0)$ RWZ $\Psi^{\rm (e)}$ mode, the
differences of data from simulations at successive resolutions are
shown in Fig.~\ref{fig:tov:even_pert_convergece}. The plot also shows
the differences MR-HR rescaled by the appropriate factor for 2nd order
convergence. 
Waveforms at all extraction radii show consistent 2nd order convergence,
compatible with the truncation error of hydrodynamics.

\begin{figure}[t]
  \centering
  \includegraphics[width=0.5\textwidth]{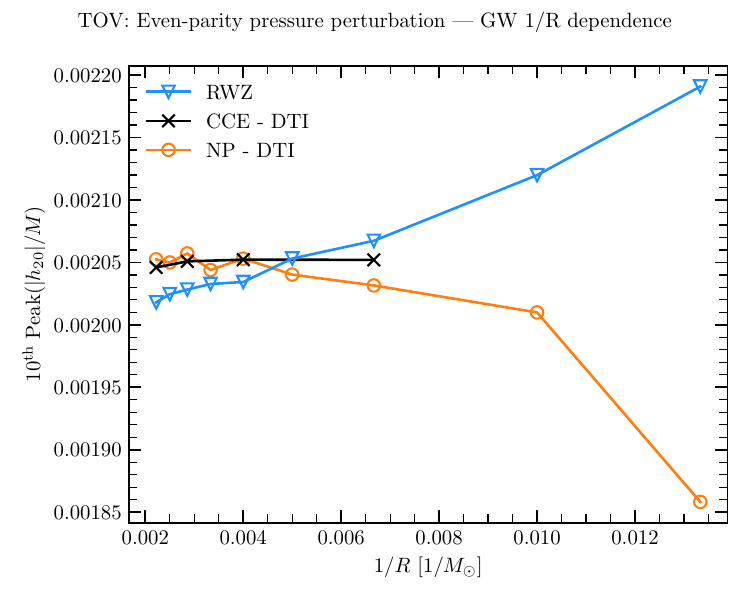}
  \caption{Extraction radius dependence of the amplitude's $10^{\rm th}$ peak of the strain
    $(2,0)$ mode for the TOV problem with pressure  perturbation
    (Eq.~\eqref{eq:tov:pert_even}).
    Different lines refer to different extraction algorithms: RWZ (blue), CCE (black) and NP 
    (orange). The dotted grey vertical line marks
    the radius with the best overall compatibility.
    Data refer to the highest resolution simulation.}
  \label{fig:tov:even_pert_1/R}
\end{figure}

Finite-radius extraction effects on the waveform amplitude are
summarized in Fig.~\ref{fig:tov:even_pert_1/R}, which shows
the $10^{\rm th}$ amplitude peak of waveforms as function of the
inverse extraction radii. The peak amplitude converges to the CCE peak
the larger the extraction radius is and for both RWZ and NP
multipoles. The maximum amplitude differences remain below the $2\%$
in the range $R\in [150,450] M_\odot$ for all extraction algorithms.
The RWZ multipole shows larger deviations from CCE than the NP
multipole, but such deviations are smaller than mesh resolution
effects. 
Inspection of the full waveform indicates that the best extraction
radius is the largest considered, at which data from the various
algorithm are most compatible with each other.

\begin{figure*}[t]
  \centering
  \includegraphics[width=\textwidth]{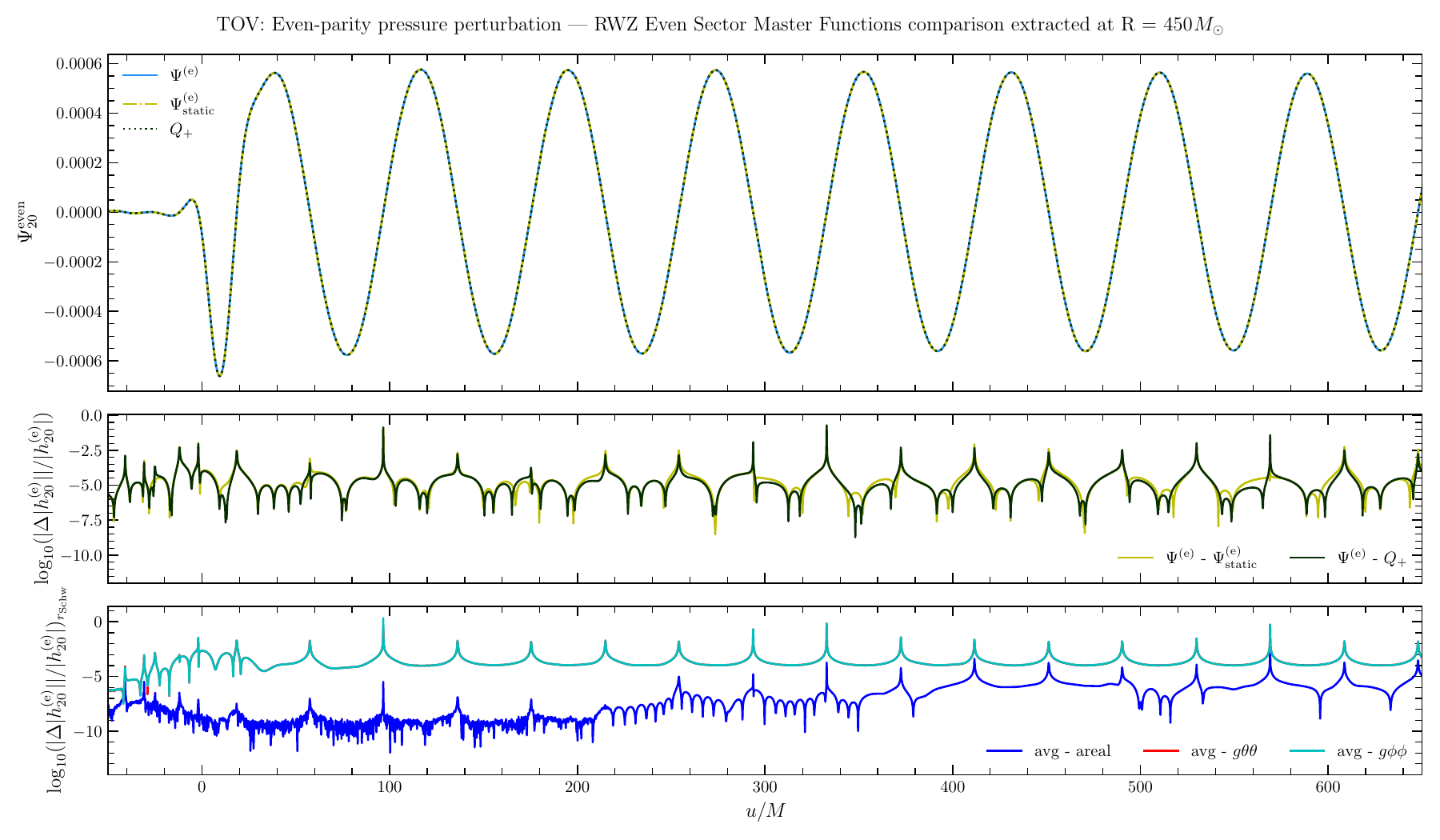}
  \caption{RWZ $(2,0)$ mode for the TOV problem with pressure 
           perturbation (Eq.~\eqref{eq:tov:pert_even})
           extracted at $R=450M_\odot$ computed using different RWZ
           master functions.
           Top:
           Even-parity master function extraction comparison $\Psi^{\rm (e)}$ 
           (solid blue), $\Psi_{\rm static}^{\rm (e)}$ (dash-dotted light green) and 
           $Q_{+}$ (dotted dark green) normalized as per
           Eq.~\eqref{eq:Q+:norm_psi}.
           Middle:
           Amplitude relative 
           differences with respect to $\Psi^{\rm (e)}$.
           Bottom: $\Psi^{\rm (e)}$ 
           amplitude relative differences with respect different
           choices of the Schwarzschild radius definitions
           with respect to the average Schwarzschild radius choice.
           Data refer to the highest resolution simulation.}
  \label{fig:tov:even_pert_master_funcs}
\end{figure*}
\begin{figure}[t]
  \centering
  \includegraphics[width=0.5\textwidth]{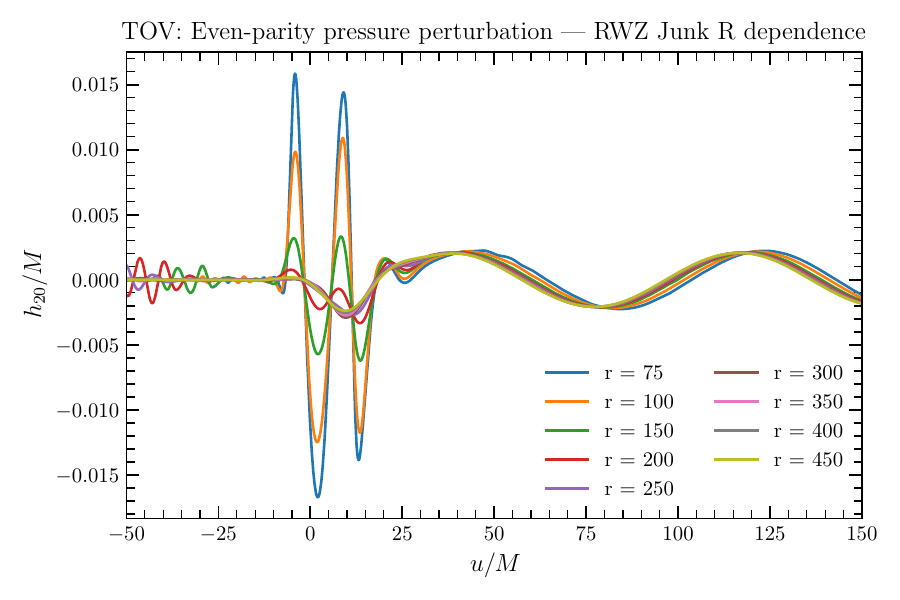}
  \caption{Strain $(2,0)$ mode for the TOV problem with pressure 
           perturbation (Eq.~\eqref{eq:tov:pert_even})
           extracted at different radii (in units of $M_\odot$) close to the initial junk radiation.}
  \label{fig:tov:even_pert_junk}
\end{figure}

We compare the RWZ different even-parity master functions
$\Psi^{\rm (e)}_{20}$ (Eq.~\ref{eq:psi:even}, blue), $\Psi_{\rm
  static\, 20}^{\rm (e)}$
(Eq.~\eqref{eq:psi:even:stat}, dash-dotted light green) and $Q_{+\, 20}$
(Eq.~\eqref{eq:Q+} normalized as per Eq.~\eqref{eq:Q+:norm_psi}, dotted dark green) in
Fig.~\ref{fig:tov:even_pert_master_funcs}.
Relative amplitude differences with respect to $\Psi^{\rm (e)}$ are around $10^{-5}$ 
for both $\Psi_{\rm static}^{\rm (e)}$ and $Q_{+}$. This sets the
level of consistency of three methods up to truncation errors for a
static background.
The bottom panel of Fig.~\ref{fig:tov:even_pert_master_funcs} 
also shows the amplitude relative differences in $\Psi^{\rm (e)}$ due to
different choices for the Schwarzschild radius used to build the
background for the master functions. The different choices are tested
by running otherwise identical simulations at resolution $N_M=128$.
The areal and average Schwarzschild radius definitions provide
waveforms compatible to ${\lesssim}10^{-5}$ level, while the
$g\theta\theta$ and $g\phi\phi$ choices (compatible with each other)
introduce larger differences.  
These differences originates from numerical errors in the different
integral expressions for the Schwazrschild radius (see below). 
The choice of Schwarzschild radius is anyway an uncertainty
significantly smaller than the choice of extraction algorithm and of the
extraction radii.

\citet{Baiotti:2008nf} reported that junk radiation in RWZ
waveforms from perturbed TOV can unphysically increase with the
extraction radius. They argue that numerical uncertainties are a
posisble cause of this behaviour, see their Fig.~13 and related
discussion. As demonstrated by 
Fig.~\ref{fig:tov:even_pert_junk}
we do not observe such an issue in our implementation: the junk
radiation converges with the radius of the extraction sphere.
The main numerical improvements in our algorithm are the interpolation
to the extraction sphere (here performed at high order) and the
computation of the projections using Gauss-Legendre quadrature, see
Appendix~\ref{app:num}. 
 
\begin{figure*}[t]
  \centering
  \includegraphics[width=\textwidth]{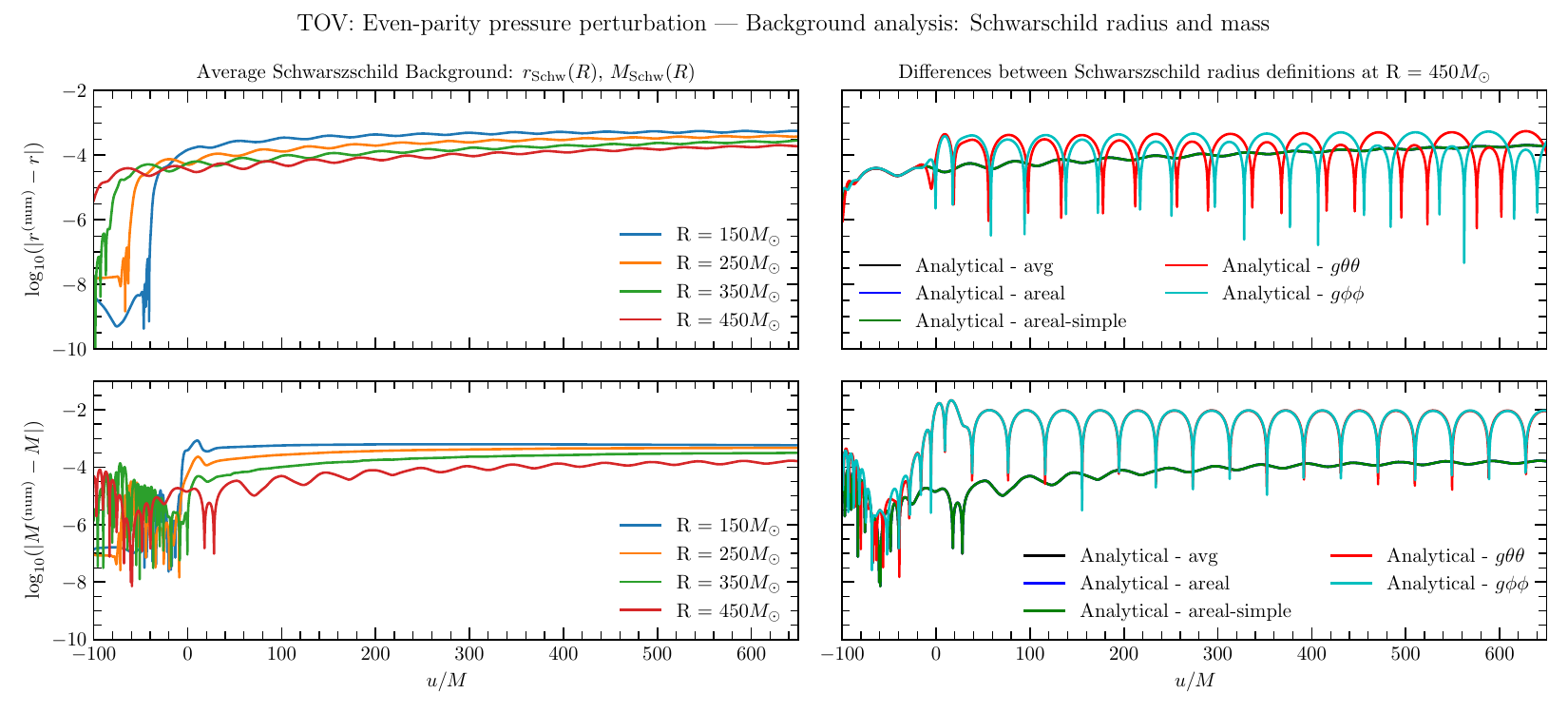}
  \caption{Background mass and Schwarzschild radius computation in
    the TOV problem with pressure perturbation (Eq.~\eqref{eq:tov:pert_even}).
    Left: Differences between the analytical and numerical 
    background Schwarzschild radius (top) and mass (bottom)
    as a function of retarded time and different isotropic extraction radii $R$.
    The average Schwarzschild radius is employed in this simulation.
    Right: Different lines show the same differences as the left
    panels but compare simulations with different choices for the 
    Schwarzschild radius.
    Data are compared at $R=450 M_\odot$.
    All data refer to the highest resolution simulation.}
  \label{fig:tov:even_pert_background}
\end{figure*}

Finally, the TOV spacetime simulations allow to investigate the numerical computation of
the spherical background in the metric extraction algorithm. In
particular, we consider a comparison of the estimated Schwarzschild
mass given by Eq.~\ref{eq:mass} with the TOV mass $M_{\rm TOV}$ and of the
analytical Schwarzschild radius 
\be
  r =   R\left(1+\frac{M_{\rm TOV}}{2R}\right)^2 \ \label{eq:rad:iso_to_schw}, 
\ee
with the possible choices discussed above.
The left panels of Fig.~\ref{fig:tov:even_pert_background} show that
both the mass and the Schwarzschild radius are affected by 
uncertainties of order ${\lesssim}10^{-4}$. The uncertainties decrease
with larger extraction radii. 
The right panels of Fig.~\ref{fig:tov:even_pert_background} show
differences in the mass and Schwarzschild radius introduced by
different choices of the latter in the metric extraction algorithm. 
The average Schwarzschild, areal and areal-simple prescriptions
provide practically identical results and perform 
better than the $g\theta\theta$ and $g\phi\phi$ prescriptions.
The latter two show deviation up to a few percent.

\subsection{Odd-parity perturbations}
\label{sec:tov:odd}

We consider a TOV star described by SLy EOS and with gravitational
mass $M_{\rm TOV}=2M_{\odot}$ and isotropic radius $R_{\rm TOV}=7.22M_\odot$.
Odd-parity modes are excited by superposing a metric perturbation
\begin{equation}
  \label{eq:tov:pert_odd}
  \delta g_{\mu\nu} = \Phi(R)\, \Re \left(
    \begin{array}{cc|cc}
      \multicolumn{2}{c|}{\multirow{2}{*}{$0$}} & 0 & 0 \\
      \multicolumn{2}{c|}{}                     &S_{\theta}^{(\ell,m)} & S_{\phi}^{(\ell,m)} \\
      \hline
      0 & S_{\theta}^{(\ell,m)} & \multicolumn{2}{c}{\multirow{2}{*}{$0$}} \\
      0 & S_{\phi}^{(\ell,m)} & \multicolumn{2}{c}{} \\
      \end{array}
  \right) \ ,
\end{equation}
similar to \citet{Pazos:2006kz}. We chose a Gaussian pulse
\be
  \Phi(R) =A e^{-\left( \frac{R-10R_{\rm TOV}}{0.15R_{\rm TOV}}  \right)^2} \ .
\ee
with $A=0.0001$ and a $(\l,m)=(2,0)$ perturbation which makes $\delta g_{r\phi}$ be the only non-vanishing component.
The initial perturbation scatters against the star and excites the
$w$-modes of the star~\cite{Kokkotas:1992ka}.
In this case we enforce the shift to be 0 during the evolution to avoid spurious signals in some modes due to gauge effects.

\begin{figure}[t]
  \centering
  \includegraphics[width=0.5\textwidth]{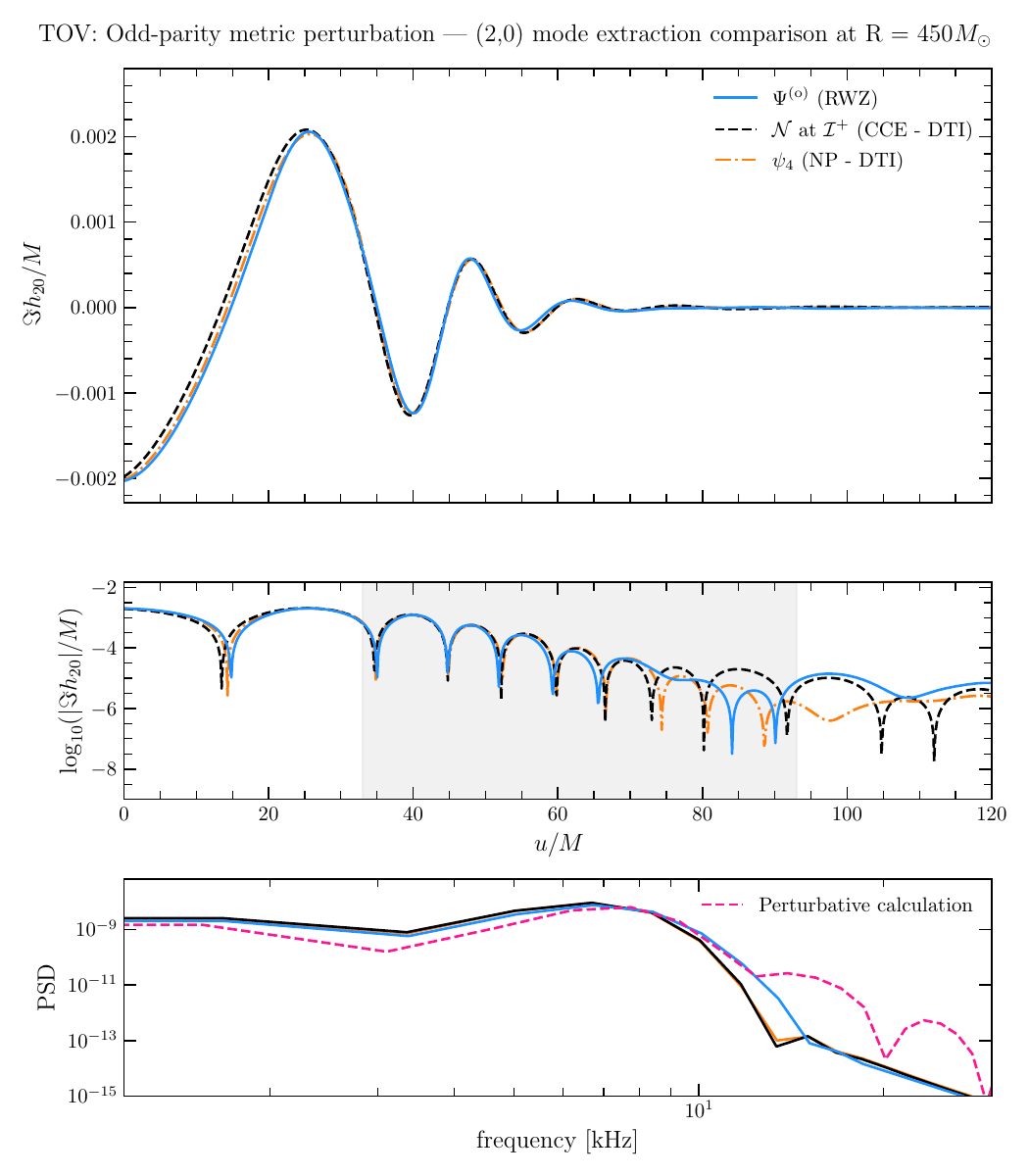}
  \caption{Strain $(2,0)$ mode for the TOV problem with spacetime 
           perturbation (Eq.~\eqref{eq:tov:pert_odd})
           extracted at $R=450M_\odot$.
           Top: Different lines refer to 
           different extraction algorithms: RWZ (solid blue), CCE (dashed black) and NP 
           (dotted-dashed orange).
           The mode is dominanted by the
           $w$-mode frequency of the star.
           Middle: same as top panel but in logarithmic scale.
           Bottom: PSD of the shaded area in the middle panel
           (ringdown part) and comparison to perturbative results (dashed pink).
           Data refer to the highest resolution simulation.}
  \label{fig:tov:odd_pert_leading}
\end{figure}

The dominant GW strain mode $(2,0)$ from these simulations is shown in
Fig.~\ref{fig:tov:odd_pert_leading}; it is computed from RWZ and NP
multipoles extracted at $R=450 M_\odot$ and compared to CCE data
evolved from the same extraction radius.
For the integration of $\psi_4$ and News function we experimented
with both FFI ($f_0=0.002/M$) and DTI to find robust results; we again
discuss results for the DTI method with a linear polynomial subtraction.

The RWZ extraction is found to be overall compatible with CCE and NP.
CCE data show again a small and convergent dephasing with respect to
the finite extraction waveforms.
The largest quantitative differences
are found in the ringdown at late times $u/M\sim 65-90$ (middle panel). %
The bottom panel show the Power Spectral Density (PSD) of the ringdown
part of the signal (shaded area from the center panel). The PSD is
compared to a perturbative simulation obtained 
using the (1+1)D code \texttt{PerBACCo} \cite{Bernuzzi:2008fu}.
Note the latter code employs a slightly different initial condition
(perturbation) than Eq.~\eqref{eq:tov:pert_odd}.
Nonetheless, the ringdown spectra are in good agreement and the
$w$-mode perturbative peak frequency at ${\sim}7749$~Hz is compatible
with the 3D numerical relativity simulation.
To our knowledge this is the first experiment obtaining $w$-modes in a 
3D simulations without symmetries; previous work has considered octant
symmetry and inverse Cowling approximation~\cite{Stergioulas:2006}. 

\begin{figure}[t]
  \centering
  \includegraphics[width=0.5\textwidth]{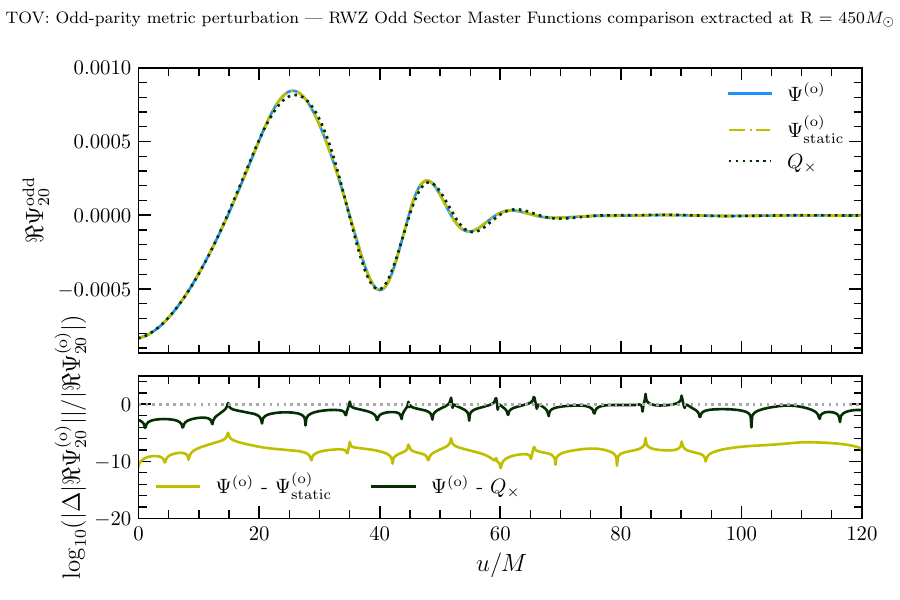}
  \caption{RWZ $(2,0)$ mode for the TOV problem with spacetime 
           perturbation (Eq.~\eqref{eq:tov:pert_odd})
           extracted at $R=450M_\odot$.
           Top:
           Odd-parity master function extraction comparison $\Psi^{\rm (o)}$ 
           (solid blue), $\Psi_{\rm static}^{\rm (o)}$ (dash-dotted light green) and 
           $Q_\times$ (dotted dark green) integrated and normalized as per
           Eq.~\eqref{eq:Qx:norm_psi}.
           Bottom:
           Amplitude relative differences with respect to $\Psi^{\rm (o)}$.
           Data refer to extraction radius $R=450 M_\odot$ and the
           highest resolution simulation.}
  \label{fig:tov:odd_pert_master_funcs}
\end{figure}

We compare the different $(2,0)$ RWZ odd-parity master functions in
Fig.~\ref{fig:tov:odd_pert_master_funcs}.
The relative differences between $\Psi^{\rm (o)}$ and $\Psi^{\rm (o)}_{\rm
  static}$ show very good consistency between the two
calculations. They are around five orders of magnitude smaller than in
the even partity case. This pattern appears to be consistent among the
all of the simulation considered, as discussed also below. 
Relative differences between $\Psi^{\rm (o)}$ and $Q_\times$ are instead
significantly larger. 
We attribute this fact to the time integration required to match 
the RWZ $\Psi^{\rm (o)}$ normalization, as per Eq.~\eqref{eq:Qx:norm_psi}.

\begin{figure}[t]
  \centering
  \includegraphics[width=0.5\textwidth]{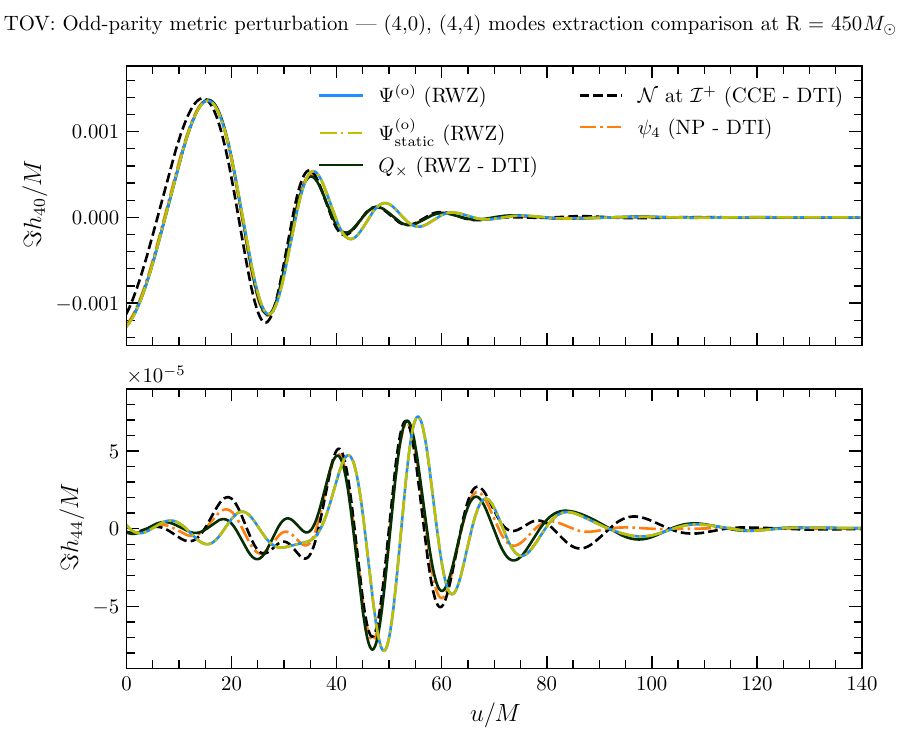}
  \caption{Strain higher modes for the TOV problem with spacetime 
           perturbation (Eq.~\eqref{eq:tov:pert_odd})
           extracted at $R=450M_\odot$.
           Different lines refer to different extraction algorithms: RWZ
           $\Psi^{\rm (o)}$  (solid blue), CCE 
           (dashed black), NP (dotted-dashed orange), RWZ $\Psi_{\rm static}^{\rm (o)}$ 
           (dash-dotted light green) and the time
           integrated/normalized of $Q_{\times}$ (dotted dark green)
           signals. 
           The top (bottom) panel refers to the $(4,0)$ ($(4,4)$)
           mode.
           Data refer to extraction radius $R=450 M_\odot$ and the
           highest resolution simulation.}
  \label{fig:tov:odd_pert_HMs}
\end{figure}

Finally, we provide a waveform comparison of the first subdominant modes
in Fig.~\ref{fig:tov:odd_pert_HMs}. The first nonzero mode is the
$(4,0)$: it has amplitude about factor two smaller than the
dominant $(2,0)$ mode and a ringdown part qualitatively very similar
to the $(2,0)$. All extraction algorithms deliver consistent waveforms,
with CCE data showing a dephasing at early times. This is, as above,
related to the finite extraction of the RWZ and NP modes.
Differences among extraction algorithms become more evident in the
$(4,4)$ mode, which is about three orders of magnitude smaller than
the $(2,0)$. The $\Psi^{\rm (o)}$ and $\Psi^{\rm (o)}_{\rm static}$ extraction
suffer of a larger dephasing with respect to $Q_\times$, possibly due to truncation
errors propagating in the calculation of Eq.~\eqref{eq:psi:even:stat}
and Eq.~\eqref{eq:psi:even} and affecting the small-amplitude signal.
Interestingyly, the $Q_\times$ master function nicely agree with CCE
and NP around the peak of the waveform.

\section{Gravitational collapse of a rotating neutron star}
\label{sec:coll}

We consider simulations of the gravitational collapse of a rotating
neutron star to a black hole, \eg~\cite{Baiotti:2007np,Reisswig:2010cd,Dietrich:2014wja,Cook:2023bag}.
\citet{Reisswig:2010cd} presented waveforms using a metric extraction
algorithm based on Schwarzschild coordinates, \ie~$Q_+$.
Here, we consider for the first time also waveforms with the covariant and
gauge-invariant metric algorithm.

The simulations use the same refinement level zero domain $\Omega$ as
for the TOV problems but this time refined with $N_L = 7$ static
refinement levels. The domain is resolved with $N_M=128$ points to
achieve a maximal resolution of $h_7 = 0.0625 M_\odot$. The CFL factor
is set to $0.2$ and the Kreiss-Oliger dissipation parameter to $\sigma
= 0.1$.
Gravitational waves are extracted at the same radii as for the TOV problems. 

We consider the uniformly rotating D4 neutron star model already
simulated in, \eg~\cite{Baiotti:2007np,Dietrich:2014wja},
with gravitational mass $M=1.86M_\odot$ and
polar-to-equatorial coordinate axis ratio of
$r_p/r_e = 0.65$. The rotation frequency is close to the Kepler limit
which sets the mass shedding and stability limit.
Initial data for the simulation are generated with the 
\texttt{rns} code~\cite{Stergioulas:1994ea}. 
We introduce a pressure perturbation in the initial data which
decreases the star's pressure a $0.17\%$, uniformly over the
pressure profile. This helps to accelerate the rotational collapse but
delays collapse enough to clearly separate the junk radiation from the
rest of the gravitational signal~\cite{Dietrich:2014wja}. 

\begin{figure}[t]
  \centering
  \includegraphics[width=0.5\textwidth]{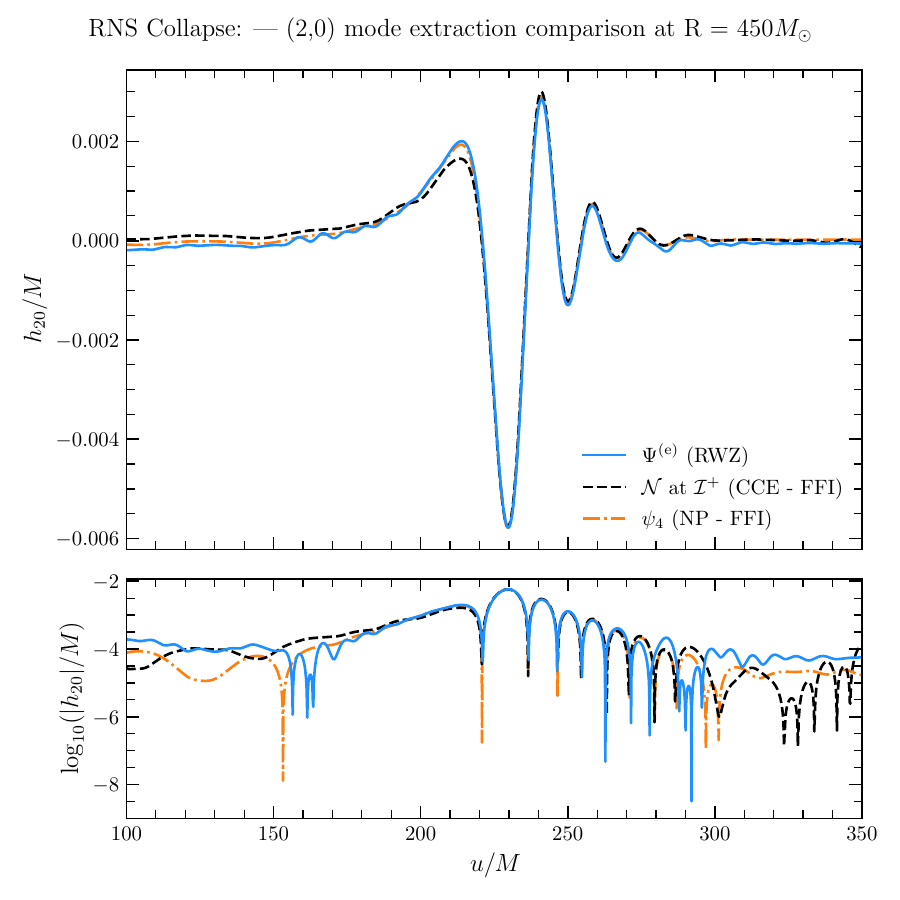}
  \caption{Strain $(2,0)$ mode for the rotating collapse problem 
           extracted at $R=450M_\odot$.
           Different lines refer to 
           different extraction algorithms: RWZ (solid blue), CCE (dashed black) and NP 
           (dotted-dashed orange).
           The top (bottom) panel shows the waveform in linear (log)
           scale. Data refer to the highest resolution simulation.}
  \label{fig:rns:20}
\end{figure}

The GWs from the rotational collapse has the well-known 
\emph{precursor-burst-ringdown} morphology~\cite{Davis:1972ud} and it
is entirely in the even-parity sector.
The background spacetime is now dynamical and in coordinates defined
by the 1+log and $\Gamma$-driver evolution equation (gauge sector of
the Z4c free-evolution scheme). 
Fig.~\ref{fig:rns:20} shows the strain $(2,0)$ mode from the RWZ,
CCE and NP extractions at $R=450 M_\odot$. 
The three methods deliver compatible waveforms with the expected
morphology. The largest differences are visibile at early times in the
precursor and at late time in the ringdown. Starting form the latter, 
the lower panel shows that the ringdown oscillations
are well captured until the $9^{\rm th}$ peak, $u/M\lesssim75$. After
that, the RWZ extraction first and then both CCE and NP become noisy
due to the small amplitude of the signal.
The differences in the precursor are due to two effects.
On the one hand, the RWZ extraction gives more noisy waveforms than
other methods, see the spurious oscillation around $u/M\sim170$. This
is compatible with what observed in \citet{Reisswig:2010cd}.
On the other hand, the integration of the CCE
$\psi_4$ or News function introduce here some uncertainty in the strain.
In absence of a physicaly motivated cutoff frequency, we experimented
with both the FFI and the DTI integration.
We find that the double integral of the CCE $\psi_4$ result in
spurious oscillations with both methods, while the integral of the
News shows an unphysical drifts for various choices of the polynomial
correction. These issues are not present in the $\psi_4$ extracted
at finite radii. Best results are obtained using FFI with
$f_0=0.00186/M$ and integrating the CCE News, which are shown in the
figure. Note, however, that the integration introduces uncertainties
of at least the same magnitude of the differences visible in the
figure at $u/M\sim150$.

\begin{figure*}[t]
  \centering
  \includegraphics[width=\textwidth]{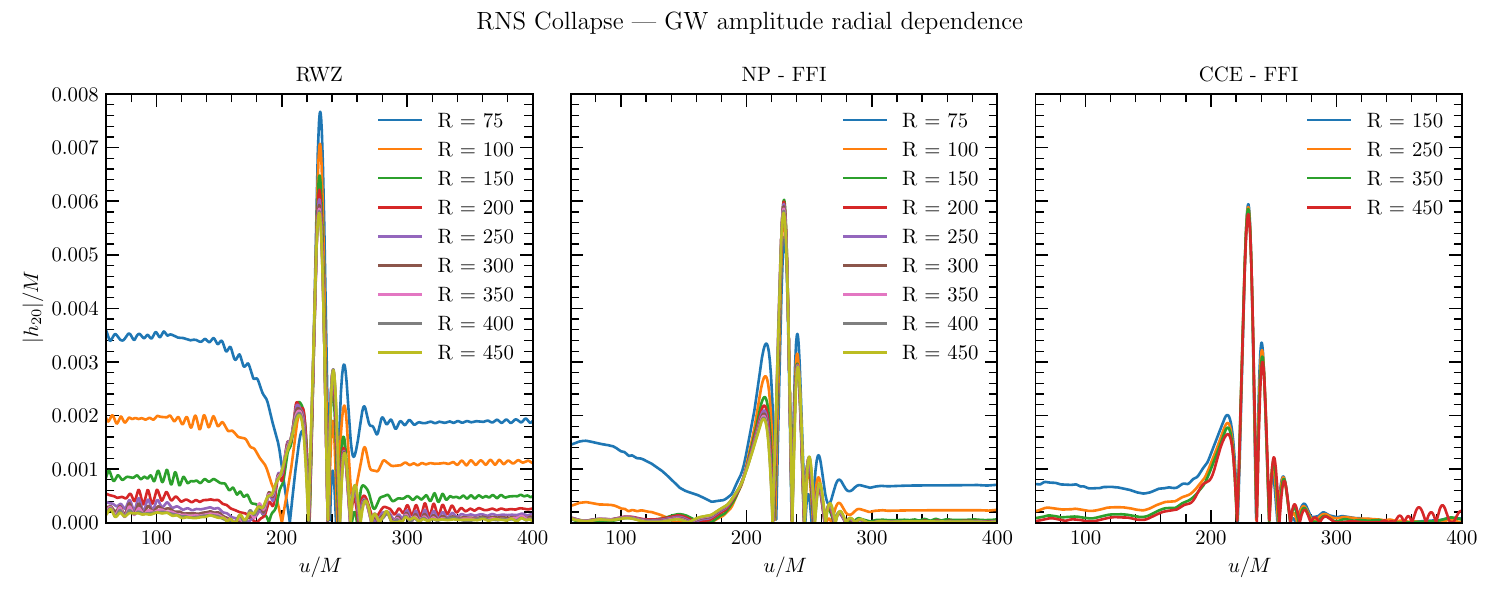}
  \caption{Extraction radius dependence of the strain $(2,0)$ mode for
    the rotating collapse problem.
    Panels from left to right: RWZ, NP and CCE extraction.
    For each panel, different lines show different extraction radii (in units of $M_\odot$).}
  \label{fig:rns:amplitude}
\end{figure*}

\begin{figure}[t]
  \centering
  \includegraphics[width=0.5\textwidth]{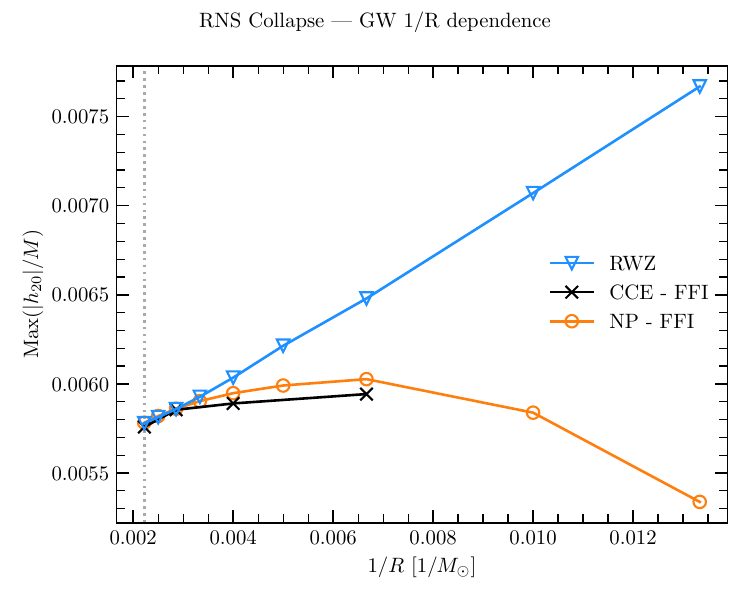}
  \caption{Extraction radius dependence of the maximum strain
    amplitude of the $(2,0)$ mode for the rotating collapse problem and
    different extraction algorithms. The dotted grey vertical line marks
    the radius with the best overall compatibility.}
  \label{fig:rns_1/R}
\end{figure}

The dependence of the waveforms on the extraction (or worldtube) radii
is shown in Fig.~\ref{fig:rns:amplitude}. Significant finite
extraction effects are evident at both early and late times for radii
$R\lesssim200\Mo$ in the RWZ and NP methods. The peak of the RWZ waveform is
clearly more sensitive to the choice of extraction sphere in this
problem. Also the precursor in the integrated CCE data is strongly
dependent on the worldtube radii. However, waveforms converge to each
other for radii $R\gtrsim350\Mo$. In order to better quantify this 
convergence,  Fig.~\ref{fig:rns_1/R} shows the GW maximum amplitude
(the GW burst) as
as a function of $1/R$: all the methods deliver a consistent waveform
peak amplitude in the large-$R$ limit.
Despite the slower $1/R$ convergence and the noise, the RWZ extraction
appears a robust choice for this problem and it can help to identify
systematics in the integration of $\psi_4$ waveforms.
Further, we checked that an extrapolation of the RWZ waveform to null
infinity significantly improves the waveform quality. The 
peak of the precursor around $u/M\sim210$ lowers and become compatible
to the CCE waveform computed with the worldtube at $R=450\Mo$. Contrary,
the precursor of the CCE data computed with the worldtube at $R=150\Mo$
remains unreliable. These results are shown in Appendix~\ref{app:rextrap}.

\begin{figure}[t]
  \centering
  \includegraphics[width=0.5\textwidth]{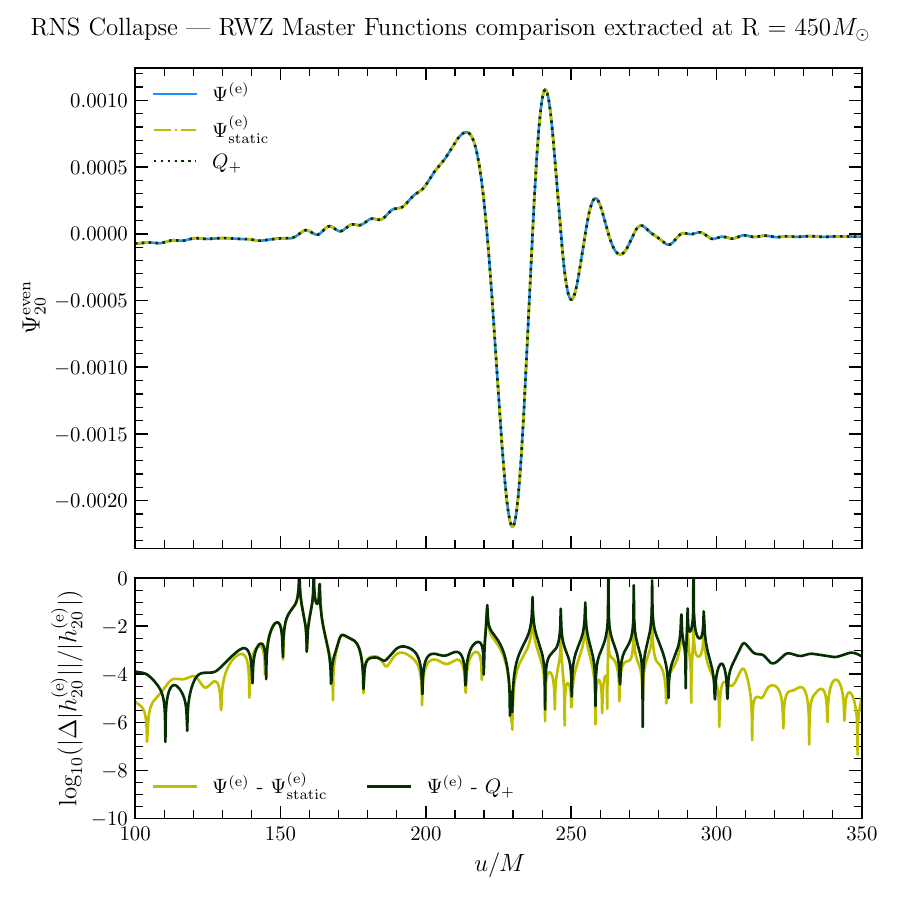}
  \caption{RWZ $(2,0)$ mode for the rotating collapse problem
           extracted at $R=450M_\odot$ and computed using different RWZ
           master functions.
           Top:
           Even-parity master function extraction comparison $\Psi^{\rm (e)}$ 
           (solid blue), $\Psi_{\rm static}^{\rm (e)}$ (dash-dotted light green) and 
           $Q_{+}$ (dotted dark green) normalized as per
           Eq.~\eqref{eq:Q+:norm_psi}.
           Bottom: 
           Relative differences with respect to $\Psi^{\rm (e)}$.
           Data refer to the highest resolution simulation.}
  \label{fig:rns_master_funcs}
\end{figure}

Finally, we compare the different even-parity RWZ extraction master
functions in Fig.~\ref{fig:rns_master_funcs}. Despite having a
dynamical background, relative differences are found of the order of
$10^{-4}$ (away from a few zero crossing). They appear in-line with
what found in the TOV problems
(\cf~Fig.~\ref{fig:tov:even_pert_master_funcs}).
Therefore, we conclude that an extraction radius at $R\sim450\Mo$ is a
robust choice for the metric extraction in this problem.

\section{Binary black hole spacetimes}
\label{sec:bbh}

In order to further test physical problems with
nontrivial background and gauge dynamics we 
consider an equal-mass, nonspinning, black hole binary 
in two different configurations: (i) a quasi-circular configuration
covering two orbits to merger, previously used as standard
code beanchmark~\cite{Brugmann:2008zz,Daszuta:2021ecf};
(ii) a dynamical capture configuration considered in 
Ref.~\cite{Albanesi:2024xus} with a close encounter followed 
then by a merger.
Gravitational waves from these configurations are radiated in either 
the  even-parity and odd-parity sector, but due to the  $1\leftrightarrow 2$ 
symmetry of the system only the $(2,2)$, $(2,0)$ $(3,2)$ and $(4,4)$ modes 
are nonzero.

Binary black hole spacetimes are simulated on a domain
$\Omega=[-1536\Mo, 1536\Mo]^3$ with $N_M=128$ points per direction. 
The domain is refined with $N_L=10$ levels using the $L_2$ 
AMR strategy detalied in~\citet{Rashti:2023wfe}.
The maximum resolution at the punctures is $h_{10} \simeq 0.023 \Mo$.
The CFL is set to  $0.25$ and the dissipation parameter is $\sigma =
0.2$. Spherical grids for 
the RWZ and NP extraction algorithms are located at
$R=(80,90,100,110,120,130,140,180,220,400)\Mo$ while the worldtubes
for CCE are at $R=(100,140,220,400)\Mo$.
Initial data are generated using the \texttt{TwoPunctures} code~\cite{Ansorg:2004ds}.
The initial data for the quasi-circular configuration (that is such to yield only
a couple of orbits up to merger) are generated with puncture ADM masses 
$M_1=M_2=M/2=0.505085 M_\odot$ and ADM energy  $E_{\rm ADM}/M = 0.996$.
For the dynamical capture the puncture ADM masses are $M_1=M_2=M/2=0.5 M_\odot$, 
ADM energy $E_{\rm ADM}/M=1.016$ and initial angular momentum 
$J_{\rm ADM}/(\nu M^2)=4.18$, where $\nu=M_1M_2/M^2=1/4$ is the symmetric mass ratio.

\subsection{Quasi-circular merger}
\label{sec:bbh:circ}

\begin{figure}[t]
  \centering
  \includegraphics[width=0.5\textwidth]{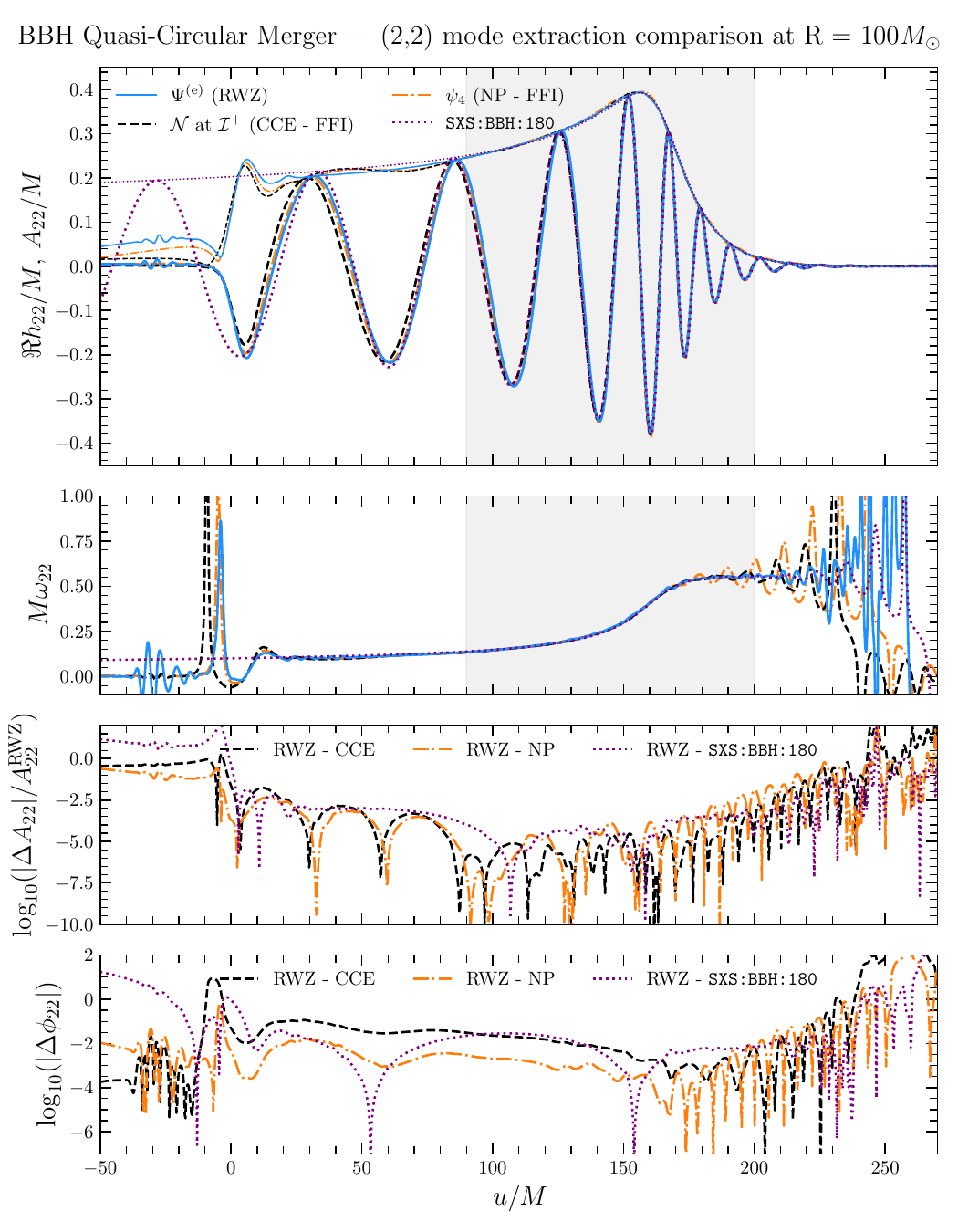}
  \caption{Strain $(2,2)$ mode for the circular BBH merger at $R=100\Mo$. 
    Different lines compare the results of the different extraction algorithms: RWZ (solid blue), 
    CCE (dashed black) and NP (dotted-dashed orange). Also shown is 
    the $\texttt{SXS:BBH:180}$ waveform (dotted purple) aligned with the RWZ; the shaded
    area marks the alignment window. From top to bottom: amplitude
    and real part, frequency, amplitude differences and phase
    differences with respect to the RWZ waveform.
    } 
  \label{fig:bbh:qc_leading}
\end{figure}

Fig.~\ref{fig:bbh:qc_leading} shows the dominant $(2,2)$ strain mode
(both real part and amplitude) from the three wave-extraction 
algorithms: (i) the RWZ metric procedure at $R_{\rm extr}=100\Mo$; (ii) $\psi_4$ at $R_{\rm extr}=100\Mo$; the News $\cal N$ at $\scri$.
The methods prove consistent among themselves. 
Note that $h_{22}$ was obtained from 
$\psi_4$ and the News ${\cal N}$ using FFI with a cutoff frequency 
$f_0=0.007/M$. The latter choice performs better than DTI due to the short 
duration of the signal and it correctly captures most of the physical features.
We further compare to
the (longer) {\tt SXS:BBH:180} \cite{SXS:catalog} waveform of the SXS
catalog. The latter is extrapolated at infinity and aligned with the RWZ on the shaded time-domain window by suitably
fixing an arbitrary relative time and phase shift.

The RWZ extracted function, $\Psi^{\rm (e)}_{22}$, correctly 
captures the expected structure of the waveform, consistently
with previous work that was using an extraction algorithm with
fixed Schwarzschild-like coordinates 
(see e.g. Ref.~\cite{Damour:2007vq} and references therein). 
The amplitude agreement with CCE and NP (and with {\tt SXS:BBH:180})
improves near the waveform peak, while some numerical artifact can 
be observed for RWZ and NP in the early part of the signal, before the junk.
The amplitude of the RWZ waveform is smoother than the CCE and NP,
especially at early times, $u/M\sim50$. Around those times, both CCE
and NP show an unphysical non-monotonic behavior which is mainly
introduced by the FFI.
The instantaneous gravitational wave frequency is also compatible 
among the different extraction algorithms. The most notable differences 
are present during the ringdown.
NP and CCE data show larger oscillation at $u/M\gtrsim200$,
which are converging away with resolution. Early results with
metric extraction ($Q_+$) also showed prominent oscillations
\cite{Damour:2007vq}, which are however significantly reduced at
the resolution considered here.
Phase differences among RWZ and NP multipoles at finite radius ($R=100\Mo$) are
consistent with truncation errors and compatible with the SXS
waveforms. 
Phase differences with respect to null infinity data can be
a factor ${\sim}2$ larger already for such a short waveform, which
indicates larger extraction radii are required in order to obtain high accuracy data.
Both RWZ and NP extractions can be indeed improved by extrapolation,
see Appendix~\ref{app:rextrap}.

\begin{figure}[t]
  \centering
  \includegraphics[width=0.5\textwidth]{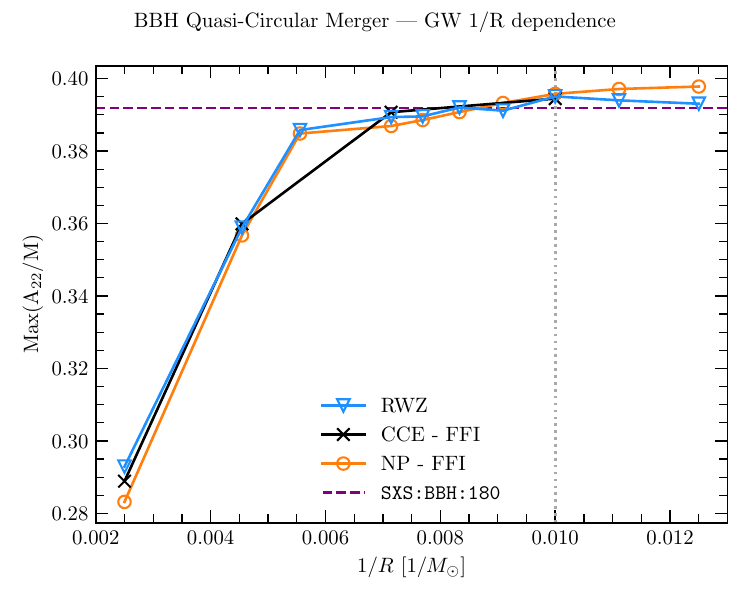}
  \caption{Extraction radius dependence of the maximum strain
    amplitude $(2,2)$ for the quasi-circular BBH merger 
    and different extraction algorithms.
    The dotted grey vertical line marks the radius with the best
    overall compatibility. The dashed purple is the value from \texttt{SXS:BBH:180}, extrapolated to $\scri$.}
  \label{fig:bbh_qc_1/R}
\end{figure}

Finite extraction uncertainties are summarized in
Fig.~\ref{fig:bbh_qc_1/R}. The figure shows the peak amplitude for all
extraction algorithms as a function of the extraction radius $1/R$.
For comparison, we also show the value $A_{22}^{\texttt{SXS}}/M=0.3919$
taken from \texttt{SXS:BBH:180} extrapolated to infinite radius with $N=3$.
The best agreement
between waveforms is obtained for $R=100\Mo$. At larger radii, the
waveform's quality decreases because the extraction spheres are placed
in refinement levels with progressively coarser resolutions. The
amplitude becomes affected by significant uncertainties at extraction
radii $R\gtrsim180\Mo$, which can decrease up to 70\% the amplitude for
$R=400\Mo$ (with the chosen mesh resolution).

\begin{figure}[t]
  \centering
  \includegraphics[width=0.5\textwidth]{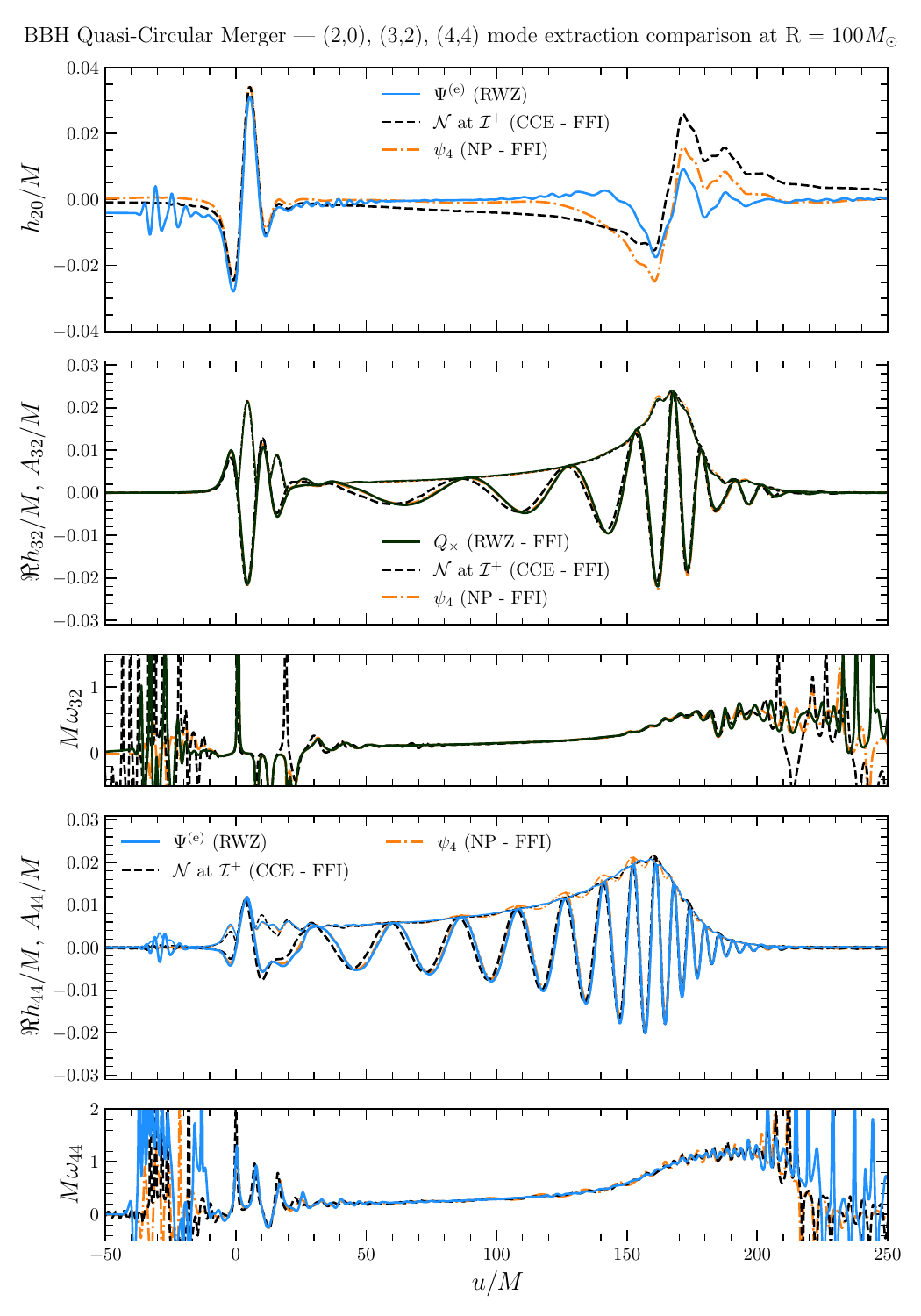}
  \caption{Subdominant modes: $(2,0)$ (top), $(3,2)$
    (middle, with frequency) and $(4,4)$ (bottom, with frequency)
    for the quasi-circular BBH merger extracted $R=100M_\odot$. 
    In each panel, different lines refer to different wave-extraction algorithms: 
    RWZ (solid blue), CCE (dashed black) and NP (dotted-dashed orange).
  }
  \label{fig:bbh:qc_HMs}
\end{figure}

Higher modes $(2,0)$, $(3,2)$ and $(4,4)$ of the strain extracted at
$R=100\Mo$ are shown in Fig.~\ref{fig:bbh:qc_HMs}
for the different extraction algorithms. CCE and NP data are
integrated with FFI using $f_{0}^{(\lm)} =
2f_{0}/\text{max}(1,m)$. This choice eliminates physical low frequency
features, in particular the non-linear memory in the $(2,0)$ mode in
CCE data, as it can be observed in the top panel of the
figure. Nonlinear memory can be captured in CCE data by using
$f_0^{(20)}=0.001/M$ but, since we are not interested in studying
this effect here, we keep for the CCE data the same integration as
used for NP data. Overall, obtaining an accurate $(2,0)$ mode from
such a short simulation appears rather challenging and the different 
extraction algorithms delivers quantitatively different predictions.
Concerning the other two modes, we find instead good agreement 
between the extraction algorithms. The figure shows some dephasing
between CCE and finite radius data and that the $(4,4)$ mode NP
data are more noisy than RWZ data.  
Notably, for the $(3,2)$ mode, we
show the RWZ waveform obtained from the master function $Q_\times$
(Schwarzschild coordinates) instead of from the $\Psi^{\rm (o)}$. 
As discussed below, this appears the most robust choice for this
particular mode.

\begin{figure*}[t]
  \centering
  \includegraphics[width=\textwidth]{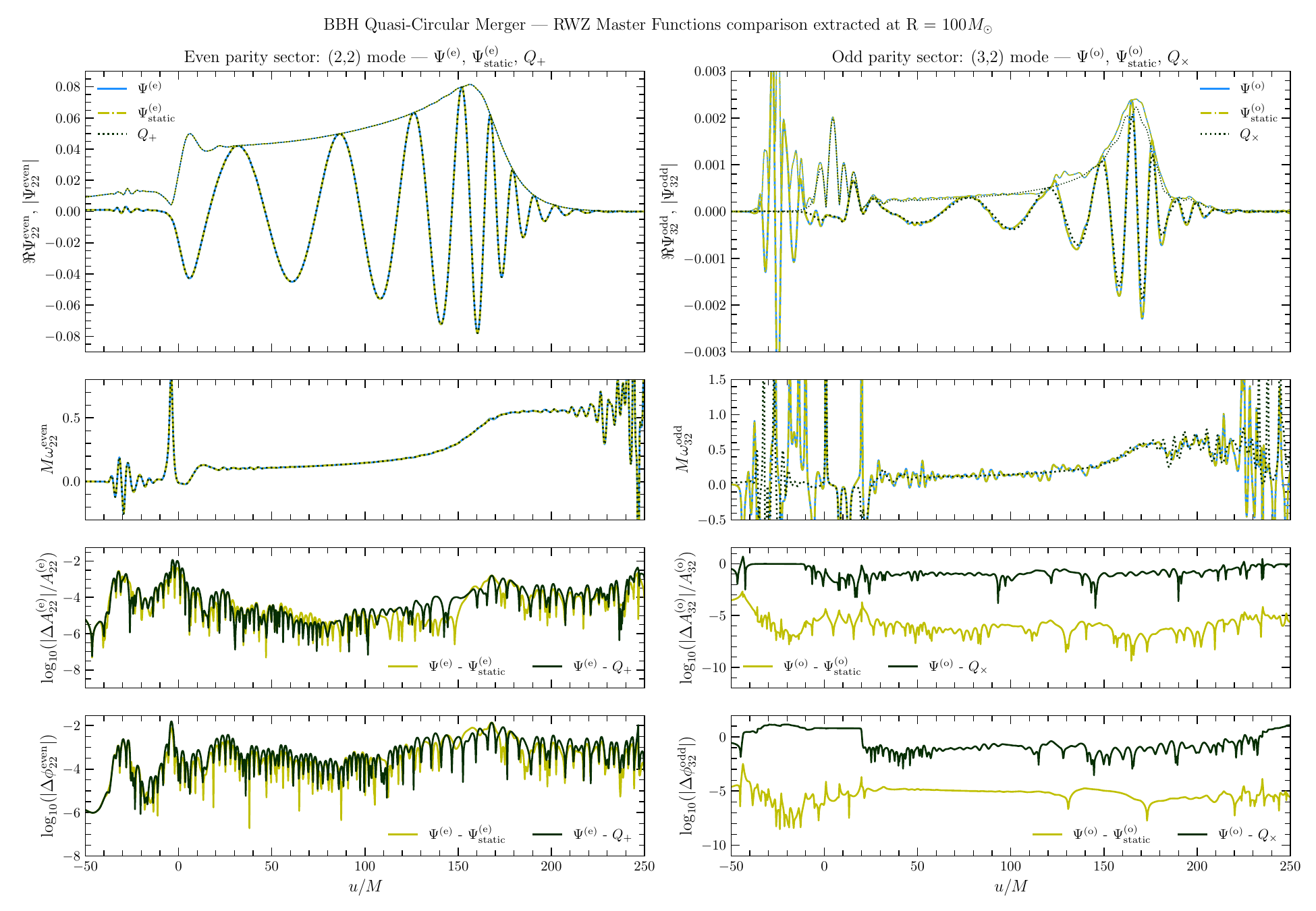}
  \caption{RWZ even- and odd-parity modes for the BBH circular merger
    at $R=100M_\odot$ and computed using different RWZ master
    functions.
    Left:
    Even-parity $(2,2)$ master function extraction comparison $\Psi^{\rm (e)}$ 
    (solid blue), $\Psi_{\rm static}^{\rm (e)}$ (dash-dotted light green) and 
    $Q_{+}$ (dotted dark green) normalized as per
    Eq.~\eqref{eq:Q+:norm_psi}.
    Right:
    Odd-parity $(3,2)$ master function extraction comparison $\Psi^{\rm (o)}$ 
    (solid blue), $\Psi_{\rm static}^{\rm (o)}$ (dash-dotted light
    green) and
     $Q_\times$ (dotted dark green) integrated and normalized as per
    Eq.~\eqref{eq:Qx:norm_psi}.
    Panels in each columns show from top to bottom:
    real part and amplitude, frequency, amplitude differences and
    phase differences. }
  \label{fig:bbh:qc_master_funcs}
\end{figure*}

We compare the different RWZ master functions for both even- and
odd-parity sectors in Fig.~\ref{fig:bbh:qc_master_funcs}. For the even
parity we show the $(2,2)$  mode and for the odd parity we show the 
$(3,2)$ mode.
In the even-parity sector, we obtain similar results as discussed in
previous problems. Amplitude relative differences with respect to
$\Psi^{\rm (e)}$ are of order ${\sim} 10^{-5}$ during the  
orbital phase and ${\lesssim} 10^{-3}$ at merger for both $\Psi^{\rm
  (e)}_{\rm static}$ and $Q_+$. Phase differences are of order ${\sim}
10^{-3}$~rad at early times to grow up to ${\sim} 10^{-2}$~rad at
merger; they remain compatible with truncation errors due to finite
resolution.

More interesting is the odd-parity sector, where significant
differences between the master functions in general coordinates
$\Psi^{\rm (o)}$ and the Schwarzschild coordinates master
function $Q_\times$. The $\Psi^{\rm (o)}$ amplitudes are larger and
more noisy, with a significant amount of junk radiation at early
times. More noise is also found in the instantaneous frequency
evolution. Phase differences with respect to $Q_\times$ are also at
the ${\sim}0.1$~rad level, comparably larger than the dephasing in
the even-parity mode discussed above.
By inspecting the data, we idenfity the origin of these features as
slowly convergent residual gauge effect due to the odd-parity $H_0^{(\lm)}$ multipole
(see Eq.~\eqref{eq:psi:odd} and Eq.~\eqref{eq:multipoles:odd}).
The latter is an integral of the shift vector which, in this example,
is effectively worsening the performance of the covariant extraction
when compared to $Q_\times$ (Schwarzschild coordinates.)
This effect is present also in other higher modes and other binary
problems, as we shall discuss below.
Similarly to the RNS problem, the effects of the dynamical background are 
practically negligible for this BBH merger problem.

We have also compared waveforms for the different choices of the
background's Schwarzschild radius prescriptions in this BBH problem. We
obtain very similar results to the TOV problem. The areal and
the average Schwarzschild radii deliver very comparable results with
relative amplitude (phase) differences of order ${\sim}10^{-6}$
(${\lesssim}10^{-5}$~rad). The $g\theta\theta$ and $g\phi\phi$
prescriptions also behaves very similarly. The relative amplitude (phase)
differences with respect to the average Schwarzschild radius are of
order ${\sim}10^{-3}$ (${\lesssim}10^{-3}$~rad).
Any of these choices appear sufficiently robust for the considered
setup and does not significantly impact the uncertainties discussed
above.

\subsection{Dynamical capture}
\label{sec:bbh:dyne}

The parameters detailed at the beginning of Sec.~\ref{sec:bbh} refering to case (ii) correspond to an initially unbound orbit which
becomes bound due to the GW emission and lead to a merger after 
two encounters.

\begin{figure}[t]
  \centering
  \includegraphics[width=0.5\textwidth]{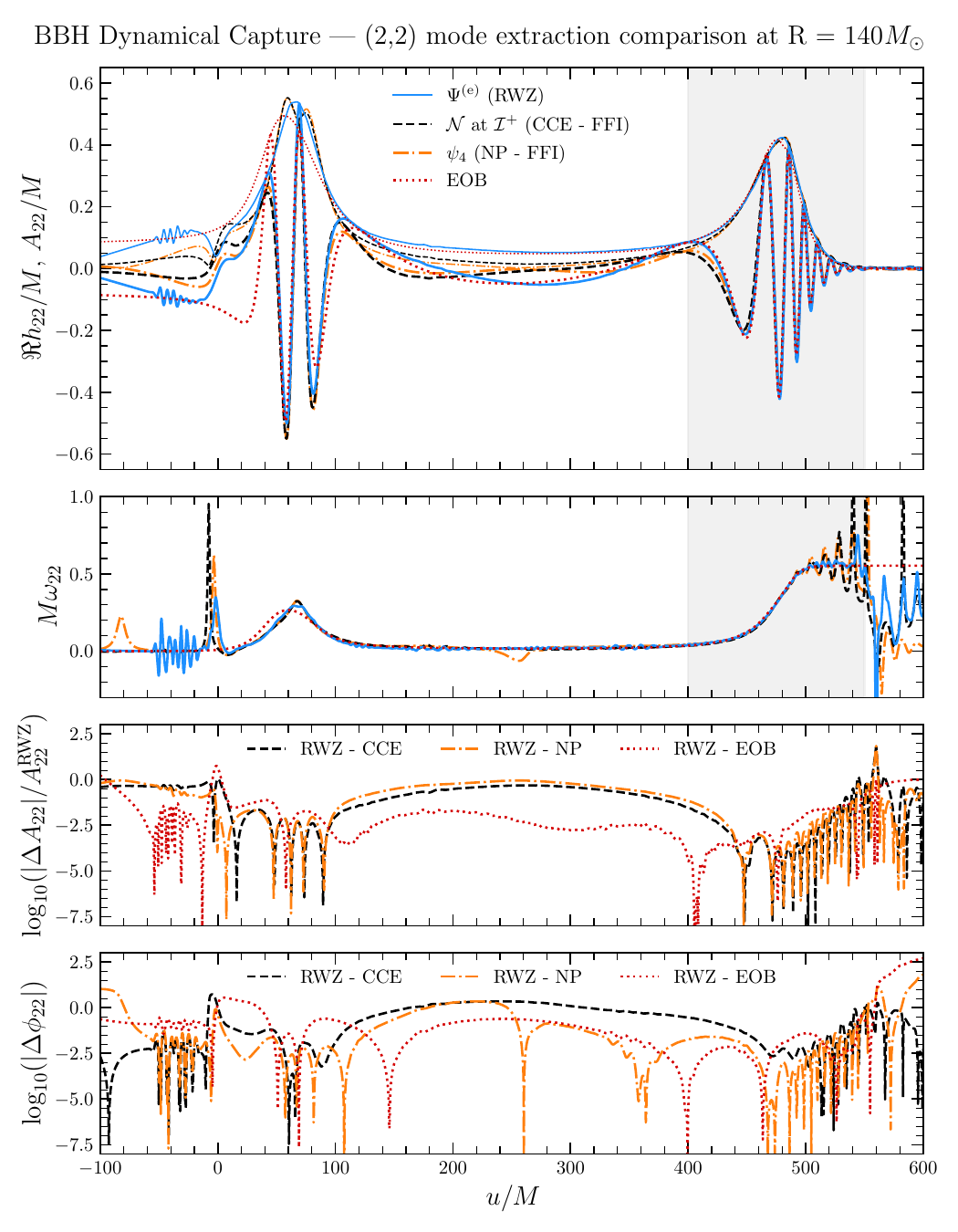}
  \caption{Strain $(2,2)$ mode for a BBH dynamical capture at
    $R=140\Mo$.
    Different lines compare the results of the different extraction algorithms: RWZ (solid blue), 
    CCE (dashed black) and NP (dotted-dashed orange). Also shown is a EOB prediction
    (dotted red) aligned with the RWZ; the shaded
      area marks the alignment window.
    From top to bottom: amplitude
    and real part, frequency, amplitude differences and phase
    differences with respect to the RWZ waveform.}
  \label{fig:bbh:cap_leading}
\end{figure}

The capture dynamics is an interesting test for GW extraction
algorithms because of the non-trivial morphology of the waveform.
Fig.~\ref{fig:bbh:cap_leading} shows the dominant $(2,2)$ mode
obtained with the different extraction algorithms at $R=140\Mo$.
The amplitude and instantaneous frequency peak at each of the two
encounters. The two peaks are separated by a low-frequency transient
signal corresponding to the punctures reaching the apastron and
returning on a bound orbit. The second peak corresponds to the merger
and it is followed by the ringdown signal of the
remnant black hole. While the waveforms qualitatively agree in reproducing
this morphology, significant differences are clearly present at the
first peak and in the subsequent transient. 
The figure also shows as reference an EOB waveform obtained with 
\dali{}~\cite{Nagar:2024oyk,Albanesi:2025txj} aligned in phase with
the RWZ signal (alignment window marked with the shaded area). The
initial data for the
EOB evolution are chosen in order to minimize the mismatch against the RWZ waveform, 
following a procedure similar to the ones used in Ref.~\cite{Andrade:2023trh}.
Such EOB waveform is an important benchmark because it delivers a
semi-analytical prediction of the low-frequency transient between the
encounters, which is captured by the Newtonian prefactor for generic 
motion~\cite{Chiaramello:2020ehz}.
Note that for the EOB waveform shown here we use a slightly different
activation function for the next-to-quasi-circular corrections adopted
in the standard model to correct the EOB plunge waveform~\cite{Nagar:2024oyk,Albanesi:2025txj}.

At the first encounter peak, both CCE and NP data show a double peak
structure which is unphysical. This artifact is induced by the 
integration of the $\psi_{4}$
converge away with increasing extraction radius (see also below).
The data in the figure use FFI integration with a cutoff frequency
$f_0 = 0.007/M$ ($f_0^{(\lm)} = 2f_0/\text{max}(1,m)$ for higher
modes), similarly to what is done in Ref.~\cite{Andrade:2023trh}.
The double peak artifact and the subsequent signal can be improved by
lowering the frequency cutoff but that introduces noise and a
drift in the ringdown part.
In absence of a physical cutoff frequency for the FFI, we also
experimented with DTI but found less robust results than those shown
in the figure.

During the low-frequency transient, the RWZ amplitude and frequency
best match the EOB, while the CCE and NP data show qualitative
differences. After merger, all the extraction algorithm become more
compatible with similar features in the late ringdown as those
discussed for the BBH circular merger. 
Overall for this type of waveform, the RWZ extraction algorithm
provides a robust alternative that can help to identify systematics in
other algorithms.

\begin{figure*}[t]
  \centering
  \includegraphics[width=\textwidth]{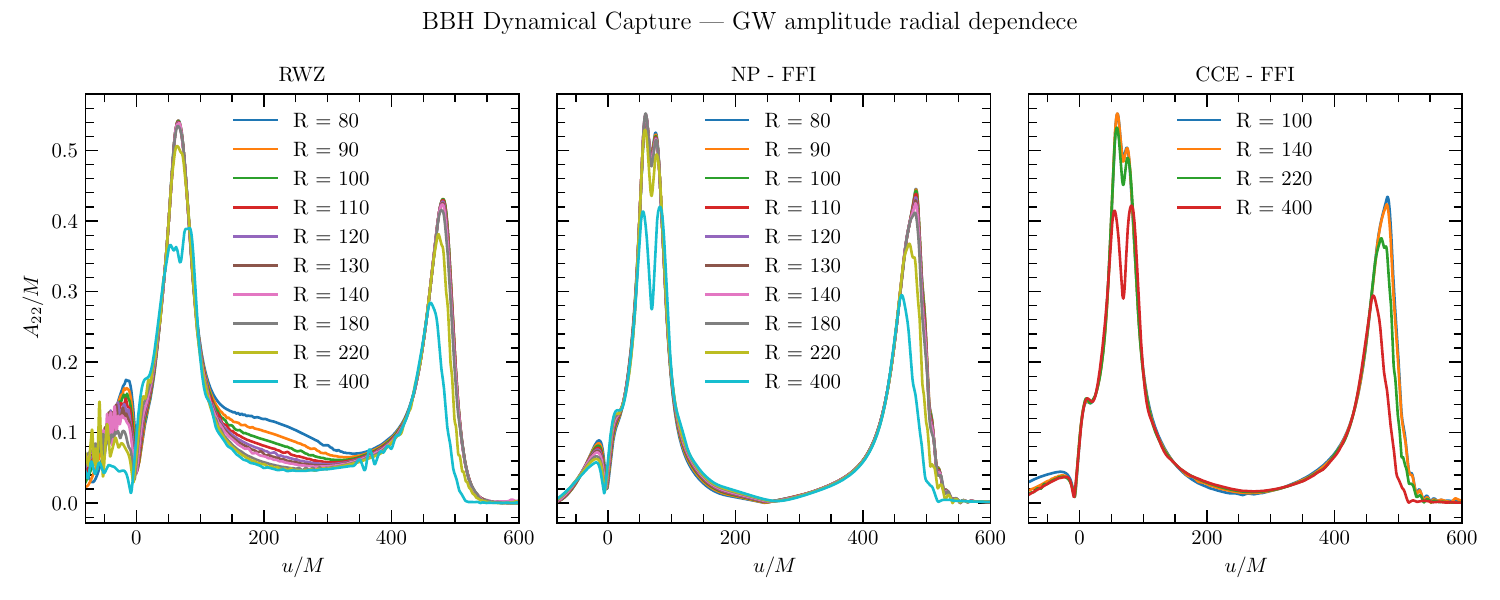}
  \caption{Extraction radius dependence of the strain $(2,2)$ mode for
    the BBH dynamical capture.
    Panels from left to right:
    RWZ, NP and CCE extraction. For each panel, different lines show
    different extraction radii (in units of $M_\odot$).}
  \label{fig:bbh:cap_amplitude}
\end{figure*}

The dependence of the waveforms on the extraction (or worldtube) radii
is illustrated by Fig.~\ref{fig:bbh:cap_amplitude} which shows the
$(2,2)$ amplitude for various algorithms. The RWZ extraction shows a
slower convergence with radius but eventually achieves the expected
structure at $R\sim140\Mo$. By constrast CCE and NP data retain the
double peak  artifact discussed above.
For larger radii $R\gtrsim 180\Mo$ the
amplitudes at the peaks starts to decrease considerably. This is again
due to the resolution of the wave zone.

\begin{figure*}[t]
  \centering
  \includegraphics[width=\textwidth]{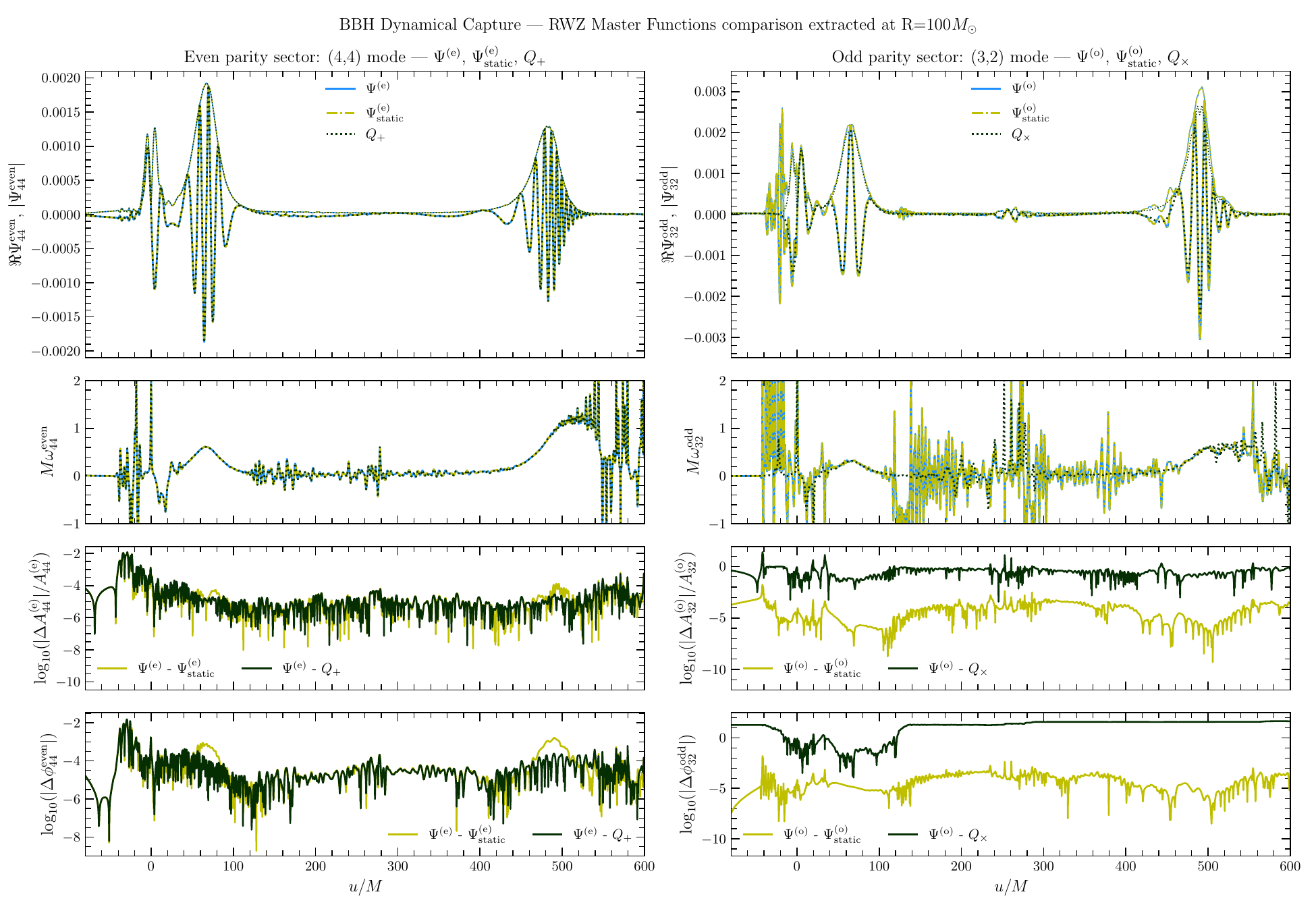}
  \caption{RWZ even- and odd-parity modes for the BBH dynamical
    encounter 
    at $R=100M_\odot$ and computed using different RWZ master
    functions.
    Left:
    Even-parity $(4,4)$ master function extraction comparison $\Psi^{\rm (e)}$ 
    (solid blue), $\Psi_{\rm static}^{\rm (e)}$ (dash-dotted light green) and 
    $Q_{+}$ (dotted dark green) normalized as per
    Eq.~\eqref{eq:Q+:norm_psi}.
    Right:
    Odd-parity $(3,2)$ master function extraction comparison $\Psi^{\rm (o)}$ 
    (solid blue), $\Psi_{\rm static}^{\rm (o)}$ (dash-dotted light
    green) and
     $Q_\times$ (dotted dark green) integrated and normalized as per
    Eq.~\eqref{eq:Qx:norm_psi}.
    Panels in each columns show from top to bottom:
    real part and amplitude, frequency, amplitude differences and
    phase differences.}
  \label{fig:bbh:cap_master_funcs}
\end{figure*}

Regarding higher modes and the choice of different RWZ master
functions, the dynamical capture simulation gives analogous results to
the circular merger. Fig. ~\ref{fig:bbh:cap_master_funcs} shows the
RWZ $(4,4)$ and $(3,2)$ modes computed from $\Psi^{\rm (e/o)}$,
$\Psi^{\rm (e/o)}_{\rm static}$ and $Q_{+/\times}$. Very consistent
results are found in the even sector for any choice of the master
function. Note that there are only subtle differences at the peaks with respect to 
$\Psi^{\rm (e)}_{\rm static}$ in the $(4,4)$ mode, which are mildly reduced at lower distance. By contrast, the odd parity waveform extraction with
$\Psi^{\rm (o)}$ and $\Psi^{\rm (o)}_{\rm static}$ are more noisy than
the $Q_\times$. Unsurprisingly, due to the same gauge employed in this
simulation and in the the circular merger, the latter choice appear
more robust for the odd modes.

\section{Binary neutron star spacetime}
\label{sec:bns}

As a final problem, we consider a ten orbits binary neutron star merger
simulation previously simulated in \cite{Radice:2016gym,Dietrich:2017aum,Dietrich:2018phi}. To our knowledge, metric waveform extraction in such long BNS 
simulations have never been previously considered.
The stars have isolation masses $M_1=M_2=M/2=1.35 M_\odot$ and are
described by the SLy EOS and irrotational fluid. 
They are initially separated by a coordinate distance $d=52.42$~km.
Initial data are generated using the \texttt{Lorene} library \cite{Gourgoulhon:2000nn}.

The simulations use a domain $\Omega=[-1024 M_\odot, 1024 M_\odot]^3$
with $N_M=160$ and $N_L=6$ refinement levels ($h_{6} = 0.2 M_\odot$)
with AMR strategy as detalied in \citet{Daszuta:2024ucu}. The CFL is
set to $0.25$ and the dissipation parameter is $\sigma = 0.5$. The
extraction spheres the RWZ and NP algorithms are located at
$R=100,200,300,400,500,600,700,800,900 M_\odot$  
while the worldtubes for the CCE are at
$R=200,400,600,800 M_\odot$.

\begin{figure*}[t]
  \centering
  \includegraphics[width=\textwidth]{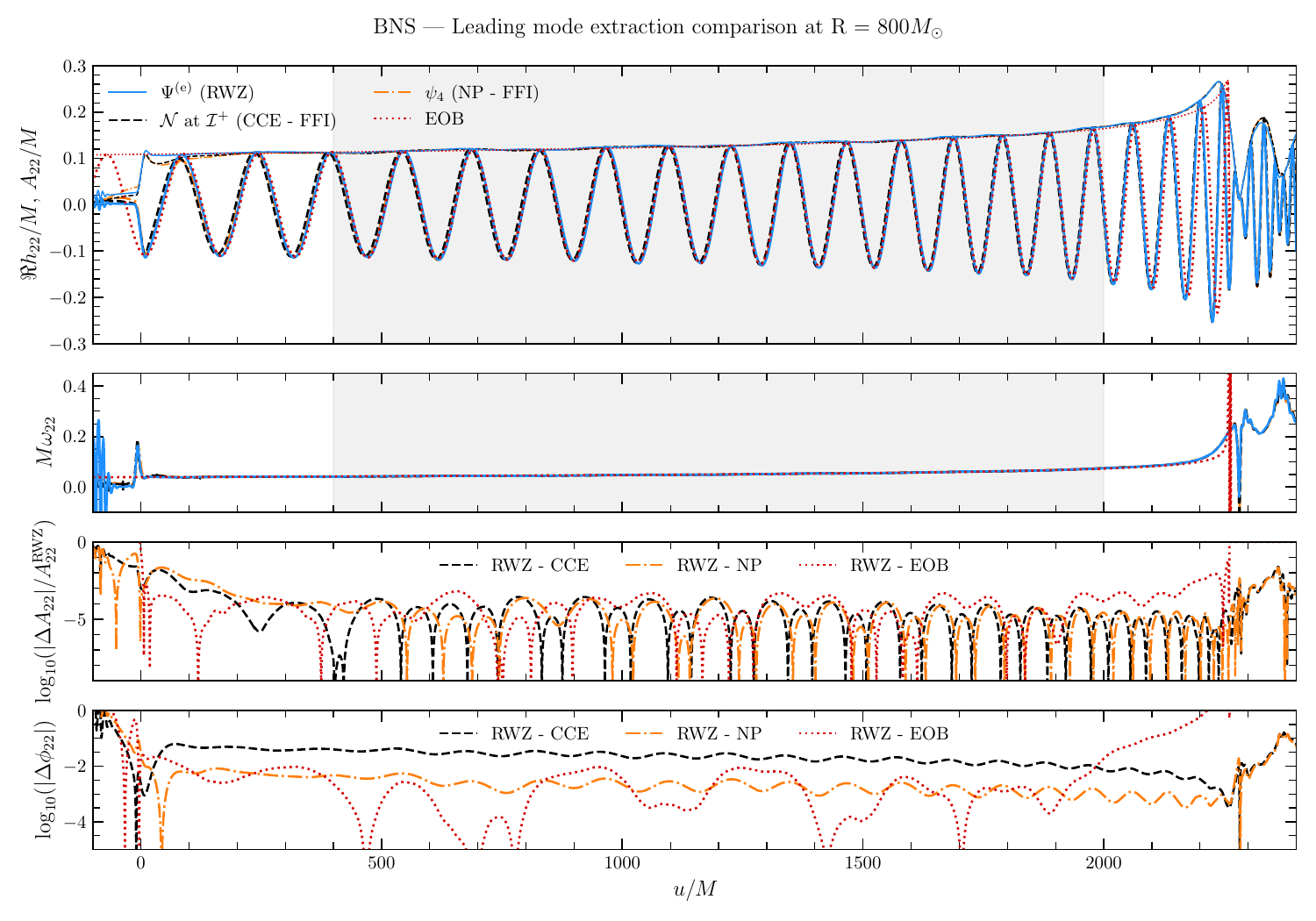}
  \caption{Strain $(2,2)$ mode for the circular BNS merger at $R=800\Mo$. 
    Different lines compare the results of the different extraction algorithms: RWZ (solid blue), 
    CCE (dashed black) and NP (dotted-dashed orange). Also shown is 
    a longer EOB waveform (dotted red) aligned with the RWZ; the shaded
    area marks the alignment window. From top to bottom: amplitude
    and real part, frequency, amplitude differences and phase
    differences with respect to the RWZ waveform.}
  \label{fig:bns:qc_leading}
\end{figure*}

The dominant $(2,2)$ mode of the GW strain is shown in
Fig.~\ref{fig:bns:qc_leading} for the three extraction
algorithms. Waveforms are also compared to a EOB waveform generated
with \dali{}. The RWZ extraction ($\Psi^{\rm (e)}_{22}$) correctly
captures the well known \emph{inspiral-merger-postmerger} structure of
the waveforms and it is entirely compatible with other extraction
algorithms. Relative amplitude differences ${\lesssim}10^{-5}$ which is
well below the uncertainty due to truncation errors. Similarly, phase
differences at finite extraction radii are of order
${\lesssim}10^{-2}$~rad and compatible with the EOB accuracy. CCE data
at $\scri$ highlight some significant dephasing with finite extraction
data, thus suggesting the need of extrapolation in order to obtain
precise data for \eg~EOB modeling~\cite{Bernuzzi:2014owa}.
Similarly to other problems, a simple extrapolation of the RWZ
multipole significantly improves the agreement with CCE data, see
Appendix~\ref{app:rextrap}. 

\begin{figure*}[t]
  \centering
  \includegraphics[width=\textwidth]{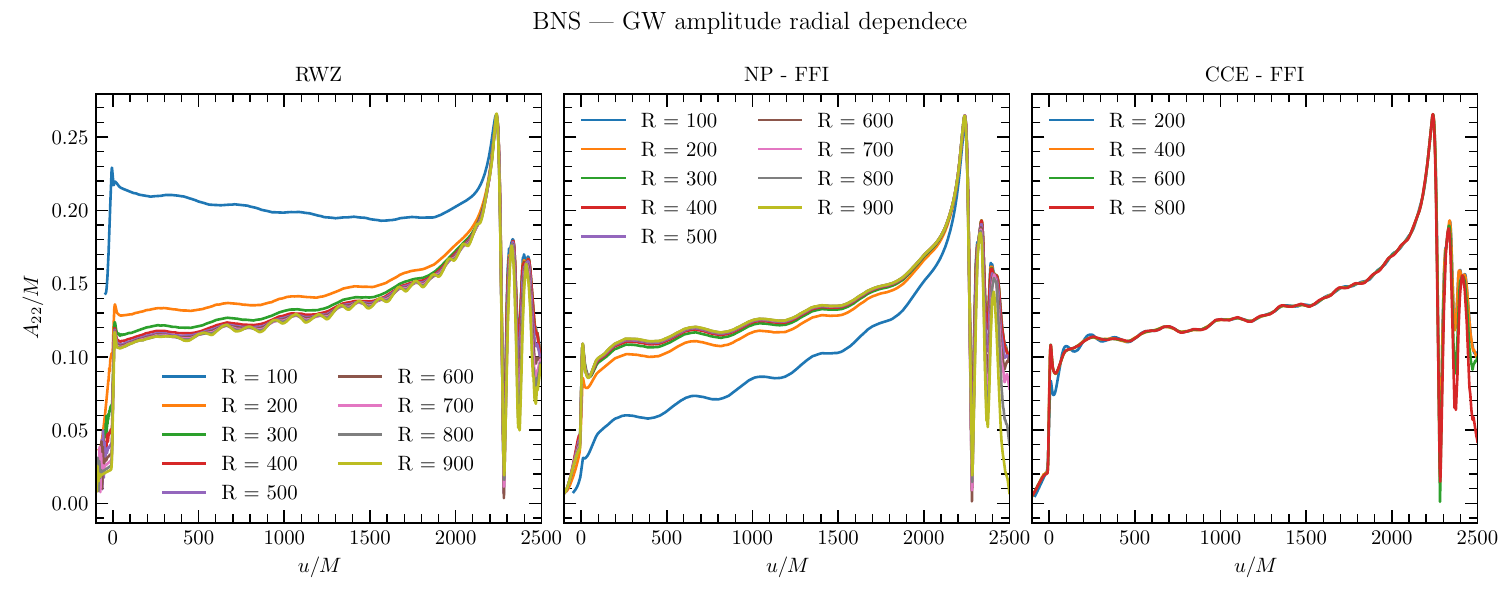}
  \caption{Extraction radius dependence of the strain $(2,2)$ mode for
    the circular BNS merger.
    Panels from left to right:
    RWZ, NP and CCE extraction. For each panel, different lines show
    different extraction radii (in units of $M_\odot$).}
  \label{fig:bns:amplitude}
\end{figure*}

Finite-radius extraction effects in the waveform's amplitude are
summarized in Fig.~\ref{fig:bns:amplitude}. The RWZ and the NP
amplitudes become consistent at radii $R>200\Mo$. Comparing to BBH
simulations, finite-radius effects in the RWZ are less
severe since in this problem it is possible to reliably extract
waveforms at larger radii. Nonetheless, RWZ data at $R=900\Mo$ start to
show unphysical oscillations due to insufficient resolution in the wave zone.

\begin{figure*}[t]
  \centering
  \includegraphics[width=\textwidth]{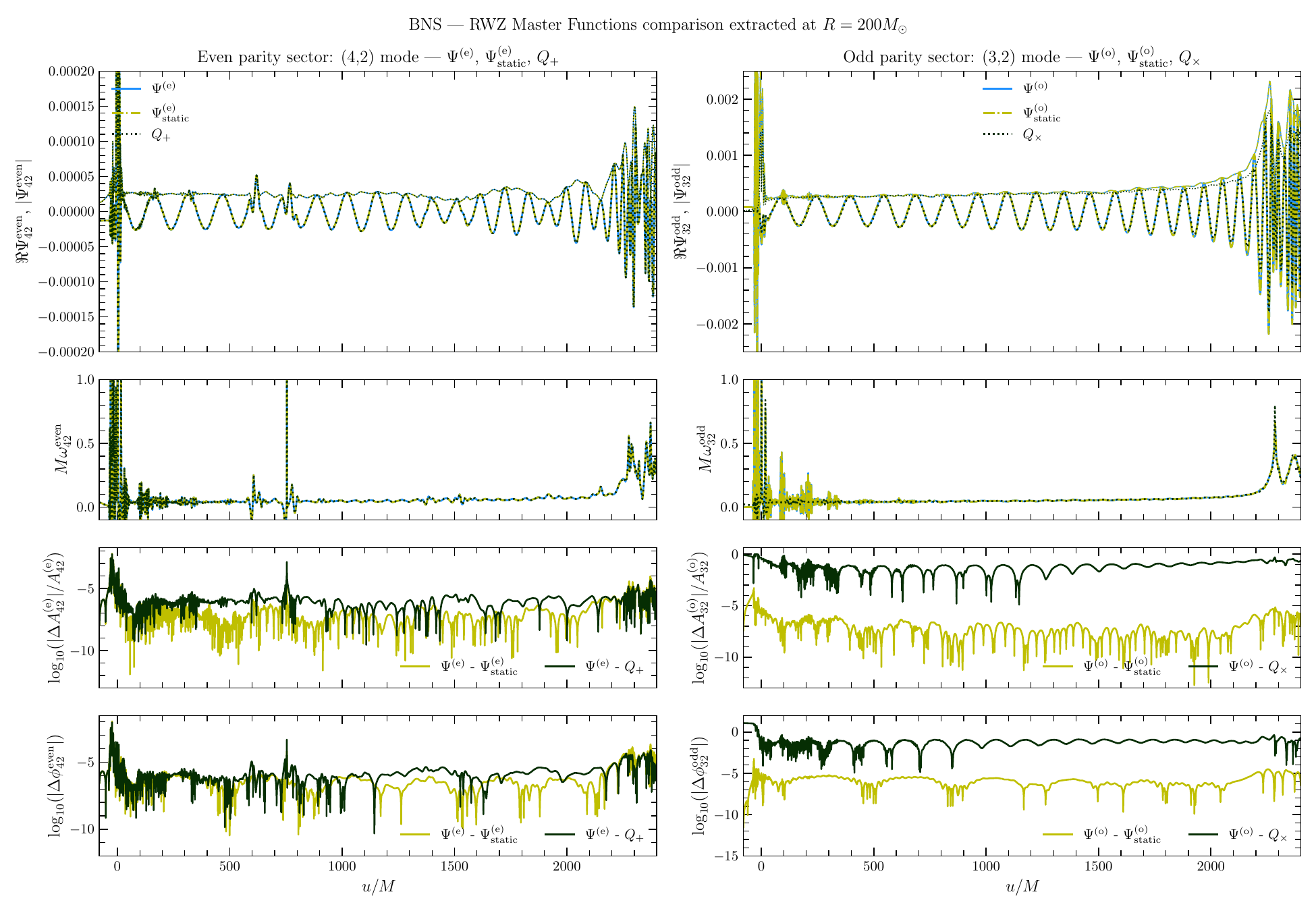}
  \caption{RWZ even- and odd-parity modes for the BNS circular merger
    at $R=800M_\odot$ and computed using different RWZ master
    functions.
    Left:
    Even-parity $(4,2)$ master function extraction comparison $\Psi^{\rm (e)}$ 
    (solid blue), $\Psi_{\rm static}^{\rm (e)}$ (dash-dotted light green) and 
    $Q_{+}$ (dotted dark green) normalized as per
    Eq.~\eqref{eq:Q+:norm_psi}.
    Right:
    Odd-parity $(3,2)$ master function extraction comparison $\Psi^{\rm (o)}$ 
    (solid blue), $\Psi_{\rm static}^{\rm (o)}$ (dash-dotted light
    green) and
     $Q_\times$ (dotted dark green) integrated and normalized as per
    Eq.~\eqref{eq:Qx:norm_psi}.
    Panels in each columns show from top to bottom:
    real part and amplitude, frequency, amplitude differences and
    phase differences.}
  \label{fig:bns:master_funcs}
\end{figure*}

Finally, we compare the different choices for the RWZ master functions.
For the $(2,2)$ mode we find that $\Psi^{\rm (e)}_{22}$, $\Psi^{\rm
  (e)}_{{\rm static}\, 22}$ and $Q_{+\,22}$ are essentially equivalent
at the optimal radius $R=800\Mo$. Relative amplitude (phase)
differences are of order ${\lesssim}10^{-6}$ ($10^{-5}$~rad).
Fig.~\ref{fig:bns:master_funcs} shows the $(4,2)$  
and $(3,2)$ modes computed with the difference master function
choices at $R=200\Mo$. Results are in-line with those discussed above for BBH, but
here the amplitude (relative) and phase differences among the master
functions are an order of magnitude smaller. Moreover, we stress again that with the $\Gamma$-driver shift
odd-parity modes are more robustly obtained from the $Q_\times$
(Schwarzschild coordinates) than from $\Psi^{\rm (o)}$.

\section{Conclusion}
\label{sec:con}

In this paper we have revisted the problem of gravitational-waves
extraction in numerical relativity using the covariant and
gauge-invariant formalism of spherical spacetime perturbations.

The first part of the paper summarizes the formalism introduced by
\citet{Moncrief:1974am,Moncrief:1974bis} and \citet{Gerlach:1979rw}
and further developed by other
authors~\cite{Gundlach:1999bt,MartinGarcia:2000ze,Sarbach:2001qq,Martel:2005ir}.
Ready-to-use expressions for the most general covariant and
gauge-invariant even- and odd-oparity master functions are reported
for the first time in terms of the (3+1) metric
multipoles. 
The presentation also connects the different
conventions and notations employed in the literature; the hope is to
provide the reader a consistent and complete summary of results that
appeared over several decades.
Technical aspects of our new implementation of the extraction
algorithm in the (3+1) numerical relativity 
code \athena{GR-} are also detailed.

The central results are given by Eq.~\eqref{eq:psi:even} and
Eq.~\eqref{eq:psi:odd} which are the even- and odd-parity
master functions for the most general choice of the spherical
background. These functions directly provide the multipoles of the
strain, Eq.~\eqref{eq:gw:rwz}. They reduce to
Eq.~\eqref{eq:psi:even:stat} and Eq.~\eqref{eq:psi:odd:stat} in case
of a static background and to \citet{Moncrief:1974am,Moncrief:1974bis}
variables via 
Eq.~\eqref{eq:Q+:norm_psi} and Eq.~\eqref{eq:Qx:norm_psi}
for a background in Schwarzschild coordinates.

The second part of the paper discusses a systematic assessment of the
metric extraction algorithm based on a comprehensive suite of
(3+1) simulations. The benchmark problems are perturbed TOV stars,
gravitational collapse of a rapidly rotating neutron star, a BBH
circular merger and a dynamical capture, and a ten-orbit BNS 
circular merger. For most of these 3D simulations, it is the first time
that a covariant and gauge-invariant metric extration algorithm is
considered and compared to curvature and CCE waveforms. Also,
to the best of our knowledge, it is the first time that TOV $w$-modes
are explored in 3D complete simulations without symmetries,
\cf~\cite{Stergioulas:2006}.  

Our numerical experiments demonstrate common features of the metric
extraction algorithm across the different benchmarks.

First, we find that metric extraction is robust in all the considered
scenarios. Metric waveforms extracted are of
comparable quality to curvature waveforms extracted at the same finite
radius. 
Metric extraction is particularly valuable
in identifying waveform systematics for problems in which the
reconstruction of the strain from the $\psi_4$ multipoles is
ambiguous. Notable examples are the rotational collapse and
the dynamical capture.
The main drawback observed in the metric extraction when compared to curvature
extraction is a slower convergence with the extraction and a slightly 
higher numerical noise. The latter is more evident in the computation of higher modes.

Second, we provide a direct comparison of the different choices for
the gauge-invariant master functions.
In the even-parity sector, any choice of the master function is robust
for all the problem. Differences in both phase and amplitude of the
RWZ multipoles are significantly below the uncertanties introduced by
the choice of extraction radius (and resolution) and generically smaller
than differences with other extraction algorithms.
In the odd-parity sector, we instead find that the gauge choice in our
simulations, in particular the $\Gamma$-driver shift, introduces some
significant systematic effect for simulations that require a
nontrivial shift (notably compact binaries.) As a consequence, 
assuming Schwarzschild coordinates for the background, \ie~considering
the $Q_\times$ multipoles, can be a better choice for such simulations.
This is an unexpected result but it can be understood by noting that
the gauge-dependent odd multipole $H_{0}^{(\lm)}$ (a projection of
the shift components) enters directly Eq.~\eqref{eq:psi:odd} and
Eq.~\eqref{eq:psi:odd:stat}, but not Eq.~\eqref{eq:Qx}.
We note this is not in contradition with the findings of
\citet{Pazos:2006kz}, since that work considered perturbations of a
stationary spacetime and employed harmonic coordinates. 
Further, the different choices of the Schwazrschild radius are all
found robust as they introduce uncertainties typically smaller than
the choice of finite extraction radius. The average Schwarzschild
radius can be used as default due to its general character.

Third, we provide a direct comparison to CCE waveforms which are
propagated to null infinity from finite radius worldtubes.
Compared to ``optimal'' extraction spheres, null infinity waveforms
show some significant phase differences, in particular for compact
binary simulations. Such differences can become a dominant source of
uncertainty if higher mesh resolutions than those considered in our
benchmnarks are employed. However, CCE waveforms also have
uncertainties associated to the choice of the
worldtube~\cite{Reisswig:2009rx,Moxon:2020gha,Rashti:2024yoc}. 
We find that a simple extrapolation to null infinity can deliver
rather precise metric waveforms compatible in phase to CCE.
These extrapolated waveforms are particularly accurate for problems
that allow extraction at sufficiently large radii as, for example, the
considered BNS merger simulation. 
A companion paper explores yet another technique to propagate 
(metric) waveforms to null infinity using perturbative simulations
with hyperboloidal foliation. That technique is naturally associated
to spherical spacetime perturbations and it employs the RWZ data
extracted at a coordinate sphere as boundary data to propagate to null
infinity.

\begin{acknowledgments}
  JF, SB, BD acknowledge support by the EU Horizon under ERC
  Consolidator Grant, no. InspiReM-101043372. 
  SA and SB acknowledge support from the Deutsche
  Forschungsgemeinschaft (DFG) project ``GROOVHY'' (BE 6301/5-1
  Projektnummer: 523180871). 
  DR and AR were supported by NASA under Awards No. 80NSSC21K1720 and
  80NSSC25K7213.

Simulations were performed on SuperMUC-NG at the Leibniz-Rechenzentrum
(LRZ) Munich and and on the national HPE Apollo Hawk at the High
Performance Computing Center Stuttgart (HLRS). 
The authors acknowledge the Gauss Centre for Supercomputing
e.V. (\url{www.gauss-centre.eu}) for funding this project by providing
computing time on the GCS Supercomputer SuperMUC-NG at LRZ
(allocations {\tt pn36go}, {\tt pn36jo} and {\tt pn68wi}). The authors
acknowledge HLRS for funding this project by providing access to the
supercomputer HPE Apollo Hawk under the grant number INTRHYGUE/44215
and MAGNETIST/44288. 
Computations were also performed on the ARA cluster at Friedrich
Schiller University Jena and on the {\tt Tullio} INFN cluster at INFN
Turin. The ARA cluster is funded in part by DFG grants INST 275/334-1
FUGG and INST 275/363-1 FUGG, and ERC Starting Grant, grant agreement
no. BinGraSp-714626. 
\end{acknowledgments}

\appendix

\section{Notation and conventions}
\label{app:not}

We use geometric units $c=G=1$ and signature $(-,+,+,+)$; partial
derivatives are indicated as $\partial$. 
A 4-metric is indicated explicitely as ${\gM}_{\mu\nu}$ with indeces 
$\mu,\nu=0,...,3$.
We employ the standard notation for the 3-metric
$\gamma_{ij}$ with indeces $i,j=1,2,3$, the shift vector $\beta^i$ and
the lapse function $\alpha$. The line element is 
\begin{equation}
ds^2=-\left(\alpha^2-\beta_i\beta^i\right)dt^2+2\beta_idt dx^i+\gamma_{ij}dx^i dx^j \ .
\end{equation}

\subsection{Schwarzschild background}
\label{app:not:metric}

Schwarzschild spacetime is given by the direct product of a
2-dimensional Lorentzian manifold $L^2$ and the 2-spheres, \ie~${\cal
  M}=L^2\times S^2$. This background spacetime is 
spherically symmetric, but the coordinates on $L^2$ are not specified.
The metric on $S^2$ is written as $\gs_{ab}$ with indeces
$a,b=2,3$. The covariant derivative is indicated as $\nabla_a$. The
volume form is indicated as $\epsilon_{cd}$ ($\epsilon_{cd}
\epsilon^{ce} = \gs^{e}_d$). On $S^2$ we use standard
latitude-longitude coordinates $(\theta,\phi)$ and thus $\gs_{ab}=
{\rm diag}(1,\sin^2\theta)$ and the non-zero components of the
Levi-Civita tensor are
$\epsilon_{\theta\phi}=\sin\theta=-\epsilon_{\phi\theta}$.
The background 2-metric on $L^2$ is written as $g_{AB}$ with indeces
$A,B=0,1$. The covariant derivative on $L^2$ is indicated as
$\nabla_A$. The (antisymmetric) volume form is
$\epsilon_{AB}=\sqrt{\det{|g_{CD}|}}\tilde{\epsilon}_{AB}$, where
$\tilde{\epsilon}_{AB}$ is the totally antisymmetric symbol and
$\epsilon^{AB}=(-\tilde{\epsilon}_{AB})/\sqrt{\det{|g_{CD}|}}\,$.
The generic coordinates on $L^2$ are written as $x^A$ where
$x^0$ is a time and $x^1$ is a radial coordinate that does not
need to be the areal radius.

The background 4-metric is written as 
\be
\gMb_{\mu\nu} = \begin{pmatrix}
  g_{AB} & 0 \\
  0 & r^2 \gs_{ab}\\
  \end{pmatrix}\,,
\ee
where $r$ is the areal radius defined from the $S^2$ volume,
\begin{equation}
  \label{eq:r2:def}
  4\pi r^2 = \int_{S^2} \sqrt{\det(\gs_{ab})} = \int_0^{2\pi}\int_0^\pi \sin\theta d \theta d\phi \ .
\end{equation}

The background spacetime for the wave-extraction algorithm is
characterized by the mass $M$ of the spacetime that can be defined
in coordinate-invariant way.
Following~\citet{Sarbach:2001qq} and~\citet{Martel:2005ir}, we 
introduce the 1-form 
\begin{equation}
dr = \de_A r\, dx^A = r_A dx^A \ ,
\end{equation}
where $r_A\equiv \de_A r$. The coordinate-invariant definition of the 
 mass of the system is 
\begin{equation}
M=\dfrac{r}{2}\left(1-N\right) \ ,
\end{equation}	
where $N:=r_A r^A$ with $r^A=g^{AB} r_A$. Note that our $N$
corresponds to \citet{Martel:2005ir} function $f$. Assuming for
example Schwarzschild coordinates $x^A=(t,r)$, where $r$ is the areal radius
of the sphere, one obtains $r_A=(0,1)$ and
$t^A=(1,0)$. The inverse metric $g^{AB}$ can be written in the frame defined by
$(t^A,r^A)$ as \begin{equation}
g^{AB}=N^{-1}\left(-t^A t^B+r^Ar^B\right) \ .
\end{equation}
As a result, the vector $r^A$ explicitly reads
\begin{equation}
r^A= \begin{pmatrix}
g^{01} \cr
 \cr
g^{11} \cr 
\end{pmatrix}\ ,
\end{equation}
and the invariant mass is 
\begin{equation}\label{eq:mass}
M=\dfrac{r}{2}\left(1-g^{11}\right) \ .
\end{equation}
This equation identifies in an coordinate-invariant way the component
$g^{11}$ of the metric,
\begin{equation}
g^{11} = 1-\dfrac{2M}{r} \ .
\end{equation}
This is not the case for the other components of the metric with upper indices that, 
in general, will depend on the chosen time coordinate.
The metric components with lower indices can depend on time too, 
in general coordinates $x^A$.
However, requiring that the metric 
satisfies Einstein equations in vacuum still implies that $M$ is time
independent~\cite{Sarbach:2001qq}.

\begin{figure*}[t]
  \centering
  \includegraphics[width=\textwidth]{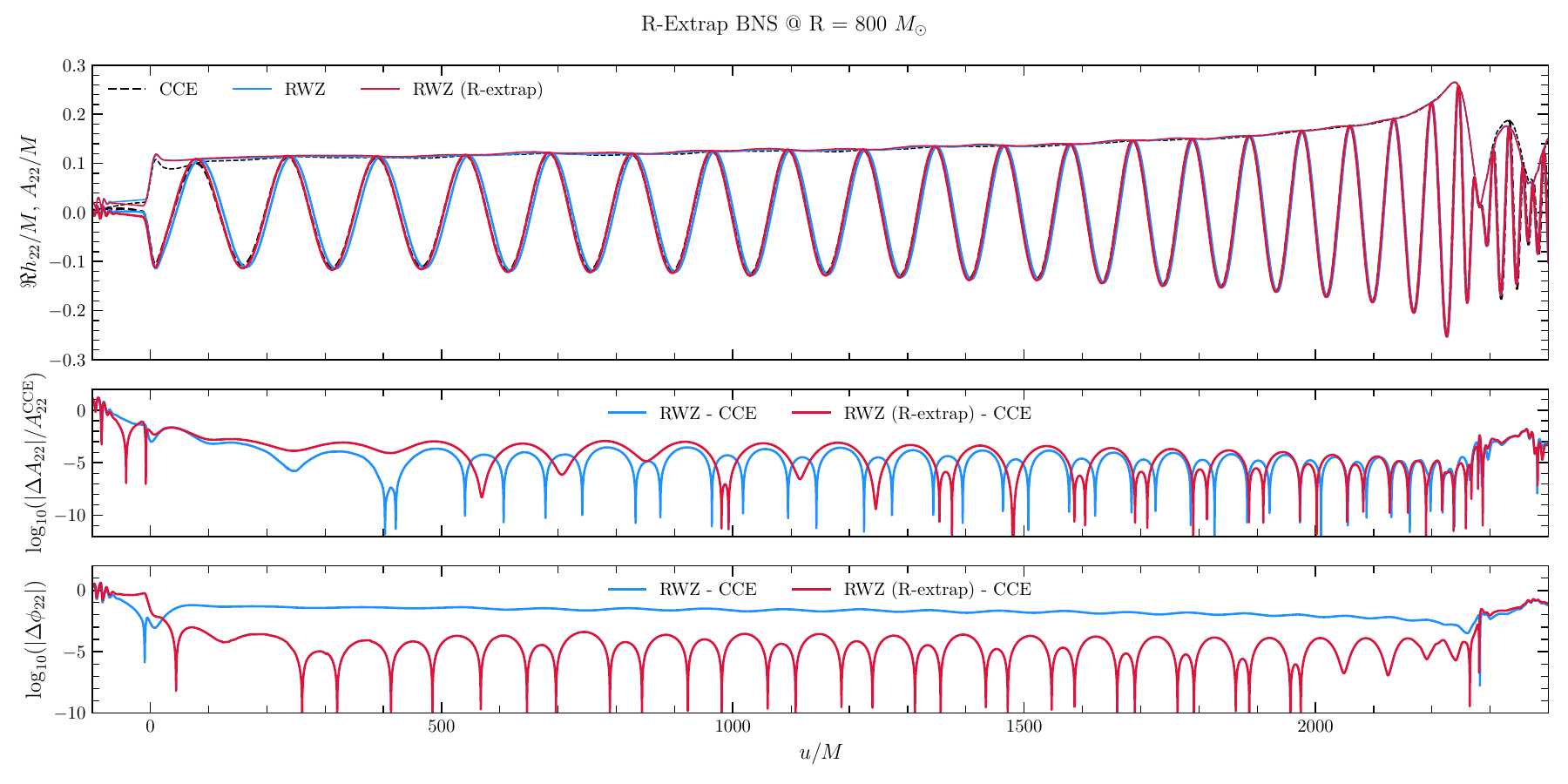}
  \caption{Strain $(2,2)$ mode for the circular BNS merger at $R=800\Mo$. 
    Different lines compare the results of the different extraction algorithms: RWZ (solid blue), RWZ R-extrapolated (solid red)
    and CCE (dashed black). From top to bottom: amplitude
    and real part, amplitude relative differences and phase
    differences with respect to the CCE waveform.}
  \label{fig:r-extrap_bns}
\end{figure*}
\begin{figure}[t]
  \centering
  \includegraphics[width=0.5\textwidth]{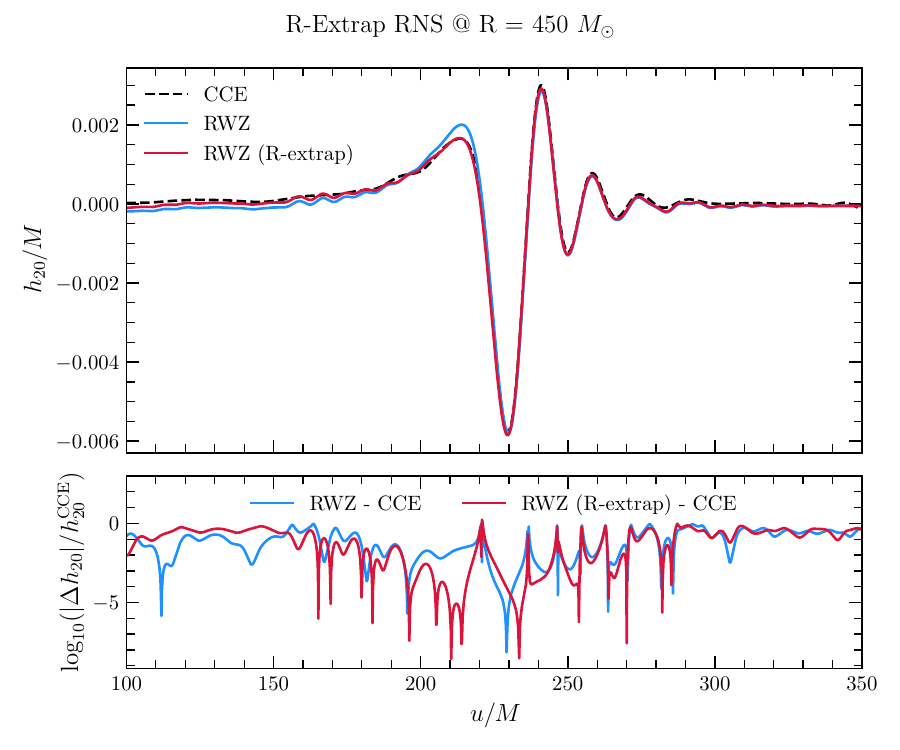}
  \caption{Strain $(2,0)$ mode for the RNS collapse at $R=450\Mo$. 
    Different lines compare the results of the different extraction algorithms: RWZ (solid blue), RWZ R-extrapolated (solid red)
    and CCE (dashed black). Top: amplitude
    and real part. Bottom: amplitude relative differences with respect to the CCE waveform.}
  \label{fig:r-extrap_rns}
\end{figure}

\subsection{Spherical Harmonics}
\label{app:not:ylm}

Scalar spherical harmonics are indicated as $Y_{\ell m}(\theta,\phi)$ and are defined by
\be\label{eq:boxY}
\gs^{ab}\nabla_a \nabla_b Y_\lm = - \lambda Y_\lm\ , 
\ee
where $\lambda=\lambda(\ell):=\ell(\ell +1)$. They are normalized such that
\begin{equation}
\int Y_{\ell m}Y_{\ell' m'}^*d\Omega = \delta_{m m'}\delta_{\ell \ell'}\ , 
\end{equation}
where $d\Omega=\sin\theta d\theta d\phi$, and so that $Y_{00}=1/\sqrt{4\pi}$.

Tensor fields on $S^2$ can be decomposed in a basis with definite parity. A 1-form on $S^2$ can be decomposed in vector spherical harmonics, 
\be
V_a = \sum_{\ell=1}^{\infty}\sum_{m=-\ell}^{\ell} \left(v_{\rm even}^{(\lm)} Y_a^{(\lm)}+v_{\rm odd}^{(\lm)} S_a^{(\lm)}\right)\ , 
\ee
where $v_{\rm even/odd}^{(\lm)}$ are the multipoles and the vector
spherical harmonics
\begin{subequations}
\label{eq:vspharm}
\begin{align}
  Y_a^{(\lm)} &:= \nabla_a Y_{\lm} \label{eq:vspharm:Y}\\
  S_a^{(\lm)} &:= \epsilon_{cd} \nabla_a Y_{\lm} = \epsilon_{ab} \gs^{bc}\nabla_c Y_{\lm} \label{eq:vspharm:S}
\end{align}
\end{subequations}
have even and odd parity, respectively.
The basis elements obey the orthogonality relations 
\begin{subequations}
\label{eq:vspharm:ortho}
\begin{align}
  \int \gs^{ab} Y_a^{(\lm)} Y_b^{(\ell'm') *}d\Omega  &= \lambda\, \delta_{\ell\ell'}\delta_{mm'}\\
  \int \gs^{ab} S_a^{(\lm)} S_b^{(\ell'm') *}d\Omega  &= \lambda\, \delta_{\ell\ell'}\delta_{mm'}\\
  \int \gs^{ab} Y_a^{(\lm)} S_b^{(\ell'm') *}d\Omega    &= 0 \ .
\end{align}
\end{subequations}
Using the $(\theta,\phi)$ coordinates on $S^2$, the above components explicitely read
\begin{subequations}
\label{eq:vspharm:coord}
\begin{align}
  Y_a^{(\lm)} &= (Y_\theta^{(\lm)},Y_\phi^{(\lm)}) = (\partial_\theta Y_{\lm},\partial_\phi Y_{\lm})\\
  S_a^{(\lm)} &= (S_\theta^{(\lm)},S_\phi^{(\lm)}) = (-\frac{1}{\sin\theta}\,\partial_\phi Y_{\lm},\sin\theta\,\partial_\theta Y_{\lm}) \ .
\end{align}
\end{subequations}
Consequently, the multipoles of a 1-form can be computed as 
\begin{subequations}
\label{eq:vspharm:mult}
\begin{align}
  v_{\rm even}^{(\lm)} = \frac{1}{\lambda}\int \left( %
     V_\theta \partial_\theta Y_{\lm}^{*} + \frac{1}{\sin^2\theta} V_\phi \partial_\phi Y_{\lm}^{*} \right) d\Omega\\
  v_{\rm odd}^{(\lm)} = \frac{1}{\lambda}\int \frac{1}{\sin\theta}  \left( %
     - V_\theta \partial_\phi Y_{\lm}^{*} + V_\phi \partial_\theta Y_{\lm}^{*}\right) d\Omega\ .
\end{align}
\end{subequations}
A 2-tensor $T_{ab}$ on $S^2$ can be decomposed in tensor spherical harmonics as
\be\label{eq:Tab_decomp}
T_{ab} = \sum_{\ell=2}^{\infty}\sum_{m=-\ell}^{\ell} \left(
K^{(\lm)} \gs_{ab} Y_{\lm} + G^{(\lm)} Y_{ab}^{\lm} + H^{(\lm)} S_{ab}^{(\lm)}\right)\ , 
\ee
where the multipoles $K^{(\lm)}$ and $G^{(\lm)}$ have even-parity and
$H^{(\lm)}$ has odd-parity.
These tensor multipoles are computed by similar projections as those
of vector multipoles. 
The tensor spherical harmonics basis of
even and odd parity are respectively~\cite{Mathews:1962,Zerilli:1970a}
\begin{subequations}
\label{eq:spharm:ten}
\begin{align}
  &Y_\lm \ \ \mbox{and} \ \  Y_{ab}^{(\lm)} := \nabla_a\nabla_b
  Y_{\lm} +
  \frac{\lambda}{2}\gs_{ab}Y_{\lm}  \label{eq:spharm:ten:even} \ , \\
  &S_{ab}^{(\lm)} := \nabla_{(a} S_{b)}^{(\lm)} = \half(\nabla_a S_b^{(\lm)} + \nabla_b S_a^{(\lm)}) \ .
\end{align}
\end{subequations}
Explicit expression in $(\theta,\phi)$ coordinates can be found in
\eg~Appendix~A of \citet{Martel:2005ir}. 
The basis elements obey the orthogonality relations 
\begin{subequations}
\label{eq:vspharm:ortho}
\begin{align}
  \int \gs^{ab} \gs^{cd} Y_{ab}^{(\lm)} Y_{cd}^{(\ell'm') *}d\Omega  &= \half\lambda(\lambda-2)\, \delta_{\ell\ell'}\delta_{mm'}\\
  \int \gs^{ab} \gs^{cd} S_{ab}^{(\lm)} S_{cd}^{(\ell'm') *}d\Omega  &= \half\lambda(\lambda-2)\, \delta_{\ell\ell'}\delta_{mm'}\\
    \int \gs^{ab} \gs^{cd} Y_{ab}^{(\lm)} S_{cd}^{(\ell'm') *}d\Omega  &= 0\ .
\end{align}
\end{subequations}
The $Y_{ab}^{(\lm)}$ basis introduced
in~\citet{Mathews:1962,Zerilli:1970a} is sometimes also indicated as $Z_{ab}^{(\lm)}$. 
The tensor spherical harmonics $Y_{ab}^{(lm)}$ and $S_{ab}^{(lm)}$ can
be related to spin-weighted spherical harmonics with spin $s=-2$
\cite{Goldberg:1966uu} and to the pure spin tensor harmonics
$T_{AB}^{{\rm E2}\,(\lm)},T_{AB}^{{\rm B2}\,(\lm)}$ of
\citet{Thorne:1980ru}, see Appendix~A of \citet{Martel:2005ir}. 
Note that \citet{Regge:1957td} used an alternative even-parity
tensor basis defined without the trace term in Eq.~\eqref{eq:spharm:ten:even}, 
\ie~$Y_{ab\, {\rm RW}}^{(\lm)} := \nabla_a\nabla_b Y_{\lm}$.
With this choice, however, the tensor harmonics are not linearly independent.

\section{Details on numerical implementation}
\label{app:num}

\subsection{Integrals on the sphere}
\label{app:num:integrals}

Integrals of functions $f(\theta,\phi)$ on the 2-spheres are computed
using Gauss-Legendre quadratures, 
\be
I_{S^2}[f] := \int f(\theta,\phi)\,d\Omega  \approx \sum_{ij} w_{ij} f(\theta_i,\phi_j)\ .
\ee
The grid in the longitude coordinate $\phi\in(0,2\pi)$ has nodes 
\be
\phi_j = \frac{2\pi}{N_\phi} \left(j+\half\right) \ ,
 \ \ \ j = 0,..., N_\phi-1 \ .
\ee
while the grid in the latitude coordinate $\theta\in(0,\pi)$
has nodes $\theta_i$ from the Gauss-Legendre nodes $x_i$
\be
\theta_i = \arccos(x_i) \ ,
 \ \ \ i = 0,..., N_\theta-1 \ .
\ee
The integral weights are computed as
\be
w_{ij} = \frac{2\pi}{N_\phi}  w_i \ ,
\ee
where $w_i$ are the Gauss-Legendre quadrature weights. Polynomials in
$x$ of degree $2N_\theta-1$ are integrated exactly by the scheme.

Numerical experiments with TOV spacetimes indicate that the background
computation requires at least $N_\theta=N_\phi/2=32$ to
obtain the Schwarzschild radius to 1\% precision. 
Nevertheless, Gauss-Legendre quadratures achieve better precision with
$N_\theta=N_\phi/2=128$ specially for more complex problems such as
binaries and higher modes.

\subsection{Metric and derivatives}
\label{app:num:drvts}

The (3+1)-metric and its spatial derivatives are computed on the
Cartesian grid and then interpolated on a sphere with given
isotropic radius $R$. First derivatives are calculated with centered
finite differencing stencils of the same order used in the evolution
algorithm ($n=4,6$) and interpolated with  
Lagrangian polynomials of order $2n-1$.
Second spatial derivatives are consistently
obtained from the interpolating polynomials. Metric's first time
derivatives are computed from the definition of the extrinsic
curvature and then interpolated.
Metric's mixed time-space derivatives are obtained
from the interpolating polynomial. For simplicity, we instead do not
implement terms proportional to second time derivatives. The choice is
justified a posteriori by the comparison $\Psi_{\rm static}$ and of
$\Psi$ that, already without these sub-dominant contributions, show
very small differences in all our simulations.

\section{$R$-extrapolated RWZ master functions}
\label{app:rextrap}

As mentioned in the above text, the quality of RWZ waveforms can be
improved by a simple extrapolation in $1/R$ and retarded time.
Our extrapolation procedure follows
Eq.~(4) of \citet{Nakano:2015pta} with a modification
analogous to their Eq.~(29), 
\begin{align}
  \label{eq:psi_rextrap}
  \Psi^{\rm (e,o)}_{\ell m}(u)\big|_{r=\infty} &= \left(1-\frac{2M}{r}\right)\bigg(\Psi^{\rm (e,o)}_{\ell m}(u,r) \nonumber \\
  & - \frac{\ell(\ell+1)}{2r} \int_0^u dt \Psi^{\rm (e,o)}_{\ell m}(t,r)\bigg) \ .
\end{align}
We show here selected results for two representative problems: 
the BNS merger and the rotational collapse.

Extrapolated waveforms for the BNS are shown in
Fig.~\ref{fig:r-extrap_bns}.
We obtain phase differences with respect to the CCE ${\sim} 2$ orders
of magnitude smaller than those obtained with finite radius extraction
With this extrapolation waveform extraction
uncertainties become negligible with respect to mesh resolution
uncertainties.

Fig.~\ref{fig:r-extrap_rns} shows extrapolated waveforms for the
collapse. Here, the main difference with respect to CCE is in the
precursor, \cf~Fig.~\ref{fig:rns:20}. The extrapolated RWZ signal
improves up to three orders of magnitude the amplitude relative differences. 

Analogous results are obtained for the $(2,0)$ waveform of the
pressure-perturbed TOV and for the BBH problem. In the latter case we
observe ${\sim} 1$ order of magnitude improvement in the agreement
between RWZ and CCE.

\end{document}